\documentclass[a4paper,fleqn,usenatbib,useAMS]{mnras}

\usepackage[T1]{fontenc}

\usepackage[english]{babel}
\usepackage{graphicx}
\usepackage{fleqn}
\usepackage{amssymb}
\usepackage{amsmath}
\usepackage{booktabs}
\usepackage{hyperref}
\renewcommand{\S}{Section}

\newcommand{\unit}[1]{\hat{\boldsymbol{#1}}}

\title[Secular dynamics of quadruple systems]{Secular dynamics of hierarchical quadruple systems: the case of a triple system orbited by a fourth body}
\author[Adrian S. Hamers, Hagai B. Perets, Fabio Antonini and Simon F. Portegies Zwart]{Adrian S. Hamers$^{1}$\thanks{E-mail: hamers@strw.leidenuniv.nl}, Hagai B. Perets$^{2}$, Fabio Antonini$^{3}$ and Simon F. Portegies Zwart$^{1}$\\
$^{1}$Leiden Observatory, Leiden University, PO Box 9513, NL-2300 RA Leiden, The Netherlands \\
$^{2}$Technion - Israel Institute of Technology, Haifa 32000, Israel \\
$^{3}$Center for Interdisciplinary Exploration and Research in Astrophysics (CIERA), and Department of Physics and Astronomy, Northwestern University, \\
2145 Sheridan Road, Evanston, IL 60208, USA}

\date{MNRAS 449, 1, 4221-4245 (2015) \\
Accepted 2015 February 27. Received 2014 February 25; in original form 2014 December 5}

\pubyear{2015}

\begin{document}
\label{firstpage}
\pagerange{\pageref{firstpage}--\pageref{lastpage}}
\maketitle

\begin{abstract}
We study the secular gravitational dynamics of quadruple systems consisting of a hierarchical triple system orbited by a fourth body. These systems can be decomposed into three binary systems with increasing semimajor axes, binaries A, B and C. The Hamiltonian of the system is expanded in ratios of the three binary separations, and orbit averaged. Subsequently, we numerically solve the equations of motion. We study highly hierarchical systems that are well described by the lowest order terms in the Hamiltonian. We find that the qualitative behaviour is determined by the ratio $\mathcal{R}_0$ of the initial Kozai-Lidov (KL) time-scales of the binary pairs AB and BC. If $\mathcal{R}_0\ll 1$, binaries AB remain coplanar if this is initially the case, and KL eccentricity oscillations in binary B are efficiently quenched. If $\mathcal{R}_0\gg 1$, binaries AB become inclined, even if initially coplanar. However, there are no induced KL eccentricity oscillations in binary A. Lastly, if $\mathcal{R}_0\sim 1$, complex KL eccentricity oscillations can occur in binary A that are coupled with the KL eccentricity oscillations in B. Even if binaries A and B are initially coplanar, the induced inclination can result in very high eccentricity oscillations in binary A. These extreme eccentricities could have significant implications for strong interactions such as tidal interactions, gravitational wave dissipation, and collisions and mergers of stars and compact objects. As an example, we apply our results to a planet+moon system orbiting a central star, which in turn is orbited by a distant and inclined stellar companion or planet, and to observed stellar quadruples. 
\end{abstract}


\section{Introduction}
\label{sect:introduction}
Hierarchical triple systems are known to be common among stellar systems. For example, a fraction of 0.076 of FG dwarfs systems in the catalogue of \citet{tokovinin_14a,tokovinin_14b} are triple systems (in the fractions cited here from \citealt{tokovinin_14b}, completeness arguments have been taken into account; the observed number of triple systems in the sample of \citealt{tokovinin_14a} is 290, with a total number of 4847 systems). The triple fraction is likely higher for more massive stars. In such hierarchical systems, the torque of the outer binary can induce high-amplitude oscillations in the inner binary over time-scales that can vary from suborbital time-scales, to time-scales exceeding Gyr. These oscillations, known as Kozai-Lidov (KL) cycles \citep{lidov_62,kozai_62}, have important implications for a large range of astrophysical systems, in particular when the effects of tidal friction are also considered. The implications include the production of short-period binaries and hot Jupiters \citep{eggleton_kiseleva-eggleton_01,wu_murray_03,eggleton_kiseleva-eggleton_06,fabrycky_tremaine_07,wu_murray_ramsahai_07,correia_ea_11,naoz_ea_11,naoz_farr_rasio_12,petrovich_14}, accelerating the merging of compact objects \citep{blaes_lee_socrates_02,thompson_11,antonini_perets_12,antonini_murray_mikkola_14}, explaining some of the blue stragglers stars \citep{perets_fabrycky_09,naoz_fabrycky_14}, affecting the formation of binary minor planets \citep{perets_naoz_09}, possibly producing a special type of type Ia supernovae through collisions of white dwarfs \citep{katz_dong_12,hamers_ea13,prodan_murray_thompson_13}, and modifying the evolution of stellar binaries that would not interact in the absence of a third star \citep{hamers_ea13}. 

Nature does not stop at $N=3$, however. Although in the catalogue of \citet{tokovinin_14a,tokovinin_14b} triple systems, with a fraction of $0.58$ (observed: 290 of 350), are most common among systems with hierarchies ($N\geq 3$), quadruple systems also constitute a considerable fraction of hierarchical systems, i.e. a fraction of $0.32$ (observed: 55 of 350). Unlike hierarchical triple systems, for which only one dynamically stable configuration is known to exist in nature, there are two different hierarchical configurations for which quadruples are known to be dynamically stable. One of these consists of two binary systems that orbit each other's barycentre, and this type of system constitutes a fraction of $0.74$ (observed: 37 of 55) of the quadruple systems in the catalogue of \citet{tokovinin_14a,tokovinin_14b}. The long-term dynamical evolution of this configuration has been studied by \citet{pejcha_antognini_shappee}, who showed, by means of direct $N$-body simulations, that eccentricity oscillations, in particular orbital flips, can be enhanced in these systems relative to triples.

\begin{figure}
\center
\includegraphics[scale = 0.4, trim = 5mm 0mm 0mm 0mm]{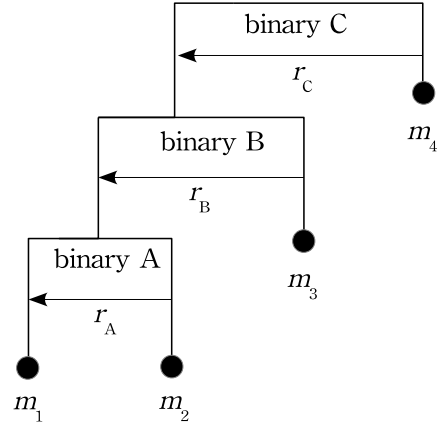}
\caption{\small A schematic depiction of the hierarchical configuration of the quadruple systems considered in this paper. }
\label{fig:hierarchy}
\end{figure}

The other configuration consists of a hierarchical triple system that is orbited by a fourth body (referred to as a {\it 3+1 quadruple} system in \citealt{tokovinin_14b}), and is the focus of this paper. In this case, three binary systems can be identified, and we will assume that they are each sufficiently separated from each other such that the quadruple system is dynamically stable. A stability analysis of these systems is beyond the scope of this work. Here, we shall always assume stability, although stability of some systems is borne out by our direct $N$-body integrations. We will refer to the binaries with the smallest, intermediate, and largest semimajor axes, as `binary A', `binary B' and `binary C', respectively. A schematic depiction of our configuration is shown in Fig. \ref{fig:hierarchy}. 

Our hierarchical configuration not only applies to stellar quadruples, but also arises in other astrophysical systems. These include, but are not limited to, multiplanet, planet-moon and binary asteroid systems in single and binary star systems. Here, we study the case of a planet+moon system (binary A) that orbits a star (binary B), which in turn is orbited by a more distant and inclined object (binary C), e.g. another planet or star. We assume that the orbit of the planet+moon system is initially coplanar with respect to that of the primary star. Therefore, in the absence of a distant body, no excitation of the eccentricity of the orbit of the planet+moon system is expected. However, we will show that, in the presence of an inclined fourth body, high-amplitude eccentricity oscillations can be induced in the planet+moon system through an intricate coupling of KL cycles. 

The structure of this paper is as follows. In \S\,\ref{sect:methods}, we describe our methods. We expand the four-body Hamiltonian in terms of the separation  ratios $r_\mathrm{A}/r_\mathrm{B}$, $r_\mathrm{B}/r_\mathrm{C}$ and $r_\mathrm{A}/r_\mathrm{C}$. In order for our method to be suitable for the study of the long-term evolution of a large number of systems, we adopt the secular approximation, i.e. we average the Hamiltonian over the three binary orbits assuming unperturbed and bound orbits for time-scales shorter than the orbital periods. Subsequently, we numerically solve the equations of motion derived from the orbit-averaged Hamiltonian. We test our method by comparing to direct $N$-body integrations. In \S\,\ref{sect:general}, we consider the general dynamics of highly hierarchical systems, i.e. systems that are well-described by the lowest-order terms in the Hamiltonian. We discuss our results in \S\,\ref{sect:discussion} and apply them to planetary and stellar systems. We give our conclusions in \S\,\ref{sect:conclusions}. 

\section{Methods}
\label{sect:methods}
\subsection{Expansion of the Hamiltonian}
\label{sect:methods:ham}
Our method to study the long-term evolution of quadruple systems is a natural extension to the orbit-averaged techniques that have been used extensively in the past to study the evolution of hierarchical triple systems, where an expansion was made in terms of the semimajor axis ratio $a_\mathrm{in}/a_\mathrm{out}$, with $a_\mathrm{in}$ and $a_\mathrm{out}$ the semimajor axes of the inner and outer orbit, respectively \citep{lidov_62,kozai_62,harrington_68,harrington_69,ford_kozinsky_rasio_00,eggleton_kiseleva-eggleton_01,laskar_boue_10,naoz_ea_13}. We note that hierarchical systems with more complex configurations have also been studied using secular methods by \citet{touma_ea_09} and \citet{boue_fabrycky_14}. We expand the Hamiltonian in terms of the separation ratios $r_\mathrm{A}/r_\mathrm{B}$, $r_\mathrm{B}/r_\mathrm{C}$ and $r_\mathrm{A}/r_\mathrm{C}$, where the separation vectors $\boldsymbol{r}_\mathrm{A}$, $\boldsymbol{r}_\mathrm{B}$ and $\boldsymbol{r}_\mathrm{C}$ are defined in terms of the position vectors of the four bodies in equation~(\ref{eq:app:rABC}). By assumption, $r_\mathrm{C}\gg r_\mathrm{B} \gg r_\mathrm{A}$; therefore, these ratios are small and such an expansion is appropriate. The expansion is carried out to up and including fourth order in the separation ratios, i.e. including terms proportional to $(r_\mathrm{A}/r_\mathrm{B})^i(r_\mathrm{B}/r_\mathrm{C})^j(r_\mathrm{A}/r_\mathrm{C})^k$, where $0\leq i+j+k\leq4$. The details are given in Appendix \ref{app:ham:circumstellar_triple}. For completeness, in addition to the configuration of a triple system orbited by a fourth body that is the focus of this paper, we have included results for the configuration of two binaries orbiting each other's barycentre in Appendix \ref{app:ham:binary_binary}.

As derived in Appendix \ref{app:ham:circumstellar_triple}, at the lowest order, $i+j+k=1$, the Hamiltonian consists of three terms that reduce to the binary binding energies of the three binaries A, B and C, assuming Kepler orbits. These terms therefore do not lead to secular orbital changes. At the next order, the `quadrupole' order ($i+j+k=2$)\footnote{The term `quadrupole' is not to be confused with `quadruple'.}, we find three terms, each of which is mathematically equivalent to the quadrupole-order Hamiltonian in the three-body problem. These three terms can also be obtained from the three-body quadrupole order Hamiltonian by appropriate substitutions of the masses and separation vectors. 

More specifically, the (non-averaged) {\it three}-body Hamiltonian at the quadrupole order is given by
\begin{align}
\label{eq:H_quad_three_body}
H_\mathrm{quad} = -\frac{Gm_1m_2m_3}{m_1+m_2} \frac{1}{r_\mathrm{out}} \left ( \frac{r_\mathrm{in}}{r_\mathrm{out}} \right )^2 \frac{1}{2} \left [ 3 \left(\unit{r}_\mathrm{in}\cdot \unit{r}_\mathrm{out} \right )^2 - 1 \right ],
\end{align}
where $\boldsymbol{r}_\mathrm{in}$ and $\boldsymbol{r}_\mathrm{out}$ are the separation vectors of the inner and outer binary, respectively. In our {\it four}-body system, the Hamiltonian, to the corresponding level of approximation, is given by three terms. These are each obtained from equation~(\ref{eq:H_quad_three_body}) by the following substitutions of separation vectors,
\begin{enumerate}
\item $\boldsymbol{r}_\mathrm{in} \rightarrow \boldsymbol{r}_\mathrm{A}$ and $\boldsymbol{r}_\mathrm{out} \rightarrow \boldsymbol{r}_\mathrm{B}$ (AB);
\item $\boldsymbol{r}_\mathrm{in} \rightarrow \boldsymbol{r}_\mathrm{B}$ and $\boldsymbol{r}_\mathrm{out} \rightarrow \boldsymbol{r}_\mathrm{C}$ (BC);
\item $\boldsymbol{r}_\mathrm{in} \rightarrow \boldsymbol{r}_\mathrm{A}$ and $\boldsymbol{r}_\mathrm{out} \rightarrow \boldsymbol{r}_\mathrm{C}$ (AC),
\end{enumerate}
and masses
\begin{enumerate}
\item (no substitutions) (AB);
\item $m_1\rightarrow m_1+m_2$, $m_2\rightarrow m_3$ and $m_3\rightarrow m_4$ (BC);
\item $m_3\rightarrow m_4$ (AC).
\end{enumerate}

In the quadrupole-order approximation, there are no terms appearing in the Hamiltonian that depend on {\it all} three position vectors $\boldsymbol{r}_\mathrm{A}$, $\boldsymbol{r}_\mathrm{B}$ and $\boldsymbol{r}_\mathrm{C}$. This is no longer the case for the next order, the `octupole' order ($i+j+k=3$). For the latter order, we find three terms that correspond to the octupole order terms in the three-body problem, and that can be obtained directly from the substitutions given above. In addition, we find a term that is a function of $\boldsymbol{r}_\mathrm{A}$, $\boldsymbol{r}_\mathrm{B}$ and $\boldsymbol{r}_\mathrm{C}$. We will refer to such terms as `cross terms'. The cross term at octupole order is given by
\begin{align}
\nonumber &H_\mathrm{oct,\,cross} = \frac{Gm_1m_2m_3m_4}{(m_1+m_2)(m_1+m_2+m_3)} \frac{1}{r_\mathrm{C}} \left (\frac{r_\mathrm{A}}{r_\mathrm{C}} \right )^2 \left (\frac{r_\mathrm{B}}{r_\mathrm{C}} \right ) \\
\nonumber &\quad \times \frac{1}{2} \left [15 \left ( \hat{\boldsymbol{r}}_\mathrm{B} \cdot \hat{\boldsymbol{r}}_\mathrm{C} \right ) \left ( \hat{\boldsymbol{r}}_\mathrm{A} \cdot \hat{\boldsymbol{r}}_\mathrm{C} \right )^2 - 3 \left( \hat{\boldsymbol{r}}_\mathrm{B} \cdot \hat{\boldsymbol{r}}_\mathrm{C} \right ) \right. \\
&\quad \quad \left. - 6\left (\hat{\boldsymbol{r}}_\mathrm{A} \cdot \hat{\boldsymbol{r}}_\mathrm{C}\right ) \left (\hat{\boldsymbol{r}}_\mathrm{A} \cdot \hat{\boldsymbol{r}}_\mathrm{B}\right ) \right ].
\label{eq:Hcross_no_av}
\end{align}

In the systems of interest here, the three terms in the Hamiltonian that can be obtained by the substitutions discussed above from the corresponding terms in the three-body problem, are generally dominated by the terms that apply to the binary combinations AB and BC. This is because, by assumption, $r_\mathrm{A}/r_\mathrm{B} \gg r_\mathrm{A}/r_\mathrm{C}$ and $r_\mathrm{B}/r_\mathrm{C} \gg r_\mathrm{A}/r_\mathrm{C}$. For the same reason, the octupole-order cross term, which is proportional to $(r_\mathrm{A}/r_\mathrm{C})^2(r_\mathrm{B}/r_\mathrm{C})$, is also typically small. However, in the three-body problem, the octupole-order term vanishes for equal masses in the inner binary (cf. equation~\ref{eq:app:H_oct_non_av}). This implies that the octupole-order terms associated with the binary combinations AB and BC vanish if $m_1=m_2$ and $m_1+m_2=m_3$, and suggests that the octupole-order cross term could be important in that case.

To investigate this further, we have also derived the terms of the next higher order, $i+j+k=4$ (henceforth `hexadecupole' order). Analogously to the lower orders, we find three terms that depend only on quantities of two of the binaries and that satisfy the substitutions given above. Their general form is given by
\begin{align}
\nonumber &H_\mathrm{hd} = -\frac{Gm m' m''  \left(m^2-m m'+m'^2\right)}{(m+m')^3} \frac{1}{r_\mathrm{out}} \left (\frac{r_\mathrm{in}}{r_\mathrm{out}} \right )^4 \\
&\quad \times \frac{1}{8} \left [ 35 \left (\hat{\boldsymbol{r}}_\mathrm{in}\cdot \hat{\boldsymbol{r}}_\mathrm{out} \right )^4 - 30  \left (\hat{\boldsymbol{r}}_\mathrm{in} \cdot \hat{\boldsymbol{r}}_\mathrm{out} \right )^2 + 3 \right ].
\label{eq:Hhd_no_av}
\end{align}
These terms do not cancel if the masses in the inner binary are equal; in fact, they do not cancel for any non-trivial combination of masses $m$ and $m'$. In addition to these terms, we find two terms that depend on quantities pertaining to all three binaries, i.e. two cross terms. Expressions for the latter terms are given in equations~(\ref{eq:app:H_hd_cross1_non_av}) and (\ref{eq:app:H_hd_cross2_non_av}). Although in this work, we do not include the hexadecupole-order terms in numerical integrations, we use our results of the hexadecupole-order Hamiltonian to evaluate the relative importance of the octupole-order cross term in \S\,\ref{sect:methods:cross_importance}.

\subsection{Orbit averaging}
\label{sect:methods:av}
We carried out an orbital averaging of the Hamiltonian expanded to up and including the hexadecupole order. For the cross terms, this entails averaging over three orbits. We assumed unperturbed Kepler orbits. 

A major advantage of the orbit-averaged approach compared to direct $N$-body integration, is the strongly reduced computational cost, in particular if the integration time is long compared to the orbital periods, and if a large number of systems is to be integrated. Furthermore, the orbit-averaged approach is a key instrument for the (semi)analytic understanding of the long-term behaviour (i.e. much longer than the orbital periods), as demonstrated e.g. below in \S\,\ref{sect:general:dep_R0:sa}. 

The main disadvantage is that the dynamics on suborbital time-scales are averaged over, therefore potentially missing important effects \citep{antonini_perets_12,antonini_murray_mikkola_14,antognini_shappee_thompson_14}. These effects can particularly be important in systems that are close to the limit of dynamical stability. However, for highly hierarchical systems, we do not expect these effects to be important, and these systems are the main focus of this work. In our numerical integrations, we check for the condition when the orbit-averaged approach likely breaks down (cf. \S\,\ref{sect:methods:EOM}). 

In the orbit-averaging procedure, we express the angular momenta and orientations of each of the three binaries in terms of the triad of perpendicular orbital state vectors $(\boldsymbol{j}_k,\boldsymbol{e}_k,\boldsymbol{q}_k)$, where $\boldsymbol{q}_k \equiv \boldsymbol{j}_k \times \boldsymbol{e}_k$ and $k \in\{\mathrm{A},\mathrm{B},\mathrm{C}\}$. Here, $\boldsymbol{j}_k$ is a vector aligned with the angular momentum vector of the orbit and which has magnitude $j_k=\sqrt{1-e_k^2}$; $\boldsymbol{e}_k$ is the eccentricity, or Laplace-Runge-Lenz vector, that is aligned with the major axis and which has magnitude $e_k$, the orbital eccentricity.

The orbit-averaged Hamiltonian is given in equation~(\ref{eq:app:H_av}). For further details, we refer to Appendix \ref{app:ham:circumstellar_triple}.  

\subsection{Equations of motion and numerical algorithm}
\label{sect:methods:EOM}
The equations of motion for the orbital vectors $\boldsymbol{j}_k$ and $\boldsymbol{e}_k$ of the three binary orbits are obtained by taking gradients of the orbit-averaged Hamiltonian $\overline{H}$ (\citealt{milankovitch_39}, see also e.g. \citealt{musen_61,allan_ward_63,allan_cook_64,breiter_ratajzcak_05,tremaine_touma_namouni_09}; see \citealt{rosengren_scheeres_14} for a recent overview),
\begin{subequations}
\label{eq:EOM}
\begin{align}
\label{eq:EOM:j}
\frac{\mathrm{d} \boldsymbol{j}_k}{\mathrm{d} t} &= -\frac{1}{\Lambda_k} \left [ \, \boldsymbol{j}_k \times \nabla_{\boldsymbol{j}_k} \overline{H} + \boldsymbol{e}_k \times \nabla_{\boldsymbol{e}_k} \overline{H} \, \right ]; \\
\label{eq:EOM:e}
\frac{\mathrm{d} \boldsymbol{e}_k}{\mathrm{d} t} &= -\frac{1}{\Lambda_k} \left [ \, \boldsymbol{e}_k \times \nabla_{\boldsymbol{j}_k} \overline{H} + \boldsymbol{j}_k \times \nabla_{\boldsymbol{e}_k} \overline{H} \, \right ].
\end{align}
\end{subequations}
Here, $\Lambda_k = mm'\sqrt{Ga_k/(m+m')}$, with $(m,m') = (m_1,m_2)$ for $k=\mathrm{A}$, $(m,m') = (m_1+m_2,m_3)$ for $k=\mathrm{B}$ and $(m,m') = (m_1+m_2+m_3,m_4)$ for $k=\mathrm{C}$. 

To solve the equations of motion, we have developed a code written in \textsc{C++}, \textsc{SecularQuadruple}, that numerically solves the system of ordinary differential equations (ODEs) equations~(\ref{eq:EOM}), to up and including octupole order. Because the ODEs are generally highly stiff, we used \textsc{CVODE} \citep{cohen_hindmarsh_96}, a library specifically designed to solve stiff ODEs. Our code is interfaced within the \textsc{AMUSE} framework \citep{pelupessy_ea_13,portegies_zwart_ea_13}. This allows for convenient comparison with direct $N$-body integration, i.e. without using the secular approximation, using any of the many $N$-body codes available in \textsc{AMUSE}. In addition, this facilitates the inclusion of effects modelled by other codes such as stellar and binary evolution. A test of the code for a hierarchical triple system is given in Appendix \ref{app:three_body_test}. 

In the integrations with \textsc{SecularQuadruple} below, we included terms up and including octupole order, but without the octupole order cross terms. Here, we consider highly hierarchical systems, and it is shown in \S\,\ref{sect:methods:cross_importance} that for these systems the octupole cross term does not dominate. Furthermore, neglect of this term is justified by the agreement with the $N$-body simulations, as shown in \S\,\ref{sect:methods:comparisons}.

As mentioned above, situations can arise in which the orbit-averaged approximation breaks down. In particular, this can occur when the time-scale for changes of the angular momentum $j_k$ is smaller than the orbital time-scale \citep{antonini_perets_12,antonini_murray_mikkola_14}. In \textsc{SecularQuadruple}, it is checked whether, at any time in the integration, any of the three binaries A, B or C satisfy this condition. This is implemented by means of a root-finding procedure: the integration is stopped whenever $t_{j,k} \leq P_{\mathrm{orb},k}$, where $P_{\mathrm{orb},k}$ is the orbital period of binary $k$ and $t_{j,k}$ is the time-scale for the angular momentum of binary $k$ to change by order itself, i.e.
\begin{align}
t_{j,k} = \left | \frac{1}{j_k} \frac{\mathrm{d} j_k}{\mathrm{d} t} \right |^{-1} = \left | \frac{e_k}{1-e_k^2} \frac{\mathrm{d} e_k}{\mathrm{d} t} \right |^{-1}.
\label{eq:t_AM}
\end{align}

Although in \textsc{SecularQuadruple} the equations of motion are solved in terms of orbital vectors for numerical reasons, below we present our results in terms of the (generally easier to interpret) orbital elements $(e_k,i_k,\omega_k,\Omega_k)$, where $i_k$ is the orbital inclination, $\omega_k$ is the argument of pericentre and $\Omega_k$ is the longitude of the ascending node. The latter quantities are defined with respect to a fixed reference frame $(x,y,z)$, and related to the orbital vectors $(\unit{e}_k,\unit{j}_k)$ according to
\begin{align}
\nonumber \unit{e}_k &= \left [ \cos(\Omega_k) \cos(\omega_k) - \sin(\Omega_k) \sin(\omega_k) \cos(i_k) \right ] \unit{x} \\
\nonumber &\quad + \left [ \sin(\Omega_k) \cos(\omega_k) + \cos(\Omega_k) \sin(\omega_k) \cos(i_k) \right ] \unit{y} \\
\nonumber &\quad + \sin(\omega_k) \sin(i_k) \, \unit{z}; \\
\unit{j}_k &= \sin(\Omega_k) \sin(i_k) \, \unit{x} -\cos(\Omega_k) \sin(i_k) \, \unit{y} + \cos(i_k) \, \unit{z}.
\end{align}
In particular, $i_k$ is defined as the angle between $\unit{j}_k$ and the $z$-axis of the fixed reference frame. It is often useful to consider {\it mutual} inclinations $i_{kl}$ between two orbits, rather than the {\it individual} inclinations $i_k$ and $i_l$. They are related according to
\begin{align}
\nonumber \cos(i_{kl}) &= \unit{j}_k \cdot \unit{j}_l \\
&= \cos(i_k) \cos(i_l) + \sin(i_k) \sin(i_l) \cos(\Omega_k-\Omega_l).
\label{eq:i_kl}
\end{align}
We note that in the hierarchical three-body problem, it is customary to define the orbital elements with respect to the {\it invariable plane}, i.e. a plane containing the total angular momentum vector (e.g. \citealt{naoz_ea_13}). This implies $\Omega_k-\Omega_l = \pi$, and therefore the simple relation $i_{kl} = i_k+i_l$ can be applied. This is not the case here, where the $z$-axis of our frame of reference is not parallel to the total angular momentum. Therefore, one must resort to the more general equation~(\ref{eq:i_kl}).

Relativistic effects are also implemented in our algorithm. An important effect is relativistic precession of the argument of pericentre, associated with the Schwarzschild metric \citep{schwarzschild_16}. The associated time-scale for precession by $2\pi$ in binary $k$ to the lowest post-Newtonian (PN) order is given by
\begin{align}
\label{eq:t_1PN}
t_{\mathrm{1PN},k} = \frac{1}{3} P_{\mathrm{orb},k} \left(1-e_k^2 \right) \frac{a_k}{r_{\mathrm{g},k}},
\end{align}
where $r_{\mathrm{g},k} \equiv Gm_{\mathrm{tot},k}$, with $m_\mathrm{tot,A}=m_1+m_2$, $m_\mathrm{tot,B}=m_1+m_2+m_3$ and $m_\mathrm{tot,C}=m_1+m_2+m_3+m_4$, is the gravitational radius. To take into account relativistic precession, the terms
\begin{align}
\label{eq:EOM_1PN}
\left. \frac{\mathrm{d} \boldsymbol{e}_k}{\mathrm{d} t} \right|_\mathrm{1PN} = e_k \frac{2\pi}{t_{\mathrm{1PN},k}} \, \unit{q}_k
\end{align}
are added to the right-hand sides in equation~(\ref{eq:EOM:e}). Here, we neglect any possible additional `interaction terms' between different binaries in the PN expansion that have been derived previously in the hierarchical three-body problem \citep{naoz_ea_13b,will_14a,will_14b}, and that could also apply, in some form, to the configuration considered here. 

\begin{figure}
\center
\includegraphics[scale = 0.45, trim = 5mm 0mm 0mm 0mm]{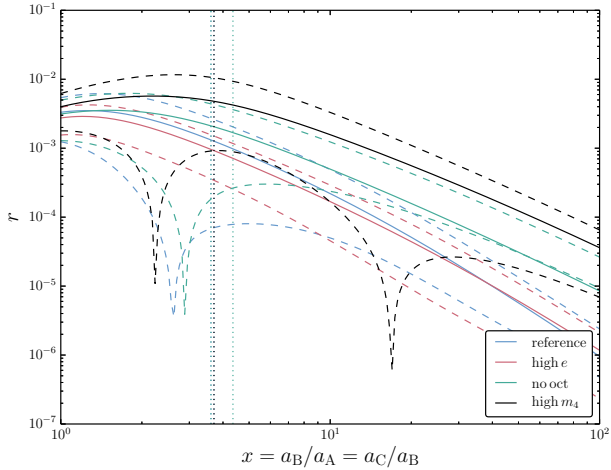}
\caption{\small The ratio $r$ of the absolute value of the orbit-averaged octupole-order cross term in the Hamiltonian to the absolute value of all other orbit-averaged terms at octupole and hexadecupole order (cf. equation~\ref{eq:rdef}), plotted as a function of $x = a_\mathrm{B}/a_\mathrm{A} = a_\mathrm{C}/a_\mathrm{B}$. An averaging over the orientations of the three binaries has been carried out (assuming random orbital orientations), and four different combinations of masses and eccentricities are assumed, which are given in Table \ref{table:cross_term_combinations}. Solid lines: mean values of $r$ for the different realizations of orbital orientations; (non-vertical) dashed lines: the same mean values, offset by the standard deviations (here, the absolute values are taken). Estimates for the minimum value of $x$ for dynamical stability (based on the \citealt{mardling_aarseth_01} criterion applied to the AB and BC binaries) for each combination of parameters are indicated with vertical dashed lines. Note that for some of the `high $m_4$' and `high $e$' combinations, these values are $>10^2$, and are therefore beyond the range of the figure. }
\label{fig:cross_term_importance}
\end{figure}

\begin{table}
\begin{tabular}{lccccccc}
\toprule
description & $m_1$ & $m_2$ & $m_3$ & $m_4$ & $e_\mathrm{A}$ & $e_\mathrm{B}$ & $e_\mathrm{C}$ \\
\midrule
reference & 2 & 1 & 1 & 1 & 0.1 & 0.1 & 0.1 \\
high $e$ & 2 & 1 & 1 & 1 & 0.99 & 0.99 & 0.99 \\
no oct & 1 & 1 & 2 & 1 & 0.1 & 0.1 & 0.1 \\
high $m_4$ & 2 & 1 & 1 & $10^6$ & 0.1 & 0.1 & 0.1 \\
\bottomrule
\end{tabular}
\caption{ Different combinations of the masses and eccentricities included in Fig. \ref{fig:cross_term_importance}. Note that $r$ depends on the masses only through their ratios, hence the mass unit is arbitrary. }
\label{table:cross_term_combinations}
\end{table}

\subsection{The importance of the octupole-order cross terms}
\label{sect:methods:cross_importance}
In \S\,\ref{sect:methods:ham} we derived a cross term in the Hamiltonian at octupole order. Here, we investigate further the importance of this term with respect to other terms at the octupole and the next higher order, the hexacupole order. Long-term effects of the cross term can only be investigated by carrying out numerical integrations in time. However, a proxy for the short-term importance of the cross term is the ratio $r$ of the absolute value of the orbit-averaged cross term, to the absolute value of all other orbit-averaged terms at octupole and hexadecupole order, i.e.
\begin{align}
\label{eq:rdef}
\nonumber & r \equiv \mathrm{abs} \left [ \, \overline{H}_\mathrm{oct,\,cross} \, \left ( \overline{H}_\mathrm{oct,AB} + \overline{H}_\mathrm{oct,BC} + \overline{H}_\mathrm{oct,AC} + \overline{H}_\mathrm{oct,\,cross} \right. \right. \\
&\left. \left. + \overline{H}_\mathrm{hd,AB} + \overline{H}_\mathrm{hd,BC} + \overline{H}_\mathrm{hd,AC} + \overline{H}_\mathrm{hd,\,cross,1} + \overline{H}_\mathrm{hd,\,cross,2} \right )^{-1} \right ].
\end{align}
Here, $\overline{H}_{\mathrm{oct},kl}$ and $\overline{H}_{\mathrm{hd},kl}$ denote the orbit-averaged octupole-order and hexadecupole order terms corresponding to pair $kl$, respectively. They can be obtained directly from the general expressions equations~(\ref{eq:app:H_av_oct}) and (\ref{eq:app:H_av_hd}), and using the substitutions discussed above in \S\,\ref{sect:methods:ham}. 

In principle, $r$ can be maximized with respect to the parameters defining the properties and state of the quadruple system, i.e. with respect to the four $m_i$ and the three $a_k$, $e_k$, $\unit{e}_k$ and $\unit{j}_k$ (with the orthogonality constraint $\unit{e}_k\cdot \unit{j}_k=0$). This would yield the largest possible contribution of the cross term. However, the dimensionality (25) of this problem is very large, and this makes it computationally very difficult to find the {\it absolute} maximum. Here, we simplify the problem by restricting the parameter space. 

In particular, we set $x \equiv a_\mathrm{B}/a_\mathrm{A} = a_\mathrm{C}/a_\mathrm{B}$, thereby reducing the dependence of the three semimajor axes to a single quantity. For given masses and eccentricities, we randomly sample the six unit vectors $\unit{e}_{k}$ and $\unit{j}_{k}$ with the orthogonality constraint $\unit{e}_k\cdot \unit{j}_k=0$. We compute $r$ for 20 of such realizations and each $x$, and subsequently, we compute the mean and standard deviations. 

In Fig. \ref{fig:cross_term_importance}, we show the resulting mean values (solid lines) and mean values offset by the standard deviations (dashed lines) of $r$ as a function of $x$. We include four different combinations of masses and eccentricities, which are enumerated in Table \ref{table:cross_term_combinations}. The minimum value of $x$ for dynamical stability of the system is estimated by computing the critical semimajor axis ratio for stability of the AB and BC systems separately using the criterion of \citet{mardling_aarseth_01}. The latter two ratios are indicated for each combination of parameters in Fig. \ref{fig:cross_term_importance} with vertical dashed lines. 

Regardless of our choice of parameters, $r$ is typically small, in the sense that for values of $x$ large enough for dynamical stability, $r \lesssim 10^{-2}$. For highly hierarchical systems, i.e. $x\gtrsim100$, $r \lesssim 10^{-4}$. This indicates that typically the cross terms do not dominate the dynamics, at least for the short-term evolution. We note, however, that the octupole-order cross terms could give rise to important dynamical effects on long time-scales in less hierarchical systems. 

\begin{figure*}
\center
\includegraphics[scale = 0.52, trim = 25mm 0mm 0mm 0mm]{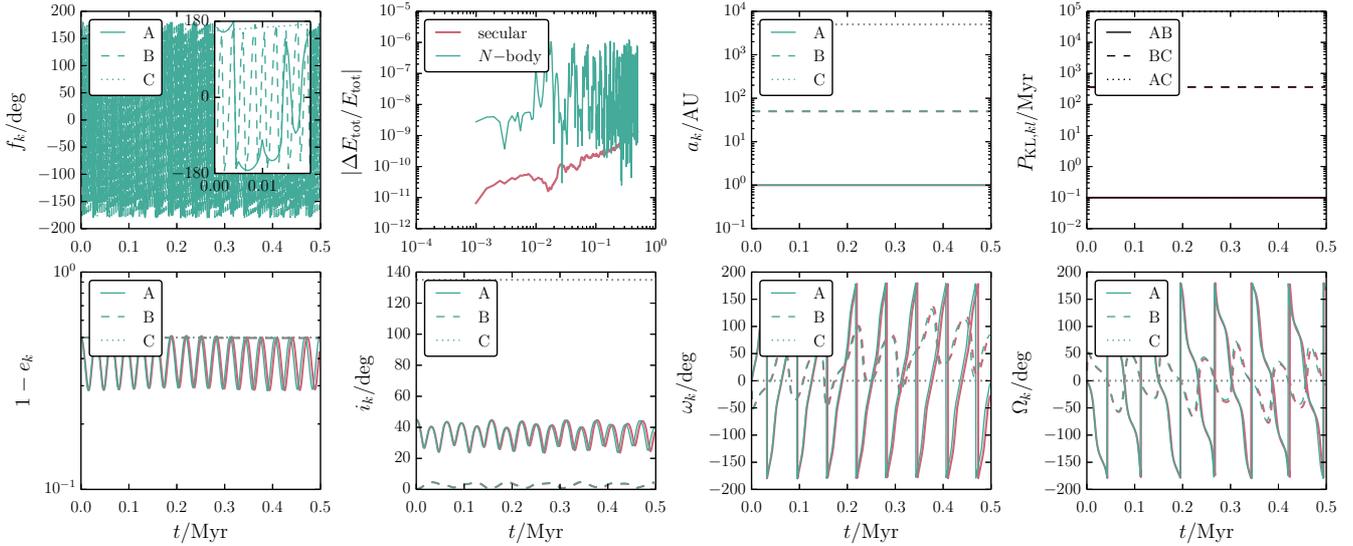}
\caption{\small Comparison between the evolution of a quadruple system as computed with the orbit-averaged code \textsc{SecularQuadruple} developed in this work (red lines) and the direct $N$-body code \textsc{Mikkola} (\citealt{mikkola_merritt_08}; green lines). The assumed initial parameters were semimajor axes $a_\mathrm{A}=1\,\mathrm{AU}$, $a_\mathrm{B}=5\times10^2\,\mathrm{AU}$ and $a_\mathrm{C}=5\times10^3\,\mathrm{AU}$, masses $m_1=m_3=m_4=1\,\mathrm{M}_\odot$ and $m_2=0.5\,\mathrm{M}_\odot$, eccentricities $e_\mathrm{A}=e_\mathrm{B}=e_\mathrm{C}=0.5$, inclinations $i_\mathrm{A} = 45^\circ$, $i_\mathrm{B}=0^\circ$, $i_\mathrm{C}=135^\circ$, arguments of pericentre $\omega_\mathrm{A}=\omega_\mathrm{B}=\omega_\mathrm{C}=0^\circ$ and longitudes of the ascending nodes $\Omega_\mathrm{A}=\Omega_\mathrm{B}=\Omega_\mathrm{C}=0^\circ$. When applicable to a single binary, solid, dashed and dotted curves correspond binaries A, B and C, respectively. When applicable to a binary pair, solid, dashed and dotted curves correspond to the binary pairs AB, BC and AC, respectively. The quantity $|\Delta E_\mathrm{tot}/E_\mathrm{tot}|$ is the absolute value of relative error in the total energy (the orbit-averaged Hamiltonian in the case of \textsc{SecularQuadruple}), and $f_k$ is the true anomaly (applicable only to the $N$-body simulations). The inset in the top-left panel shows a magnification between $t=0$ and 0.02 Myr. Note that the orbital period of binary A, $P_\mathrm{A}\approx 0.8\,\mathrm{yr}$, is too short compared to the output resolution ($\approx 500 \, \mathrm{yr}$) for $f_\mathrm{A}$ to be resolved. Also note that in the orbit-averaged code, the semimajor axes are constant by assumption, whereas the KL time-scales $P_{\mathrm{KL},kl}$ in principle depend on time through the time-dependence of $e_l$ (cf. equation~\ref{eq:PK}). However, in this case, the dependence is extremely weak and not visible in the top-right panel. }
\label{fig:comparison_short}
\end{figure*}

\begin{figure*}
\center
\includegraphics[scale = 0.52, trim = 25mm 0mm 0mm 0mm]{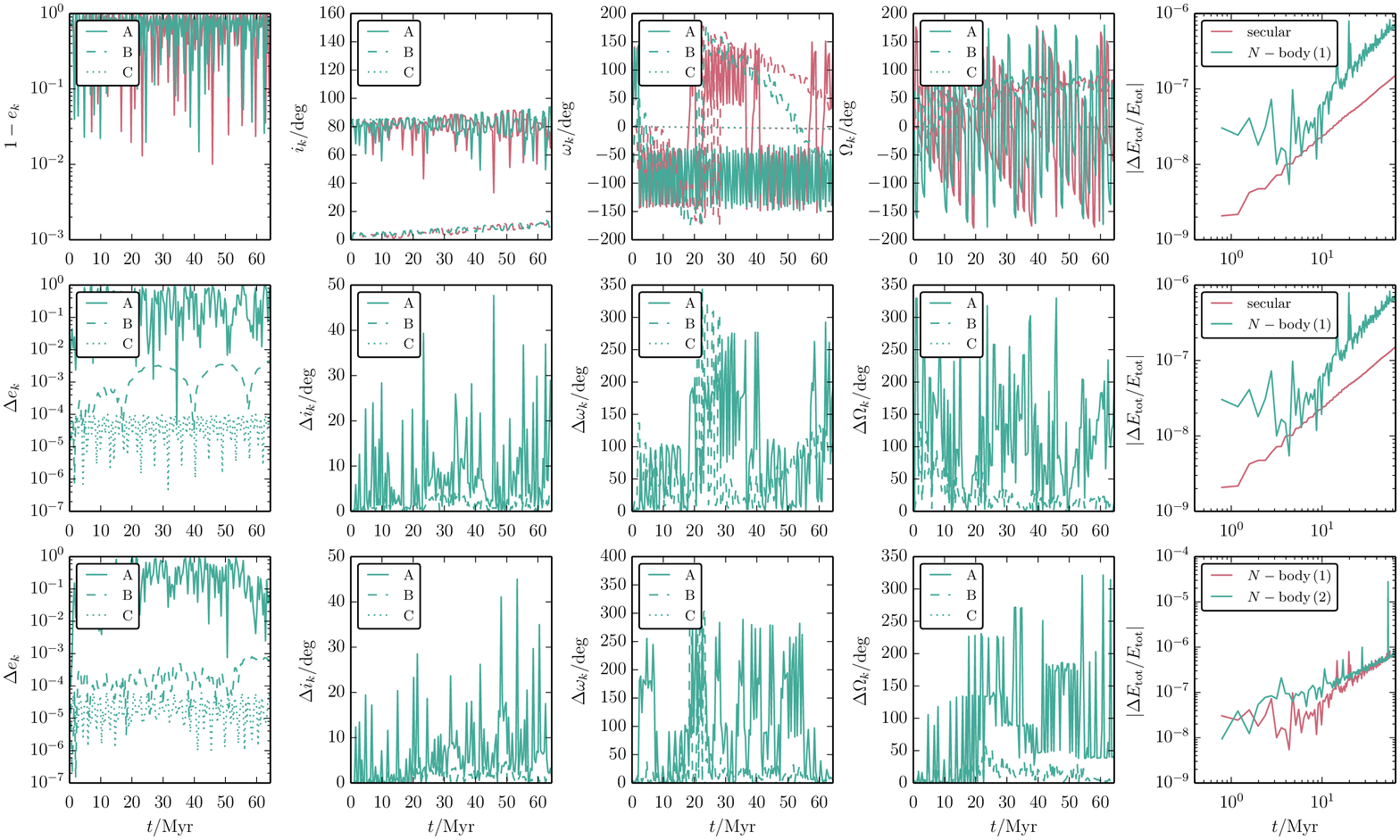}
\caption{\small Top row: comparison between the evolution as computed with the orbit-averaged code \textsc{SecularQuadruple} (red lines) and the direct $N$-body code \textsc{Mikkola} (\citealt{mikkola_merritt_08}; green lines). The system parameters are $a_\mathrm{A} = 1 \mathrm{AU}$, $a_\mathrm{B}=10^2\,\mathrm{AU}$ and $a_\mathrm{C}=5\times10^3\,\mathrm{AU}$, $m_1=m_3=m_4=1\,\mathrm{M}_\odot$ and $m_2=0.5\,\mathrm{M}_\odot$, $e_\mathrm{A}=e_\mathrm{B}=e_\mathrm{C}=0.01$, $i_\mathrm{A}=i_\mathrm{C}=85^\circ$ and $i_\mathrm{B}=0^\circ$, $\omega_\mathrm{A}=\omega_\mathrm{B}=\omega_\mathrm{C}=0^\circ$ and $\Omega_\mathrm{A}=\Omega_\mathrm{B}=\Omega_\mathrm{C}=0^\circ$ (the same as in panels 1-6 of Fig. \ref{fig:inclined:ex}).  Middle row: the differences in $e_k$, $i_k$, $\omega_k$ and $\Omega_k$ between the secular code and one realization of the $N$-body code, as a function of time. Bottom row: the differences in $e_k$, $i_k$, $\omega_k$ and $\Omega_k$ between two realizations with the $N$-body code with different initial orbital phases $f_k$. Relative energy errors are shown in the last column. }
\label{fig:comparison_long}
\end{figure*}

\begin{table*}
\begin{tabular}{ccccccccccccccccccc}
\toprule
& \multicolumn{6}{c}{$e_k$} & \multicolumn{6}{c}{$\omega_k$} & \multicolumn{6}{c}{$\Omega_k$} \\
& \multicolumn{2}{c}{A} & \multicolumn{2}{c}{B} & \multicolumn{2}{c}{C} & \multicolumn{2}{c}{A} & \multicolumn{2}{c}{B} & \multicolumn{2}{c}{C} & \multicolumn{2}{c}{A} & \multicolumn{2}{c}{B} & \multicolumn{2}{c}{C} \\
K-S pair & $D$ & $p$ & $D$ & $p$ & $D$ & $p$ & $D$ & $p$ & $D$ & $p$ & $D$ & $p$ & $D$ & $p$ & $D$ & $p$ & $D$ & $p$ \\
\midrule
SN1  &  0.1  &  0.38  &  0.16  &  0.03  &  0.94  &  0.0  &  0.38  &  0.0  &  0.19  &  0.01  &  0.07  &  0.0  &  0.04  &  1.0  &  0.06  &  0.9  &  0.02  &  1.0 \\
$\langle \mathrm{SN}\rangle$  &  0.08  &  0.65  &  0.18  &  0.02  &  0.73  &  0.0  &  0.34  &  0.11  &  0.17  &  0.02  &  0.09  &  0.58  &  0.05  &  0.95  &  0.07  &  0.85  &  0.02  &  1.0 \\
NN1 &  0.09  &  0.46  &  0.11  &  0.31  &  0.1  &  0.38  &  0.45  &  0.0  &  0.04  &  1.0  &  0.08  &  0.38  &  0.04  &  1.0  &  0.06  &  0.96  &  0.01  &  1.0 \\
$\langle \mathrm{NN}\rangle$ &  0.09  &  0.55  &  0.12  &  0.43  &  0.35  &  0.14  &  0.49  &  0.08  &  0.05  &  0.95  &  0.12  &  0.36  &  0.05  &  0.97  &  0.06  &  0.88  &  0.02  &  1.0 \\
\bottomrule
\end{tabular}
\caption{ Results of two-sided K-S tests (statistic $D$ and the $p$-value) for time series in $e_k$, $\omega_k$ and $\Omega_k$ for the integrations shown in Fig. \ref{fig:comparison_long}. In the first row, the secular code is compared to one realization of the $N$-body code, \textsc{Mikkola} \citep{mikkola_merritt_08}. In the second row, the secular code is compared to five realizations of the $N$-body code (i.e. with different initial mean anomalies), and given are the resulting values of $D$ and $p$ averaged over individual K-S tests. In the third row, two realizations of the $N$-body code are compared, and in the fourth row, K-S tests are carried out for all combinations of the five realizations of the $N$-body codes, and the quoted values of $D$ and $p$ are averaged over these combinations. }
\label{table:nbody_comparisons}
\end{table*}

\subsection{Comparisons to direct $N$-body integrations}
\label{sect:methods:comparisons}

As a first demonstration of our algorithm, we show in Fig. \ref{fig:comparison_short} a comparison of a short-term integration with \textsc{SecularQuadruple} (red lines) and \textsc{Mikkola} \citep{mikkola_merritt_08}, a highly accurate direct $N$-body code that uses chain regularization (green lines)\footnote{We remark that for this type of systems, it is essential to use a highly accurate $N$-body code because a large number of orbits, in particular in binary A, needs to be integrated very accurately.}. The assumed initial parameters were semimajor axes $a_\mathrm{A}=1\,\mathrm{AU}$, $a_\mathrm{B}=5\times10^2\,\mathrm{AU}$, $a_\mathrm{C}=5\times10^3\,\mathrm{AU}$, masses $m_1=m_3=m_4=1\,\mathrm{M}_\odot$ and $m_2=0.5\,\mathrm{M}_\odot$, eccentricities $e_\mathrm{A}=e_\mathrm{B}=e_\mathrm{C}=0.5$, inclinations $i_\mathrm{A} = 45^\circ$, $i_\mathrm{B}=0^\circ$ and $i_\mathrm{C}=135^\circ$, arguments of pericentre $\omega_\mathrm{A}=\omega_\mathrm{B}=\omega_\mathrm{C}=0^\circ$ and longitudes of the ascending nodes $\Omega_\mathrm{A}=\Omega_\mathrm{B}=\Omega_\mathrm{C}=0^\circ$. Initially, i.e. during the first few KL oscillations in the AB pair, the two methods show very good agreement. However, as time progresses, noticeable deviations develop.

This poses a problem when comparing the two methods in longer integrations, i.e. for time-scales $\gg P_\mathrm{KL,AB}$, where $P_\mathrm{KL,AB}$ is the KL time-scale for the AB binary pair (cf. equation~\ref{eq:PK} below). To illustrate this, we show in the top row in Fig. \ref{fig:comparison_long} another example, where the integration time is $\sim 60 \, P_\mathrm{KL,AB}$. In this case, we set $a_\mathrm{A} = 1 \mathrm{AU}$, $a_\mathrm{B}=10^2\,\mathrm{AU}$ and $a_\mathrm{C}=5\times10^3\,\mathrm{AU}$, $m_1=m_3=m_4=1\,\mathrm{M}_\odot$ and $m_2=0.5\,\mathrm{M}_\odot$, $e_\mathrm{A}=e_\mathrm{B}=e_\mathrm{C}=0.01$, $i_\mathrm{A}=i_\mathrm{C}=85^\circ$ and $i_\mathrm{B}=0^\circ$, $\omega_\mathrm{A}=\omega_\mathrm{B}=\omega_\mathrm{C}=0^\circ$ and $\Omega_\mathrm{A}=\Omega_\mathrm{B}=\Omega_\mathrm{C}=0^\circ$ (note that this system is the same as in panels 1-6 of Fig. \ref{fig:inclined:ex}). In the top row of Fig. \ref{fig:comparison_long}, the quantities $e_k$, $i_k$, $\omega_k$, $\Omega_k$ and the relative energy errors, pertaining to integrations with \textsc{SecularQuadruple} (\textsc{Mikkola}), are shown with red (green) lines. The differences in $e_k$, $i_k$, $\omega_k$ and $\Omega_k$ between the integrations with these codes are shown as a function of time in the middle row in Fig. \ref{fig:comparison_long}. In this case, there is clearly no longer a one-to-one agreement between the two methods. 

When comparing these results on long time-scales (i.e. long compared to $P_\mathrm{KL,AB}$), it is important to take into account that for this system, the phase of the KL cycle in binary A becomes inherently chaotic on a time-scale that is shorter than $P_\mathrm{KL,AB}$. To establish this, we determined the Lyapunov time-scale by carrying out pairs of integrations where in one realization, the initial value of $e_\mathrm{A}=0.01$, was increased by $\Delta e_\mathrm{A}(0)=10^{-4}$. We found that the difference $\Delta e_\mathrm{A}(t)$ between the two integrations initially shows an exponential behaviour as a function of time. Subsequently, we fitted $\log[\Delta e_\mathrm{A}(t)/\Delta e_\mathrm{A}(0)]$ with time assuming a linear relation, i.e. $\log[\Delta e_\mathrm{A}(t)/\Delta e_\mathrm{A}(0)] = C + \lambda t$ where $C$ is a constant, and we determined the Lyapunov time-scale $t_\mathrm{Ly}$ from the inverse of the resulting slope, i.e. $t_\mathrm{Ly} = \lambda^{-1}$.

We find a Lyapunov time-scale of $t_\mathrm{Ly}\approx0.11 \, \mathrm{Myr}$ for this system, and this value is the same for either the secular and direct codes. Reducing the accuracy in the secular integrations does not affect the result, unless the accuracy is reduced such that the relative energy errors increase to $>0.1$. We also verified that $t_\mathrm{L}\approx 0.11 \, \mathrm{Myr}$ for integrations with another $N$-body code, \textsc{Sakura} \citep{ferrari_ea_14}.  

This value of $0.11 \, \mathrm{Myr}$ is shorter than $P_\mathrm{KL,AB}\approx 1 \, \mathrm{Myr}$, which suggests that the system is chaotic on a short time-scale. However, we find that this chaoticity arises from a slightly different phase of the KL cycle between the integrations with $\Delta e_\mathrm{A}(0)=10^{-4}$, whereas the shape of the $e_\mathrm{A}(t)$ remains essentially the same. This result suggests that for long time-scales, it is not meaningful to compare the secular and $N$-body integrations on a one-to-one basis. However, given that the chaotic behaviour discussed above is associated with the phase of the KL cycle, it should still be appropriate to compare the two methods statistically. 

We also note that when comparing the two methods, it is important to take into account that in the $N$-body integrations, there is an additional dependence on the three initial orbital phases. We have also carried out $N$-body integrations with different initial orbital phases, where the initial mean anomaly was sampled randomly. We show the differences between two different $N$-body realizations as a function of time in the bottom row in Fig. \ref{fig:comparison_long}. These differences are typically at least as large as the differences between the secular code, and a single realization with the $N$-body code. Furthermore, we have determined the Lyapunov time-scale as described above, where now $\Delta e_\mathrm{A}(t)$ was determined from two short-term $N$-body integrations with different random mean anomalies. Again, we find that $\Delta e_\mathrm{A}$ increases exponentially with time, with a Lyapunov time-scale of $t_\mathrm{Ly}\approx 0.1 \, \mathrm{Myr}$. Therefore, the differences between the direct $N$-body integrations with different initial orbital phases can be ascribed to the chaotic nature of the phase of the KL cycles. 

In Table \ref{table:nbody_comparisons}, we show results of two-sided Kolmogorov-Smirnov (K-S) tests \citep{Kolmogorov_33,smirnov_48} between time series in $e_k$, $\omega_k$ and $\Omega_k$ obtained from the integration carried out with the secular code, and the integration of five different realizations with the $N$-body code (i.e. with different initial mean anomalies). For K-S tests between the secular and $N$-body integrations, and K-S tests between $N$-body integrations with different realizations, the $D$-values are generally low and the $p$-values are typically high. This shows that the integrations between the secular and $N$-body integrations are statistically consistent, and that the same applies to the $N$-body integrations with different realizations. 

We conclude that, for the highly hierarchical systems considered here, the secular code gives results that are statistically consistent with the direct $N$-body code. The much greater speed makes the former highly suited for the long-term study of a large number of systems. For example, the integration with \textsc{SecularQuadruple} for one of the systems in Fig. \ref{fig:comparison_long} is $\sim10^4$ times faster compared to \textsc{Mikkola}.

\section{Global evolution of highly hierarchical systems}
\label{sect:general}
In principle, the \textsc{SecularQuadruple} algorithm can be used to perform a systematic parameter space study. Instead, here we choose to focus in detail on particular configurations to get insight into the typically complex dynamics that can arise. We consider the following two cases: (1) binaries A and B are initially coplanar ($i_\mathrm{AB,0}=0^\circ$) and highly inclined with respect to binary C ($i_\mathrm{BC,0}=85^\circ$), and (2) binaries A and B are initially highly inclined ($i_\mathrm{AB,0}=85^\circ$) while binary B is also highly inclined with respect to binary C ($i_\mathrm{BC,0}=85^\circ$). In both cases, we assume that the quadruple system is highly hierarchical at all times, i.e. $r_\mathrm{p,A} \ll r_\mathrm{p,B} \ll r_\mathrm{p,C}$, where $r_{\mathrm{p},k}$ is the pericentre distance in binary $k$.

For both cases (1) and (2), we performed a sequence of integrations in which $a_\mathrm{A}$ was varied between $10^{-3}$ and 1 AU, and all other initial parameters were kept fixed. The latter were assumed to be semimajor axes $a_\mathrm{B}=10^2\,\mathrm{AU}$ and $a_\mathrm{C}=5\times10^3\,\mathrm{AU}$, masses $m_1=m_3=m_4=1\,\mathrm{M}_\odot$ and $m_2=0.5\,\mathrm{M}_\odot$, eccentricities $e_\mathrm{A}=e_\mathrm{B}=e_\mathrm{C}=0.01$, arguments of pericentre $\omega_\mathrm{A}=\omega_\mathrm{B}=\omega_\mathrm{C}=0^\circ$ and longitudes of the ascending nodes $\Omega_\mathrm{A}=\Omega_\mathrm{B}=\Omega_\mathrm{C}=0^\circ$. The integration time for each system was set to $20 \,P_\mathrm{KL,BC,0}$, where $P_\mathrm{KL,BC,0}$ is the initial KL time-scale applied to binaries B and C, which we approximate by \citep{innanen_ea_97}
\begin{align}
P_{\mathrm{KL},kl} = \frac{P_l^2}{P_k} \frac{m_{k,\mathrm{p}} + m_{k,\mathrm{s}} + m_{l,\mathrm{s}}}{m_{l,\mathrm{s}}} \left ( 1-e_l^2 \right )^{3/2},
\label{eq:PK}
\end{align}
where $m_{k,\mathrm{p}}=m_1$, $m_{k,\mathrm{s}} = m_2$ and $m_{l,\mathrm{s}}=m_3$ in the case of $P_\mathrm{KL,AB}$, and $m_{k,\mathrm{p}}=m_1+m_2$, $m_{k,\mathrm{s}} =m_3$ and $m_{l,\mathrm{s}}=m_4$ in the case of $P_\mathrm{KL,BC}$ (cf. \S\,\ref{sect:methods:ham}). Note that, contrary to triple systems and at the quadrupole-order approximation, the `outer' orbit eccentricity $e_l$ in equation~(\ref{eq:PK}) can change in time if this equation is applied to binaries A and B. This is addressed in more detail below. 

For hierarchical triple systems, the octupole parameter
\begin{align}
\label{eq:eps_oct}
\epsilon_\mathrm{oct} \equiv \frac{m_1-m_2}{m_1+m_2} \frac{a_\mathrm{in}}{a_\mathrm{out}} \frac{e_\mathrm{out}}{1-e_\mathrm{out}^2}
\end{align}
is a useful proxy for the importance of octupole-order effects, in particular, orbital flips. The latter can occur if $\epsilon_\mathrm{oct} \gtrsim 10^{-3}$, and are typically associated with very high eccentricities \citep{lithwick_naoz_11,katz_ea_11,teyssandier_13,li_ea_14}. In the systems considered here, the initial octupole parameters $\epsilon_\mathrm{oct}$ range between $\approx3.3\times 10^{-8}$ and $\approx3.3\times 10^{-5}$ for binary pair AB; for binary pair BC, $\epsilon_\mathrm{oct} \approx4.0\times 10^{-5}$. This indicates that octupole-order terms are not important. Furthermore, the initial ratio $r_0$ of the orbit-averaged octupole-order cross term to all other orbit-averaged terms at octupole and hexadecupole order (cf. \S\,\ref{sect:methods:cross_importance}) ranges between $\approx 2 \times 10^{-12}$ and $\approx 3 \times 10^{-7}$, indicating that the orbit-averaged octupole-order cross term can similarly be neglected. The results presented below therefore demonstrate the dynamics that are manifested at the lowest possible, i.e. quadrupole, order. 

\subsection{Examples: A and B initially coplanar}
\label{sect:general:examples_coplanar}

\begin{figure*}
\center
\includegraphics[scale = 0.48, trim = 25mm 5mm 0mm 0mm]{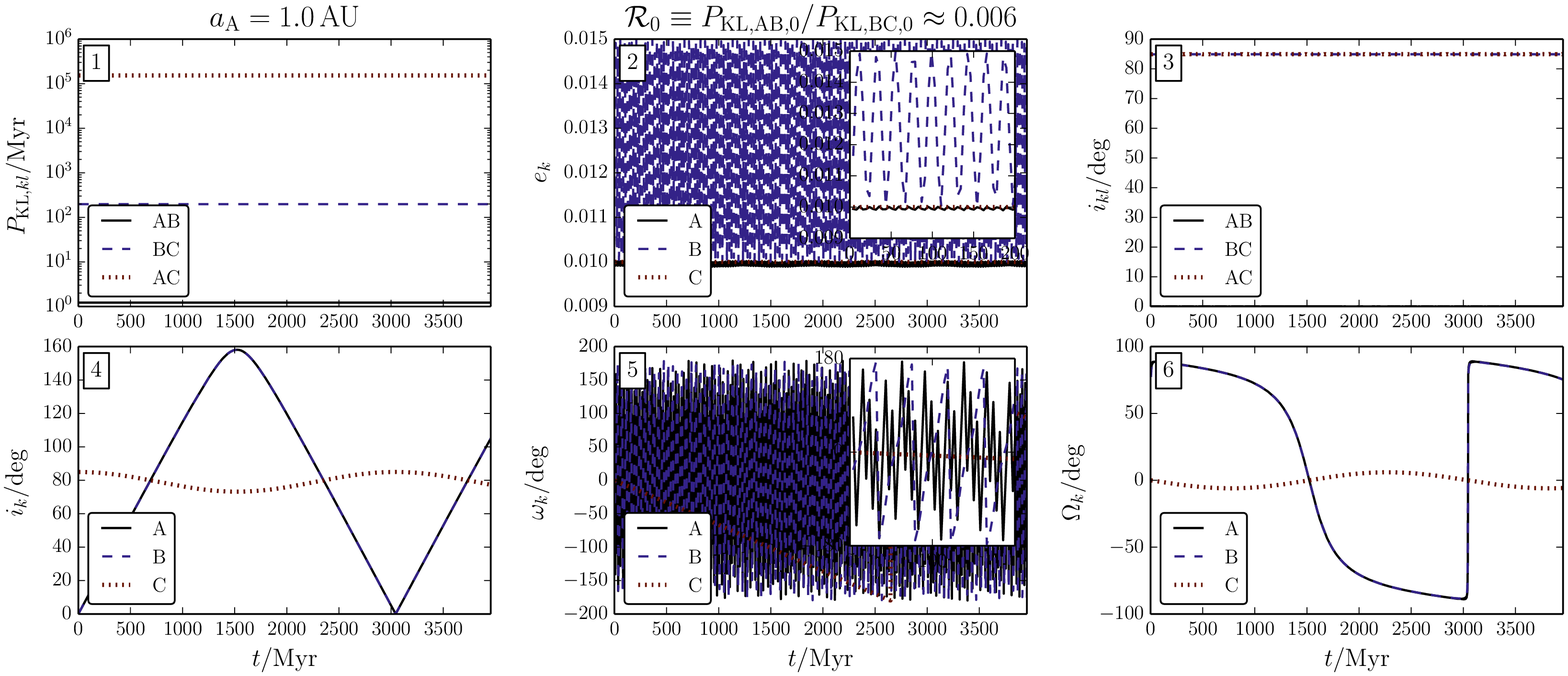}
\includegraphics[scale = 0.48, trim = 25mm 5mm 0mm 0mm]{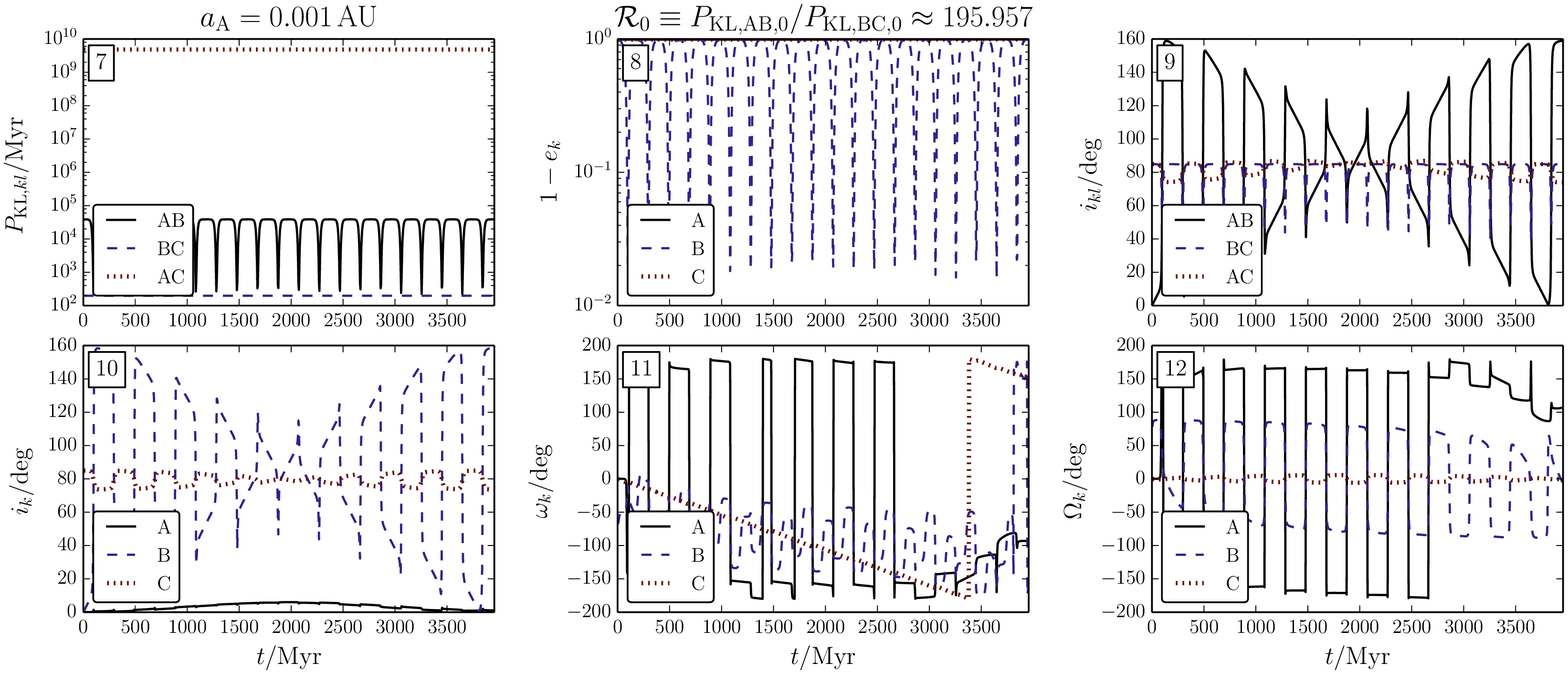}
\includegraphics[scale = 0.48, trim = 25mm 5mm 0mm 0mm]{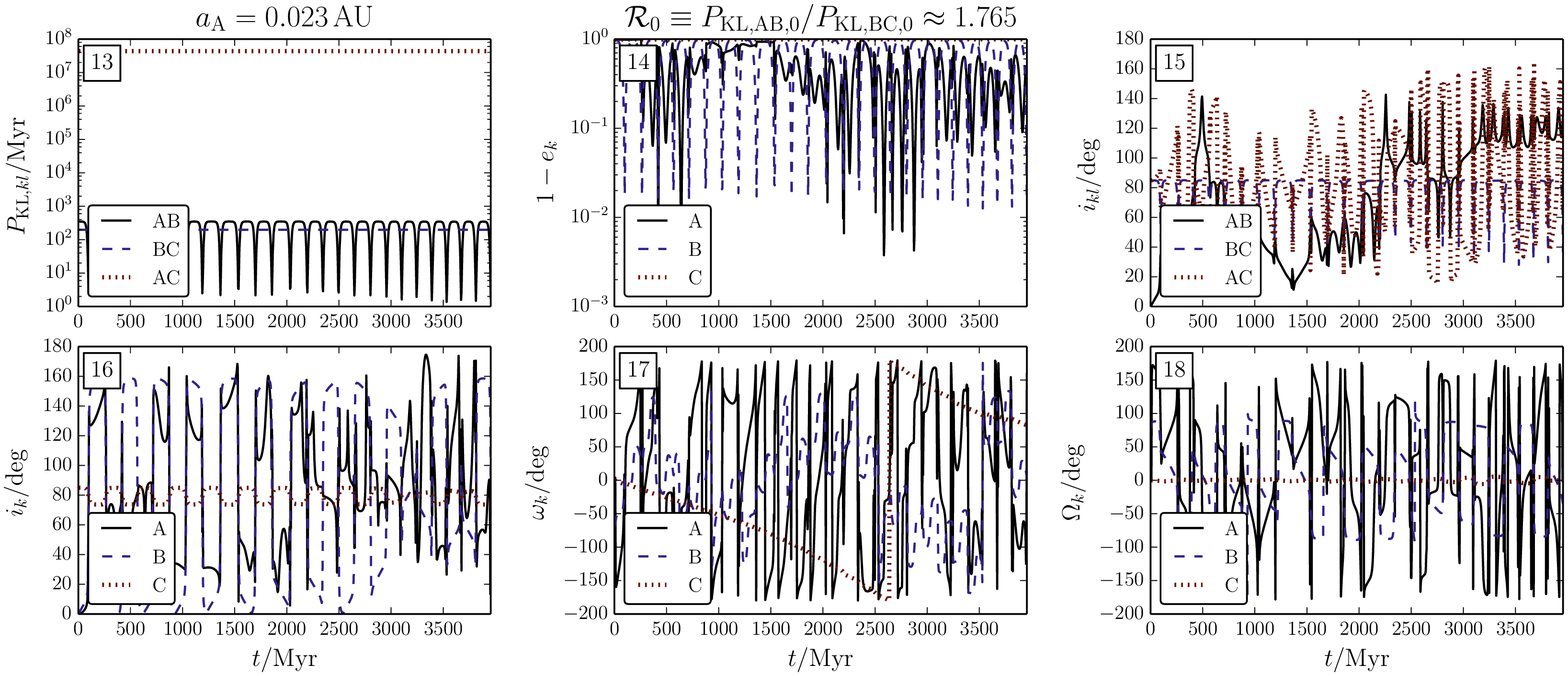}
\caption{\small Evolution of three quadruple systems as discussed in \S\,\ref{sect:general:examples_coplanar}, computed with \textsc{SecularQuadruple}. Binaries A and B are initially coplanar (as opposed to highly inclined in Fig. \ref{fig:inclined:ex}). Reference numbers are shown in the top left of each panel. Panels 1-6, 7-12 and 13-18 correspond to semimajor axes of binary A of 1, 0.001 and 0.023 AU, respectively. The other initial parameters are the same for these groups of panels, and are given by semimajor axes $a_\mathrm{B}=10^2\,\mathrm{AU}$ and $a_\mathrm{C}=5\times10^3\,\mathrm{AU}$, masses $m_1=m_3=m_4=1\,\mathrm{M}_\odot$ and $m_2=0.5\,\mathrm{M}_\odot$, eccentricities $e_\mathrm{A}=e_\mathrm{B}=e_\mathrm{C}=0.01$, inclinations $i_\mathrm{A}=i_\mathrm{B}=0^\circ$, $i_\mathrm{C}=85^\circ$, arguments of pericentre $\omega_\mathrm{A}=\omega_\mathrm{B}=\omega_\mathrm{C}=0^\circ$ and longitudes of the ascending nodes $\Omega_\mathrm{A}=\Omega_\mathrm{B}=\Omega_\mathrm{C}=0^\circ$. In panels 2 and 5, the abscissae in the inset range between $t=0$ and 200 Myr. }
\label{fig:coplanar:ex}
\end{figure*}

\begin{figure*}
\center
\includegraphics[scale = 0.48, trim = 25mm 5mm 0mm 0mm]{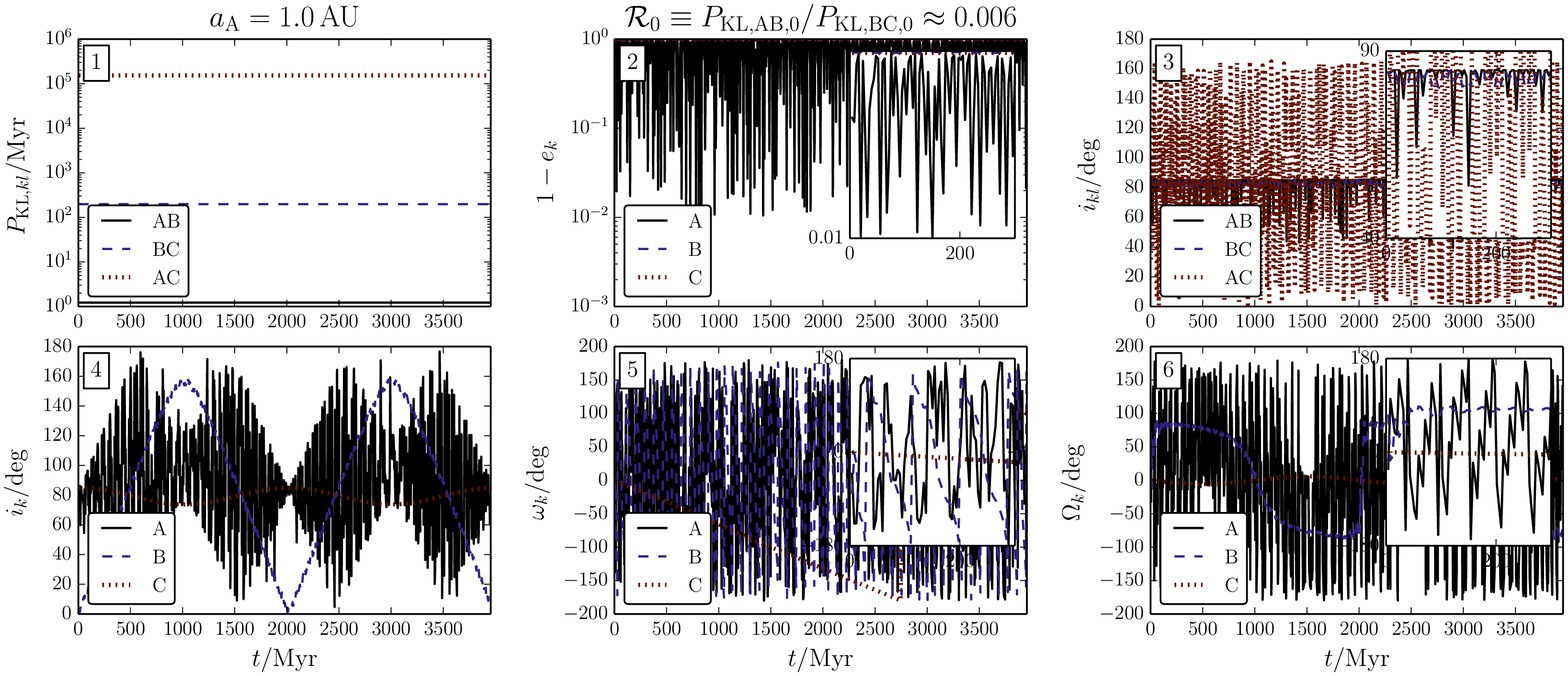}
\includegraphics[scale = 0.48, trim = 25mm 5mm 0mm 0mm]{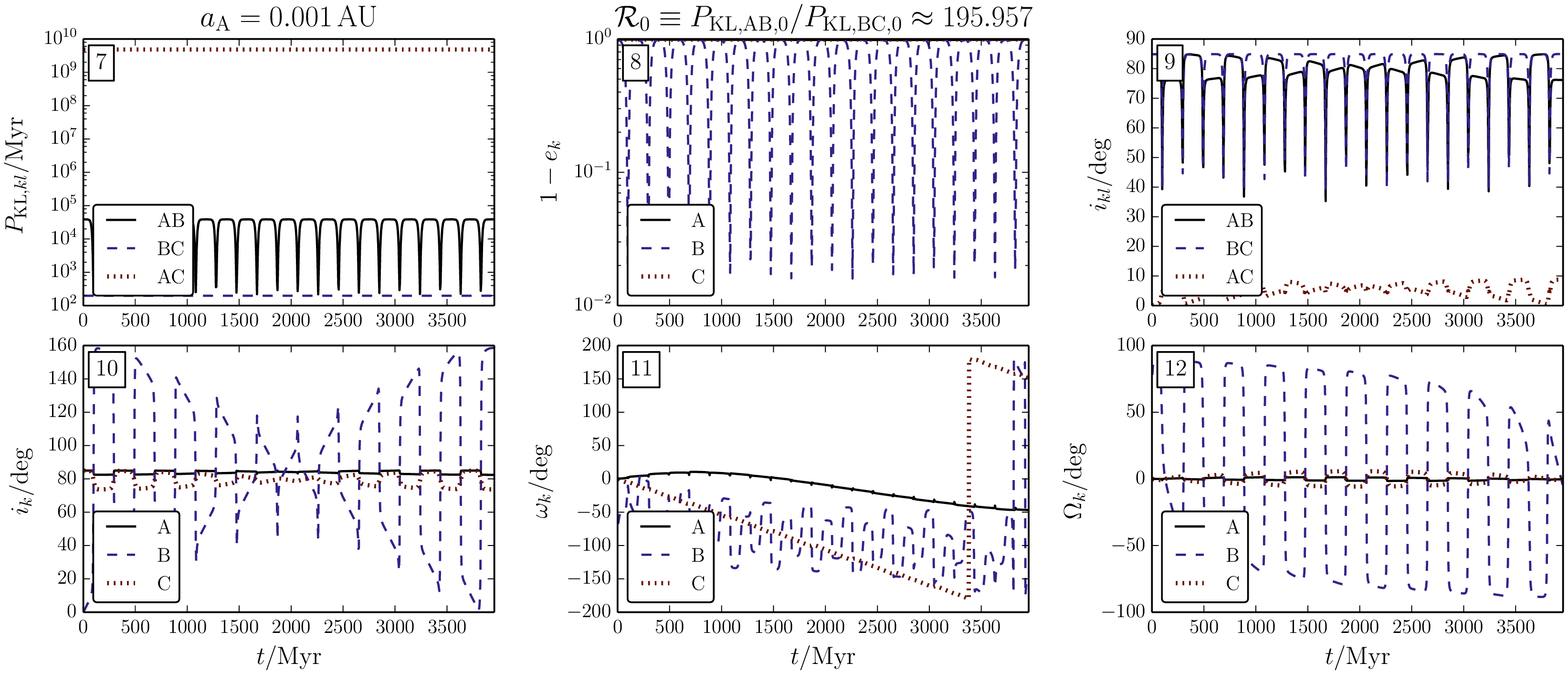}
\includegraphics[scale = 0.48, trim = 25mm 5mm 0mm 0mm]{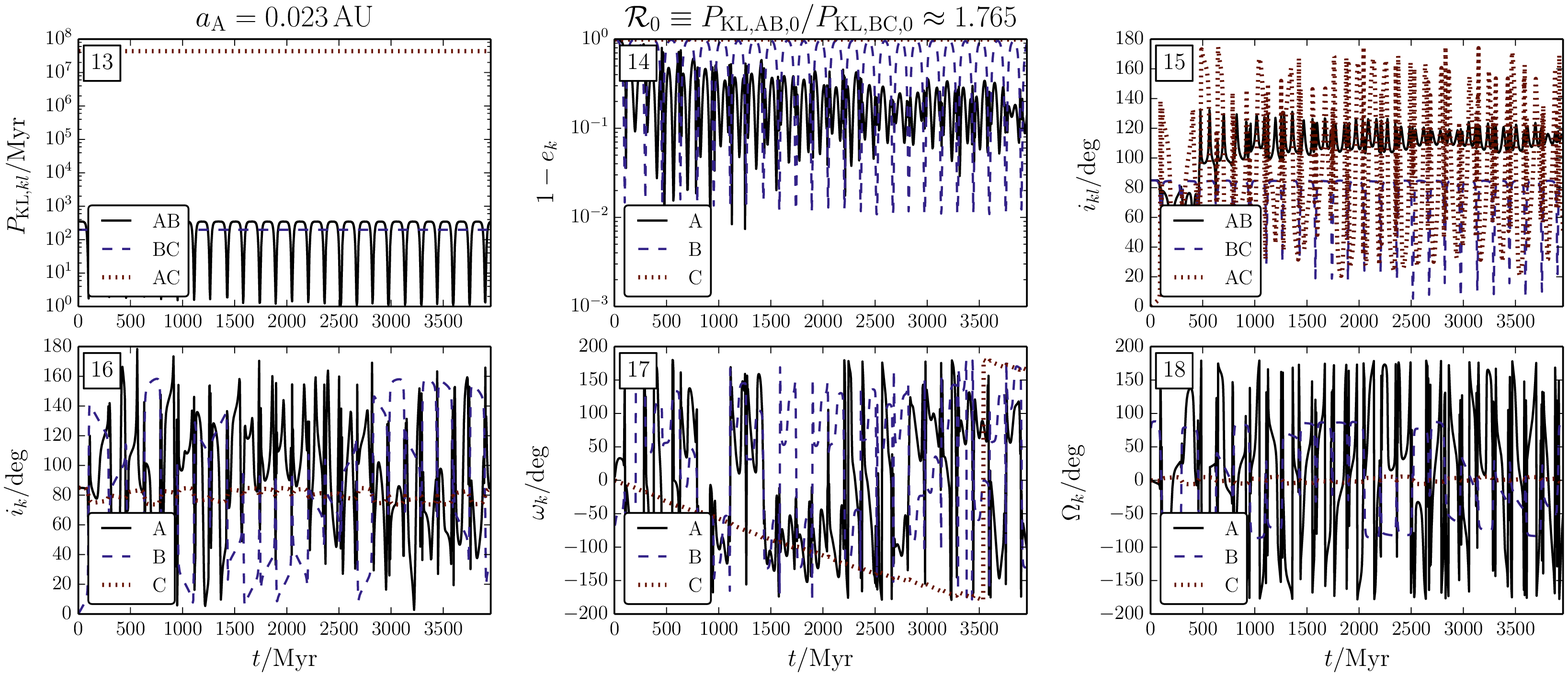}
\caption{\small Evolution of three quadruple systems as discussed in \S\,\ref{sect:general:examples_inclined}, computed with \textsc{SecularQuadruple}. Binaries A and B are initially inclined by $85^\circ$ (as opposed to $0^\circ$ in Fig. \ref{fig:coplanar:ex}). Panels 1-6, 7-12 and 13-18 correspond to semimajor axes of binary A of 1, 0.001 and 0.023 AU, respectively. The other parameters are the same for these groups of panels, and are given by semimajor axes $a_\mathrm{B}=10^2\,\mathrm{AU}$ and $a_\mathrm{C}=5\times10^3\,\mathrm{AU}$, masses $m_1=m_3=m_4=1\,\mathrm{M}_\odot$ and $m_2=0.5\,\mathrm{M}_\odot$, eccentricities $e_\mathrm{A}=e_\mathrm{B}=e_\mathrm{C}=0.01$, inclinations $i_\mathrm{A}=i_\mathrm{C}=85^\circ$, $i_\mathrm{B}=0^\circ$, arguments of pericentre $\omega_\mathrm{A}=\omega_\mathrm{B}=\omega_\mathrm{C}=0^\circ$ and longitudes of the ascending nodes $\Omega_\mathrm{A}=\Omega_\mathrm{B}=\Omega_\mathrm{C}=0^\circ$. In panels 2, 3, 5 and 6, the abscissae in the insets range between $t=0$ and 300 Myr. }
\label{fig:inclined:ex}
\end{figure*}

In our first case, $i_\mathrm{AB,0}=0^\circ$ and $i_\mathrm{BC,0}=85^\circ$, which is achieved by setting the initial $i_\mathrm{A}=i_\mathrm{B}=0^\circ$ and $i_\mathrm{C}=85^\circ$ (note that, initially, $\Omega_\mathrm{A}=\Omega_\mathrm{B}=\Omega_\mathrm{C}$). In the absence of the fourth body, there would not be any excitation of the eccentricity in binaries A and B because they are not mutually inclined and only the quadrupole-order terms are important. We note that if the initial $e_\mathrm{B}=0.01$ were much larger (and therefore $\epsilon_\mathrm{oct}$ would be much higher, cf. equation~\ref{eq:eps_oct}), owing to the greater importance of the octupole-order terms, orbital flips and very high eccentricity oscillations in binary A would be possible in certain conditions, even if $i_\mathrm{AB,0}$ is close to zero \citep{li_naoz_kocsis_loeb_13}. We show in Fig. \ref{fig:coplanar:ex} three examples of numerically integrated systems, in which $a_\mathrm{A}$ is either 1 (panels 1-6), 0.001 (panels 7-12) or 0.023 AU (panels 13-18). 

For $a_\mathrm{A} = 1 \, \mathrm{AU}$, the {\it mutual} inclination between binaries A and B, $i_\mathrm{AB}$, remains zero (cf. the solid line in panel 3 of Fig. \ref{fig:coplanar:ex}). However, the {\it individual} inclinations of binaries A and B, $i_\mathrm{A}$ and $i_\mathrm{B}$, which are initially zero, do change (cf. the solid and dashed lines in panel 4 of Fig. \ref{fig:coplanar:ex}; note that these curves overlap). This can be understood from the large torque of binary B on binary A, compared to the torque of binary C on binary B. More quantitatively, the KL time-scales can be interpreted as proxies for the importance of these torques, and the initial KL time-scale for binaries A and B, $P_\mathrm{KL,AB,0} \approx 1.2\, \mathrm{Myr}$, is much shorter (i.e. corresponding to a larger torque) than the initial KL time-scale for binaries B and C, $P_\mathrm{KL,BC,0} \approx 2\times10^2\, \mathrm{Myr}$. The large torque of binary B on binary A enforces that zero {\it mutual} inclination between these binaries is maintained, despite the torque from binary C on binary B. The latter torque changes the {\it individual} inclination of binary B on the time-scale of $P_\mathrm{KL,BC} \gg P_\mathrm{KL,AB}$. Note that the mutual inclination is determined by the individual inclinations $i_k$ and longitudes of the ascending nodes $\Omega_k$ (cf. equation~\ref{eq:i_kl}). Therefore, both these angles for binaries A and B follow each other very closely (cf. panels 4 and 6 of Fig. \ref{fig:coplanar:ex}). 

If binary A were replaced by a point mass, the eccentricity in binary B would oscillate as a result of the torque from binary C, with maxima of $1-e_\mathrm{B,\max} \approx 10^{-2}$. However, in the case of a quadruple system, the short KL time-scale in binary A with binary B causes rapid precession in both binaries A and B, on roughly the same time-scale (cf. the black solid and blue dashed lines in panel 5 of Fig. \ref{fig:coplanar:ex}). Consequently, the rapid precession in binary B quenches any KL oscillations induced by the torque of binary C. This effect is analogous to the quenching of KL oscillations in triple systems due to additional sources of periapse precession. Here, the additional precession is due to the extended nature of one of the components in the inner binary, rather than due to e.g. relativistic precession or tidal bulges. This quenching effect is discussed more quantitatively below, in \S\,\ref{sect:general:dep_R0}.

In panels 7-12 of Fig. \ref{fig:coplanar:ex}, we show the evolution of an example system with $a_\mathrm{A} = 10^{-3} \, \mathrm{AU}$. The initial KL time-scale for binaries A and B is $P_\mathrm{KL,AB,0} \approx 39 \, \mathrm{Gyr} \gg P_\mathrm{KL,BC,0} \approx 2\times10^2 \, \mathrm{Myr}$. Therefore, there is no induced precession of binary A on binary B, and KL eccentricity oscillations occur in binary B with maxima of $1-e_\mathrm{B,\max} \approx 10^{-2}$ (cf. the dashed lines in panel 8 of Fig. \ref{fig:coplanar:ex}). Furthermore, the torque of binary C on binary B dominates compared to the torque of binary B on binary A. Consequently, the inclination of binary B changes rapidly, whereas the inclination of binary A hardly changes (cf. the solid and dashed lines in panel 10 of Fig. \ref{fig:coplanar:ex}). However, this also changes the {\it mutual} inclination $i_\mathrm{AB}$ between binaries A and B. The latter increases very rapidly (cf. the solid line in panel 9 of Fig. \ref{fig:coplanar:ex}). Nevertheless, binaries A and B are only highly mutually inclined ($i_\mathrm{AB}$ close to $90^\circ$) for short periods of time, and therefore, no significant eccentricity oscillations occur in binary A. In other words, the latter oscillations are impeded by rapid changes of the mutual inclination between binaries A and B, because of KL oscillations induced by binary C.

Finally, in panels 13-18 of Fig. \ref{fig:coplanar:ex}, $a_\mathrm{A} \approx 0.023 \, \mathrm{AU}$. The initial KL time-scales for the binary pairs AB and BC are comparable, i.e. $P_\mathrm{KL,AB,0} \approx 3\times10^2 \, \mathrm{Myr} \sim P_\mathrm{KL,BC,0} \approx 2\times10^2 \, \mathrm{Myr}$, and therefore the torques of binary B on binary A and of binary C on binary B are also comparable. Binaries A and B become mutually inclined, and the KL time-scale for the AB pair is short enough for large excitation of the eccentricity of binary A. The result is a complex evolution in which the oscillations in $e_\mathrm{A}$ are highly non-regular and strongly coupled with the oscillations of $e_\mathrm{B}$. Interestingly, although binaries A and B started out with a mutual inclination of $i_\mathrm{AB,0}=85^\circ<90^\circ$, the orientation between binaries A and B at $t\sim400 \, \mathrm{Myr}$ changes from prograde to retrograde. Such orbital flips also occur at later times, and are associated with high eccentricities in binary A. The evolution of the eccentricity of binary B is also affected, although the effect is much smaller and the oscillations can still be considered as regular. In \S\,\ref{sect:general:dep_R0}, we study the effect of the eccentricity of binary B in more detail.

\subsection{Examples: A and B initially highly inclined}
\label{sect:general:examples_inclined}

In our second case, we assume that both binaries A and B and binaries B and C are initially highly inclined, i.e. $i_\mathrm{AB,0}=85^\circ$ and $i_\mathrm{BC,0}=85^\circ$, which is achieved by setting $i_\mathrm{A}=i_\mathrm{C}=85^\circ$ and $i_\mathrm{B}=0^\circ$. The evolution of three example systems, with other parameters identical to those in \S\,\ref{sect:general:examples_coplanar}, is shown in Fig. \ref{fig:inclined:ex}. In the absence of the fourth body, high-eccentricity KL oscillations would be induced in binary A. 

For $a_\mathrm{A} = 1 \, \mathrm{AU}$ (panels 1-6 of Fig. \ref{fig:inclined:ex}), $P_\mathrm{KL,AB} \ll P_\mathrm{KL,BC}$, and for time-scales comparable to $P_\mathrm{KL,AB}$, KL eccentricity oscillations in binaries A and B are hardly affected by the torque of binary C. On much longer time-scales comparable to $P_\mathrm{KL,BC}$, $i_\mathrm{B}$ changes because of the torque of binary C (cf. the blue dashed line in panel 4 of Fig. \ref{fig:inclined:ex}). However, the KL eccentricity oscillations between binaries A and B are not noticeably affected (note that in panels 2-6 of Fig. \ref{fig:inclined:ex}, the KL oscillations associated with binaries A and B are undersampled). Consequently, $i_\mathrm{A}$, $i_\mathrm{B}$, $\Omega_\mathrm{A}$ and $\Omega_\mathrm{B}$ are modulated on the $P_\mathrm{KL,BC}$ time-scale. We note that, as a consequence of KL oscillations in the AB pair, there is still short-time-scale precession induced on binary B, preventing any eccentricity excitation in binary B. This is similar to the previous case when binaries A and B are initially coplanar. 

For $a_\mathrm{A} = 0.001 \, \mathrm{AU}$ (panels 7-12 of Fig. \ref{fig:inclined:ex}), the evolution is qualitatively very similar to the case when $i_\mathrm{AB,0}=0^\circ$. This may be surprising, given the high initial mutual inclination between binaries A and B. However, the latter changes strongly on the much shorter time-scale of $P_\mathrm{KL,BC}$, and this prevents any eccentricity excitation in binary A. Note that in this case, the quenching of KL eccentricity oscillations in binary A is not due to induced precession. As can be seen in panel 11 of Fig. \ref{fig:inclined:ex}, $\omega_\mathrm{A}$ is not much affected on the $P_\mathrm{KL,BC}$ time-scale, although there is also a trend on a much longer time-scale of $\sim 4 \times 10^3$ Myr. The KL time-scale for the AB pair changes periodically as $e_\mathrm{B}$ oscillates (cf. panel 7 of Fig. \ref{fig:inclined:ex}). Therefore the time-scale of  $\sim 4 \times 10^3$ Myr can, in this case, be interpreted as an effective KL time-scale for the AB pair. 

When the KL time-scales for the AB and BC pairs are similar (cf. panels 13-18 of Fig. \ref{fig:inclined:ex}), the evolution of $e_\mathrm{A}$ is complex and high eccentricities are attained, similarly to the case when $i_\mathrm{AB,0}=0^\circ$. Again, an orbital flip occurs around $t\sim400$ Myr. Interestingly, subsequently there are no orbital flips, and the amplitude of the oscillations in $e_\mathrm{A}$ and $i_\mathrm{AB}$ gradually decreases. 

\subsection{Qualitative trends}
\label{sect:general:trends}
The above examples suggest that the ratio of the (initial) KL time-scales for the AB and BC pairs, 
\begin{align}
\label{eq:R0_def}
\nonumber \mathcal{R}_0 &\equiv \frac{P_\mathrm{KL,AB,0}}{P_\mathrm{KL,BC,0}} \\
&= \left ( \frac{a_\mathrm{B}^3}{a_\mathrm{A} a_\mathrm{C}^2} \right )^{3/2} \left ( \frac{m_1+m_2}{m_1+m_2+m_3} \right )^{1/2} \frac{m_4}{m_3} \left( \frac{1-e_\mathrm{B,0}^2}{1-e_\mathrm{C,0}^2} \right )^{3/2},
\end{align}
is an indication of the global trend of the inclination and eccentricity oscillations. We identify the following three regimes.
\begin{enumerate}
\item $\mathcal{R}_0\ll 1$: binaries A and B remain coplanar if this was initially the case. If they are initially inclined, KL eccentricity oscillations in binary A are not much affected by the presence of the fourth body. In either case, KL eccentricity oscillations in binary B are quenched.
\item $\mathcal{R}_0\gg 1$: binaries A and B become inclined if they are initially coplanar. However, there are no eccentricity oscillations in binary A, even if binaries A and B are initially highly inclined. This is because the mutual inclination between binaries A and B is large only for a small fraction of the KL time-scale for the AB pair, i.e. for a time of $<P_\mathrm{KL,BC,0} = P_\mathrm{KL,AB,0}/\mathcal{R}_0 \ll P_\mathrm{KL,AB,0}$. Furthermore, KL eccentricity oscillations are not quenched in binary B.
\item $\mathcal{R}_0\sim 1$: binaries A and B become inclined if they are initially coplanar; complex KL eccentricity oscillations arise in binary A that are coupled with the -- much less affected -- KL eccentricity oscillations in binary B.
\end{enumerate}
These three regimes correspond to panels 1-6, 7-12 and 13-18 in Figs \ref{fig:coplanar:ex} and \ref{fig:inclined:ex}.

A complication in the above, is that $P_\mathrm{KL,AB}$ can change periodically with time because of KL eccentricity oscillations in binary B (cf. equation~\ref{eq:PK}). Periodically higher values of $e_\mathrm{B}$ reduce $P_\mathrm{KL,AB}$ at the same times, therefore potentially increasing the range of $\mathcal{R}$ for which the eccentricity in binary A can be excited. Furthermore, for large enough values of $e_\mathrm{B}$, higher order terms in the Hamiltonian become more important, and in extreme cases, the orbit-averaged approach could break down. 

In principle, the time-dependence of $e_\mathrm{B}$ could be taken into account by e.g. averaging $P_\mathrm{KL,AB}$ over a KL cycle in binary B. However, except for a few simple cases, there are no analytic solutions for $e_\mathrm{B}(t)$. Therefore, this would require numerical integration and hence not be of much practical use for predicting the behaviour {\it without} resorting to such integration. Nevertheless, because of the very peaked nature of $e_\mathrm{B}(t)$ and the small width (in time) of the peaks, we expect the averaged value of $P_\mathrm{KL,AB}$ typically not to be very different from the value computed from $e_\mathrm{B,0}$, at least in systems in which the lowest order (quadrupole-order) terms dominate.

\begin{figure*}
\center
\includegraphics[scale = 0.45, trim = 10mm 0mm 0mm 0mm]{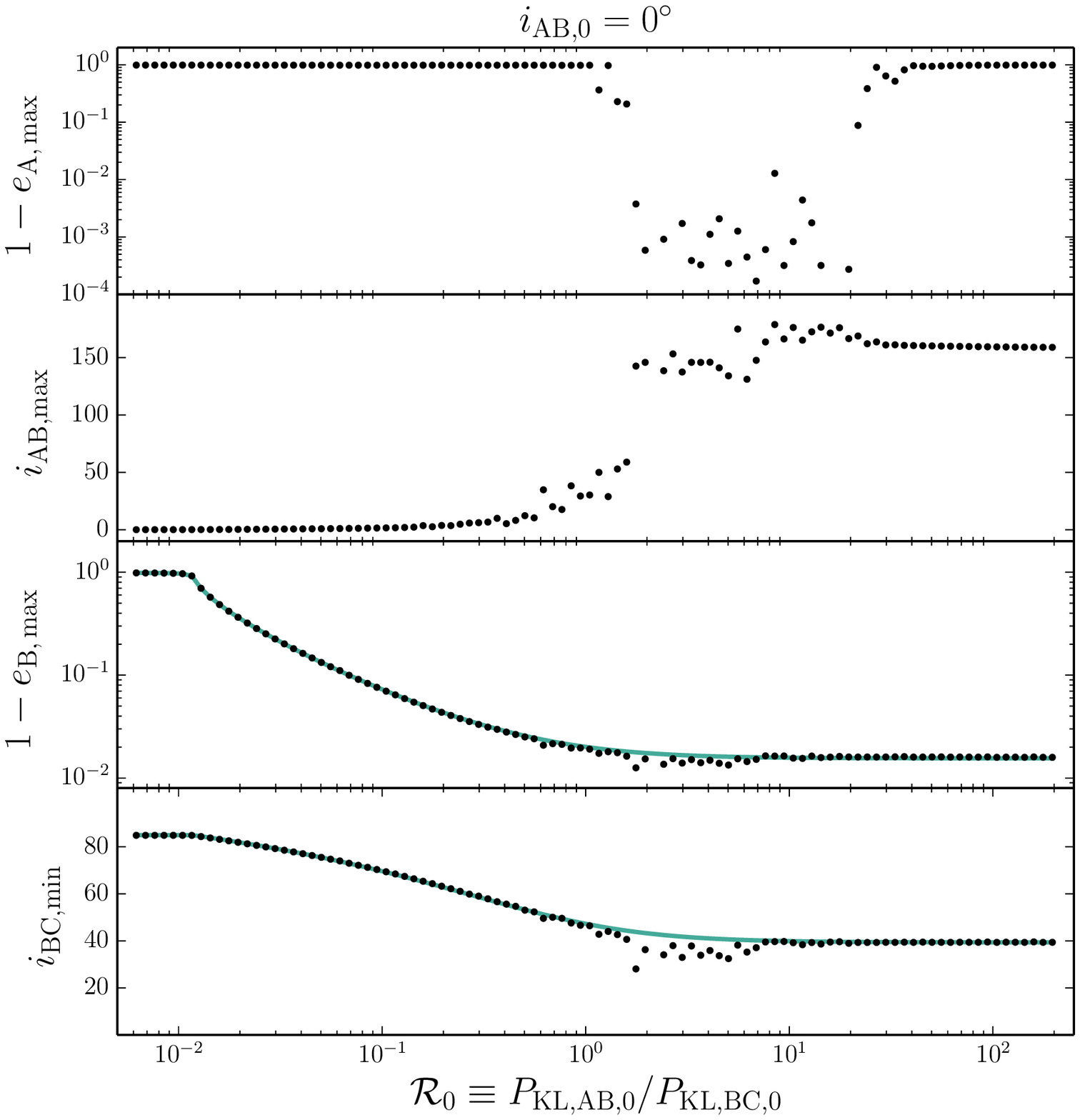}
\includegraphics[scale = 0.45, trim = 10mm 0mm 0mm 0mm]{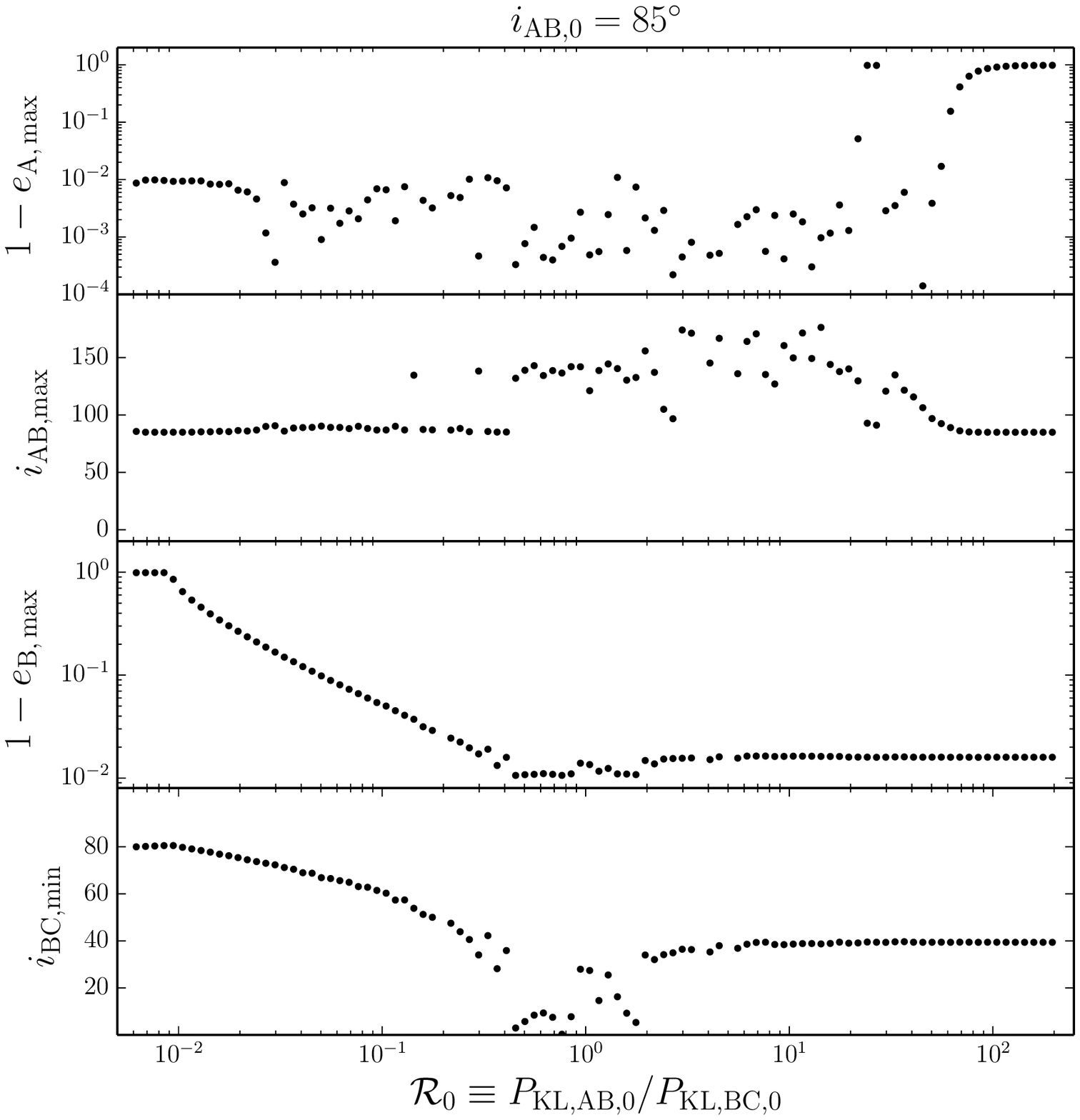}
\caption{\small The maximum eccentricities in binaries A and B (first and third panels from the top), the maximum inclination between binaries A and B (second panels from the top), and the minimum inclination between binaries B and C (fourth panels from the top), as a function of the ratio $\mathcal{R}_0$ of the KL time-scales for the AB and BC pairs (cf. equation~\ref{eq:R0_def}). Here, $\mathcal{R}_0$ is varied by changing the initial $a_\mathrm{A}$ and keeping the other initial semimajor axes, masses and eccentricities fixed. In both left- and right-hand panels, the initial conditions are the same as in \S\,\ref{sect:general:examples_coplanar} and \S\,\ref{sect:general:examples_inclined}, i.e. semimajor axes $a_\mathrm{B}=10^2\,\mathrm{AU}$ and $a_\mathrm{C}=5\times10^3\,\mathrm{AU}$, masses $m_1=m_3=m_4=1\,\mathrm{M}_\odot$ and $m_2=0.5\,\mathrm{M}_\odot$, eccentricities $e_\mathrm{A}=e_\mathrm{B}=e_\mathrm{C}=0.01$, arguments of pericentre $\omega_\mathrm{A}=\omega_\mathrm{B}=\omega_\mathrm{C}=0^\circ$ and longitudes of the ascending nodes $\Omega_\mathrm{A}=\Omega_\mathrm{B}=\Omega_\mathrm{C}=0^\circ$. In the left- (right-hand) panels, binaries A and B are assumed to be initially coplanar (inclined by $85^\circ$), i.e. in the left-hand panel, $i_\mathrm{A}=i_\mathrm{B}=0^\circ$ and $i_\mathrm{C}=85^\circ$, whereas in the right-hand panel, $i_\mathrm{A}=i_\mathrm{C}=85^\circ$ and $i_\mathrm{B}=0^\circ$. Black dots: computed with \textsc{SecularQuadruple} (the integration time was $20 \, P_\mathrm{KL,BC}$); solid lines: computed using the semianalytic method discussed in \S\,\ref{sect:general:dep_R0:sa}. }
\label{fig:sa}
\end{figure*}

\subsection{Quantitative dependence on $\mathcal{R}_0$}
\label{sect:general:dep_R0}
\subsubsection{Results from numerical integrations}
\label{sect:general:dep_R0:num_time}
Here, we describe the dynamics outlined in \S\,\ref{sect:general:trends} more quantitatively, focusing in particular on the effect of the quenching of KL eccentricity oscillations in binary B by the induced precession of binary A, and on the excitation of the eccentricity in binary A in the regime $\mathcal{R}_0 \sim 1$.

In Fig. \ref{fig:sa}, we show with black dots the maximum eccentricities in binaries A and B, the maximum inclination between binaries A and B, and the minimum inclination between binaries B and C as a function of $\mathcal{R}_0$, as determined from numerical integrations with \textsc{SecularQuadruple}. Here, $\mathcal{R}_0$ is varied by changing $a_\mathrm{A}$ (cf. equation~\ref{eq:R0_def}) in the sequence of integrations described in the beginning of \S\,\ref{sect:general}. In the left (right) panels, results are shown assuming that binaries A and B are initially coplaner (highly inclined). 

For $i_\mathrm{AB,0}=0^\circ$, $i_\mathrm{AB,max}$ is zero for $\mathcal{R}_0\lesssim 1$ and rapidly increases for $\mathcal{R}_0\gtrsim 1$; $e_\mathrm{A,max}$ is equal to the initial value for $\mathcal{R}_0\lesssim 1$ and for $\mathcal{R}_0\gtrsim 20$. This is consistent with the trend that was outlined in \S\,\ref{sect:general:trends}. Furthermore, if $\mathcal{R}_0\lesssim 10^{-2}$, $e_\mathrm{B,max} \approx 0$, demonstrating that the induced precession of system A on B in this regime can completely quench any KL oscillations in binary B. Consequently, the minimum inclination between binaries B and C is constant and $\approx 85^\circ$, the initial value (note that for the regular KL oscillations in binary B, a maximum eccentricity corresponds to minimum inclination with respect to binary C). If $1\lesssim \mathcal{R}_0 \lesssim 20$, $i_\mathrm{AB,max}$ is non-zero; $e_\mathrm{A,max}$ is also non-zero and reaches high values of up to $\approx 1-10^{-4}$. Although the behaviour of these two quantities as a function of $\mathcal{R}_0$ is non-regular, there is a general trend in which $i_\mathrm{AB,max}$ asymptotes to $\approx 160^\circ$. A general trend is also apparent in $e_\mathrm{A,max}$. 

If binaries A and B are initially inclined by $85^\circ$ (cf. the right-hand panels in Fig. \ref{fig:sa}), the dependence of $e_\mathrm{A,max}$ as a function of $\mathcal{R}_0$ is more complicated. For a large range in $\mathcal{R}_0$, $3\times10^{-2}\lesssim \mathcal{R}_0 \lesssim 50$, $e_\mathrm{A,max}$ fluctuates strongly with $\mathcal{R}_0$, reaching high values of $1-e_\mathrm{A,max} \sim 10^{-4}$ for $\mathcal{R}_0$ already as low as $\mathcal{R}_0 \approx 3\times 10^{-2}$. For $\mathcal{R}_0 \gtrsim 50$, $e_\mathrm{A,max}$ approaches $e_\mathrm{A,0}$, as was observed previously in \S\,\ref{sect:general:examples_inclined}. Furthermore, binary B is more affected compared to the coplanar case, in the sense that $i_\mathrm{BC,min}$ decreases more strongly in the regime $1\lesssim \mathcal{R}_0 \lesssim 20$. The maximum eccentricity in binary B is similar to the coplanar case, however. 

\subsubsection{Semianalytic description}
\label{sect:general:dep_R0:sa}
The maximum eccentricity (and hence minimum inclination) reached in binary B can be computed approximately using a semianalytic method based on conservation of the total energy (i.e. the Hamiltonian) and the total angular momentum. This method is similar to that used by \citet{miller_hamilton_02,blaes_lee_socrates_02,fabrycky_tremaine_07,naoz_ea_13}. We neglect any changes in binary A between the initial and final states, where the final state corresponds to a maximum eccentricity in binary B. To our knowledge, it is not possible to predict (i.e. without resorting to `brute-force' numerical integrations as in \S\,\ref{sect:general:dep_R0:num_time}) these changes in system A, and this is likely related to the generally chaotic nature of the evolution of binary A, in particular in the regime $1\lesssim \mathcal{R}_0 \lesssim 20$ (cf. \S\,\ref{sect:general:R1}). Stated more mathematically, conservation of total energy and angular momentum and the condition that $e_\mathrm{B}$ is stationary, do not generally provide enough constraints to solve both for $e_\mathrm{B,max}$ and the corresponding $e_\mathrm{A}$. 

In the Hamiltonian to quadrupole order and for the hierarchy considered here, the term corresponding to binaries A and C in the Hamiltonian can safely be neglected. This can readily be seen from equation~(\ref{eq:app:H_av_quad}): the three terms at quadrupole order scale with the semimajor axes according to
\begin{align}
\nonumber \overline{H}_\mathrm{quad,AB} & \propto \frac{1}{a_\mathrm{B}} \left ( \frac{a_\mathrm{A}}{a_\mathrm{B}} \right )^2; \quad \overline{H}_\mathrm{quad,BC} \propto \frac{1}{a_\mathrm{C}} \left ( \frac{a_\mathrm{B}}{a_\mathrm{C}} \right )^2; \\
\overline{H}_\mathrm{quad,AC} &\propto \frac{1}{a_\mathrm{C}} \left ( \frac{a_\mathrm{A}}{a_\mathrm{C}} \right )^2.
\end{align}
Because, by assumption, $a_\mathrm{C} \gg a_\mathrm{B} \gg a_\mathrm{A}$, this implies that $\overline{H}_\mathrm{quad,AC}$ can be neglected compared to $\overline{H}_\mathrm{quad,AB}$ and $\overline{H}_\mathrm{quad,BC}$. The Hamiltonian to quadrupole order is therefore well approximated by (cf. equation~\ref{eq:app:H_av_quad})
\begin{align}
\label{eq:H_quad_appr}
\nonumber \overline{H}_0 &= C_\mathrm{AB} j_\mathrm{B}^{-5} \left [ \left(1-6e_\mathrm{A}^2\right ) j_\mathrm{B}^2 + 15 \left ( \boldsymbol{e}_\mathrm{A} \cdot \boldsymbol{j}_\mathrm{B} \right )^2 - 3\left ( \boldsymbol{j}_\mathrm{A} \cdot \boldsymbol{j}_\mathrm{B} \right )^2 \right ] \\
& + C_\mathrm{BC} j_\mathrm{C}^{-5} \left [ \left(1-6e_\mathrm{B}^2\right ) j_\mathrm{C}^2 + 15 \left ( \boldsymbol{e}_\mathrm{B} \cdot \boldsymbol{j}_\mathrm{C} \right )^2 - 3\left ( \boldsymbol{j}_\mathrm{B} \cdot \boldsymbol{j}_\mathrm{C} \right )^2 \right ],
\end{align}
where
\begin{align}
\nonumber C_\mathrm{AB} &= \frac{1}{8} \frac{Gm_1m_2m_3}{m_1+m_2} \frac{1}{a_\mathrm{B}} \left ( \frac{a_\mathrm{A}}{a_\mathrm{B}} \right )^2; \\
C_\mathrm{BC} &= \frac{1}{8} \frac{G(m_1+m_2)m_3 m_4}{m_1+m_2+m_3} \frac{1}{a_\mathrm{C}} \left ( \frac{a_\mathrm{B}}{a_\mathrm{C}} \right )^2.
\end{align}
The equation of motion for $\boldsymbol{e}_\mathrm{B}$ that follows from equation~(\ref{eq:H_quad_appr}) is given by (cf. equation~\ref{eq:EOM})
\begin{align}
\label{eq:quad_eom}
\nonumber &\frac{\mathrm{d}\boldsymbol{e}_\mathrm{B}}{\mathrm{d} t} = \frac{6}{\Lambda_\mathrm{B}} \\
\nonumber &\quad \times \left [  C_\mathrm{AB} j_\mathrm{B}^{-5} \left \{ \left ( \boldsymbol{j}_\mathrm{A} \cdot \boldsymbol{j}_\mathrm{B} \right) \left ( \boldsymbol{e}_\mathrm{B} \times \boldsymbol{j}_\mathrm{A} \right) + 5 \left ( \boldsymbol{e}_\mathrm{A} \cdot \boldsymbol{j}_\mathrm{B} \right) \left ( \boldsymbol{e}_\mathrm{A} \times \boldsymbol{e}_\mathrm{B} \right) \right \} \right. \\
\nonumber &\quad \quad \left. + C_\mathrm{BC} j_\mathrm{C}^{-5} \left \{ \left ( \boldsymbol{j}_\mathrm{B} \cdot \boldsymbol{j}_\mathrm{C} \right) \left ( \boldsymbol{e}_\mathrm{B} \times \boldsymbol{j}_\mathrm{C} \right) - 5 \left ( \boldsymbol{e}_\mathrm{B} \cdot \boldsymbol{j}_\mathrm{C} \right) \left ( \boldsymbol{j}_\mathrm{B} \times \boldsymbol{j}_\mathrm{C} \right) \right\} \right ].
\end{align}

A stationary value of $e_\mathrm{B}$ corresponds to
\begin{align}
0= \frac{\mathrm{d}e_\mathrm{B}}{\mathrm{d}t}=\unit{e}_\mathrm{B} \cdot \frac{\mathrm{d}\boldsymbol{e}_\mathrm{B}}{ \mathrm{d} t}.
\end{align}
Neglecting the terms proportional to $C_\mathrm{AB}$ in equation~(\ref{eq:quad_eom}), this condition implies $\boldsymbol{e}_\mathrm{B} \cdot \boldsymbol{j}_\mathrm{C}=0$, and/or $(\boldsymbol{j}_\mathrm{B} \times \boldsymbol{j}_\mathrm{C}) \cdot \boldsymbol{e}_\mathrm{B}=0$. The former cannot be generally true in the case of a maximum eccentricity, therefore the second condition must apply. The latter can be rewritten using the vector identity equation~(\ref{eq:vec_id}) as
\begin{align}
\left ( \unit{e}_\mathrm{B}\cdot \unit{j}_\mathrm{C} \right )^2 = 1 - \left ( \unit{j}_\mathrm{B}\cdot \unit{j}_\mathrm{C} \right )^2.
\end{align}
The mutual inclination between binaries B and C can be related to $e_\mathrm{B}$ using conservation of the total angular momentum vector,
\begin{align}
\label{eq:tot_AM}
\boldsymbol{L}_\mathrm{tot} = \Lambda_\mathrm{A} \boldsymbol{j}_\mathrm{A} + \Lambda_\mathrm{B} \boldsymbol{j}_\mathrm{B} + \Lambda_\mathrm{C} \boldsymbol{j}_\mathrm{C}.
\end{align}
At this level of approximation, 
\begin{align}
\nabla_{\boldsymbol{e}_\mathrm{C}} \overline{H}_0=\boldsymbol{0},
\end{align}
therefore
\begin{align}
\frac{\mathrm{d}e_\mathrm{C}}{\mathrm{d}t}=\unit{e}_\mathrm{C} \cdot \frac{\mathrm{d}\boldsymbol{e}_\mathrm{C}}{ \mathrm{d} t} = -\frac{1}{\Lambda_\mathrm{C}} \left [ \unit{e}_\mathrm{C} \cdot \left (\boldsymbol{e}_\mathrm{C} \times \nabla_{\boldsymbol{j}_\mathrm{C}} \overline{H}_0 \right ) \right ] = 0,
\end{align}
and $e_\mathrm{C}$ is constant. Neglecting the term corresponding to binary A and writing $e_\mathrm{C}=e_\mathrm{C,0}$, equation~(\ref{eq:tot_AM}) gives
\begin{align}
\nonumber \unit{j}_\mathrm{B}\cdot \unit{j}_\mathrm{C} &= \frac{1}{2\sqrt{1-e_\mathrm{B}^2} \sqrt{1-e_\mathrm{C,0}^2}} \left [ 2 \sqrt{1-e_\mathrm{B,0}^2} \sqrt{1-e_\mathrm{C,0}^2} \right. \\
&\quad \times \left. \left (\unit{j}_\mathrm{B}\cdot \unit{j}_\mathrm{C}  \right )_0 + \frac{\Lambda_\mathrm{B}}{\Lambda_\mathrm{C}} \left (e_\mathrm{B}^2 - e_\mathrm{B,0}^2 \right ) \right ].
\label{eq:tot_AM_appr}
\end{align}
Furthermore, if any changes in binary A between the initial and final state are neglected, then the remaining unknown terms in equation~(\ref{eq:H_quad_appr}) are simply given by $e_\mathrm{A} = e_\mathrm{A,0}$, $\boldsymbol{j}_\mathrm{A} \cdot \boldsymbol{j}_\mathrm{B} = (\boldsymbol{j}_\mathrm{A} \cdot \boldsymbol{j}_\mathrm{B} )_0$ and $\boldsymbol{e}_\mathrm{A} \cdot \boldsymbol{j}_\mathrm{B} = (\boldsymbol{e}_\mathrm{A} \cdot \boldsymbol{j}_\mathrm{B} )_0$. 

With these simplifications, equation~(\ref{eq:H_quad_appr}) only contains the single unknown quantity $e_\mathrm{B}$ corresponding to stationary points. In general, this equation cannot be solved analytically. A notable exception is when the term proportional to $C_\mathrm{AB}$ in equation~(\ref{eq:H_quad_appr}) is neglected (i.e. neglecting the contribution from binary A), as is the term proportional to $\Lambda_\mathrm{B}/\Lambda_\mathrm{C}$ in equation~(\ref{eq:tot_AM_appr}) (i.e. assuming a highly hierarchical system). In that case, the solution corresponding to the maximum eccentricity is
\begin{align}
e_\mathrm{B,max} = \sqrt{1 - \frac{5}{3} \left ( \unit{j}_\mathrm{B}\cdot \unit{j}_\mathrm{C} \right )_0^2},
\end{align}
which is a well-known result for hierarchical triple systems applied to binaries B and C, and where binary A is essentially replaced by a point mass (note that $\unit{j}_\mathrm{B}\cdot \unit{j}_\mathrm{C} = \cos[i_\mathrm{BC}]$). More general numerical solutions are shown in the bottom two panels of Fig. \ref{fig:sa} with the solid lines, where $i_\mathrm{BC,min}$ is computed using equation~(\ref{eq:tot_AM_appr}). Although the semianalytic curves do not capture the detailed behaviour of $e_\mathrm{B,max}$ and $i_\mathrm{BC,min}$ in the regime $1\lesssim \mathcal{R}_0 \lesssim 20$, for other $\mathcal{R}_0$ they agree well with the results obtained from the numerical integrations with \textsc{SecularQuadruple}. 

\begin{figure*}
\center
\includegraphics[scale = 0.45, trim = 10mm 0mm 0mm 0mm]{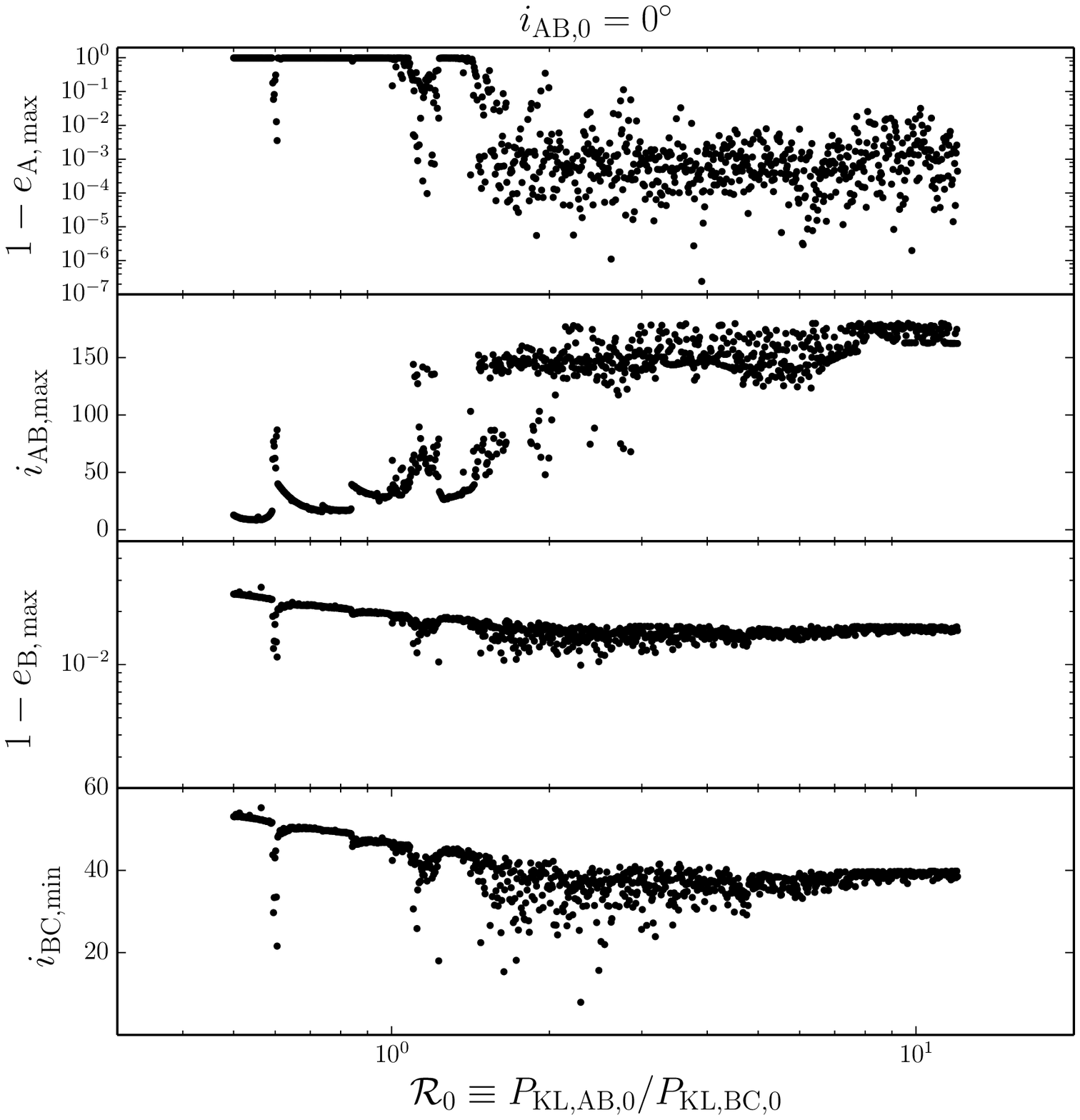}
\includegraphics[scale = 0.45, trim = 10mm 0mm 0mm 0mm]{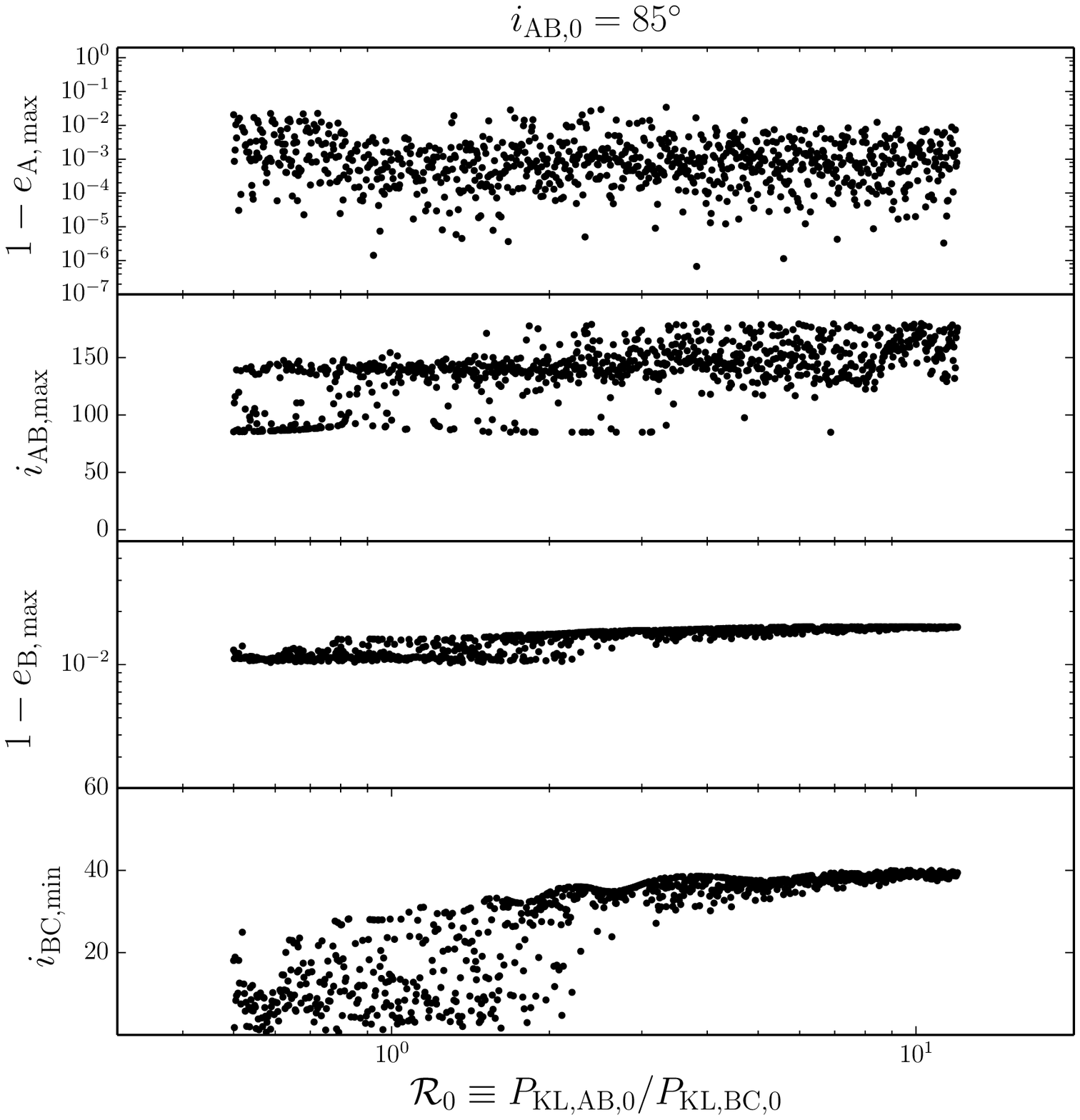}
\caption{\small Similar to Fig. \ref{fig:sa}, showing greater detail near $\mathcal{R}_0=1$.}
\label{fig:sa:R0_1}
\end{figure*}

\subsection{Behaviour near $\mathcal{R}_0=1$}
\label{sect:general:R1}

It is apparent from Fig. \ref{fig:sa} that near $\mathcal{R}_0=1$, the behaviour of the maximum eccentricities of binaries A and B as a function of $\mathcal{R}_0$ is complex and non-regular. Here, we briefly discuss in more detail the behaviour in this regime based on numerical integrations with \textsc{SecularQuadruple}. 

In Fig. \ref{fig:sa:R0_1}, we show the same quantities as in Fig. \ref{fig:sa}, now based on 1000 numerical integrations within a smaller interval of $\mathcal{R}_0$ near $\mathcal{R}_0=1$. In the coplanar case and for $\mathcal{R}_0\lesssim 1$, there are distinct peaks corresponding to enhanced eccentricities in both binaries A and B at specific values of $\mathcal{R}_0$. For $\mathcal{R}_0\gtrsim 1$, individual peaks are harder to distinguish. We speculate that the peaked behaviour is due to resonances in the arguments of pericentre of binaries A and B that occur at specific integer ratios of the KL time-scales for the AB and BC pairs. In addition, for $\mathcal{R}_0\gtrsim 1$ there may be an overlap of many resonances, thereby producing a chaotic behaviour as a function of $\mathcal{R}_0$ \citep{chirikov_79}. Interestingly, the peaks for $\mathcal{R}_0\lesssim1$ are much less pronounced, if not completely absent, in the highly inclined case. 

These phenomena merit further study, but are beyond the scope of this work.

\subsection{General relativistic effects}
\label{sect:general:GR}

\begin{figure*}
\center
\includegraphics[scale = 0.45, trim = 10mm 0mm 0mm 0mm]{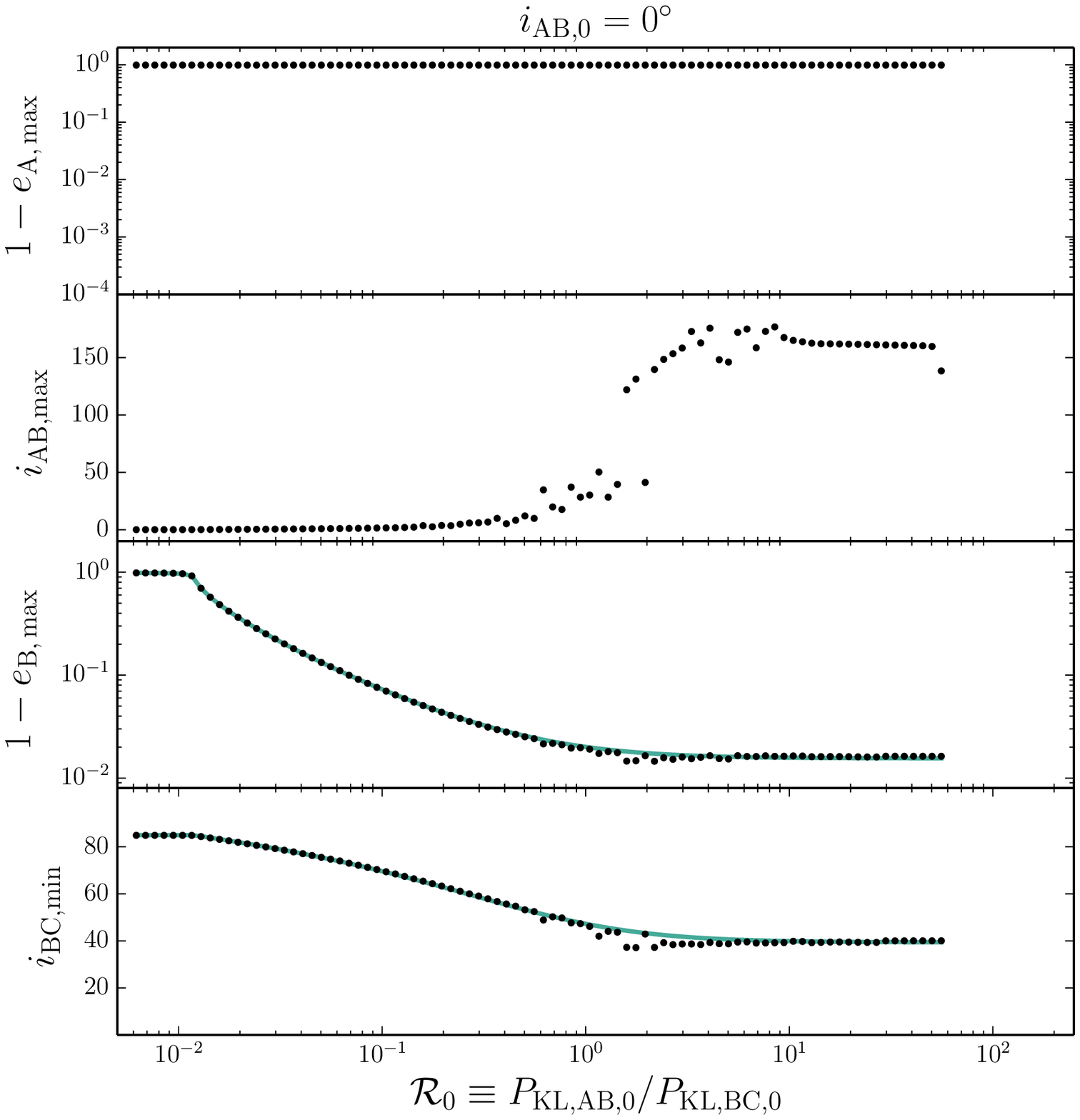}
\includegraphics[scale = 0.45, trim = 10mm 0mm 0mm 0mm]{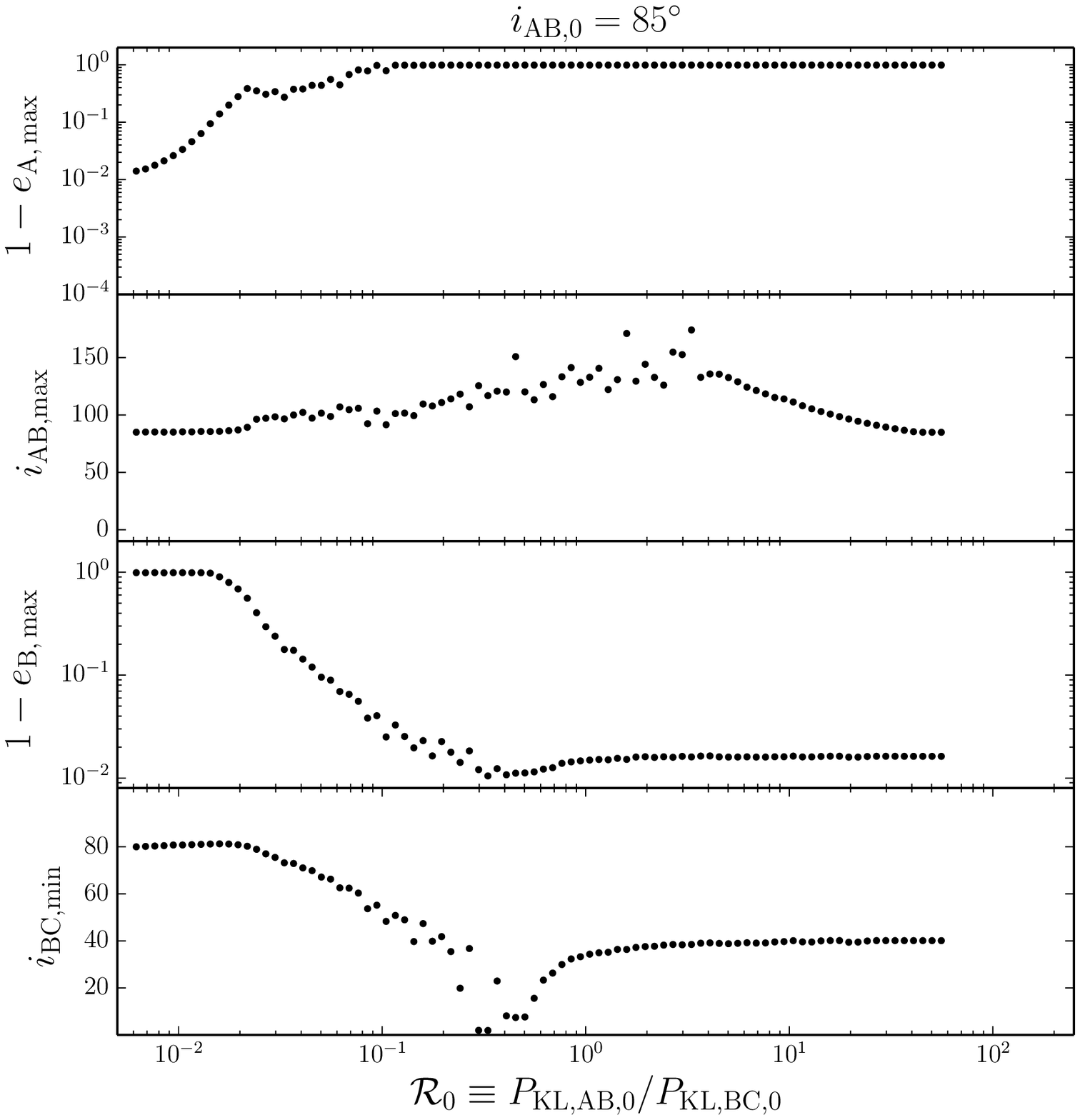}
\caption{\small Similar to Fig. \ref{fig:sa}, but here with the inclusion of relativistic precession to 1PN order in all three binary systems. }
\label{fig:sa_PN}
\end{figure*}

In the results presented above, all four bodies were assumed to be point masses and general relativistic effects were not included. In Fig. \ref{fig:sa_PN}, we show the results of integrations with \textsc{SecularQuadruple} similar to those presented in Fig. \ref{fig:sa}, but now including 1PN precession in the equations of motion for all three binaries (cf. equation~\ref{eq:EOM_1PN}). We note that in the sequence of integrations shown in the left- and right-hand panels of Fig. \ref{fig:sa_PN}, only $a_\mathrm{A}$ is varied; consequently, both $t_\mathrm{1PN,A}$ (cf. equation~\ref{eq:t_1PN}) and $P_\mathrm{KL,AB}$ (cf. equation~\ref{eq:PK}) are affected. For the smallest value of $\mathcal{R}_0$ in Fig. \ref{fig:sa_PN} (largest value of $a_\mathrm{A} = 1 \, \mathrm{AU}$, cf. panels 1-6 in Figs \ref{fig:coplanar:ex} and \ref{fig:inclined:ex}), the initial $t_\mathrm{1PN,A} \approx 18.4 \, \mathrm{Myr}$ and $P_\mathrm{KL,AB} \approx 1.2 \, \mathrm{Myr}$. For the largest value of $\mathcal{R}_0$ in Fig. \ref{fig:sa_PN} (smallest value of $a_\mathrm{A} = 0.001 \, \mathrm{AU}$, cf. panels 7-12 in Figs \ref{fig:coplanar:ex} and \ref{fig:inclined:ex}), the initial $t_\mathrm{1PN,A} \approx 5.8\times 10^{-7} \, \mathrm{Myr}$ and $P_\mathrm{KL,AB} \approx 3.9 \times 10^4 \, \mathrm{Myr}$. 

In the coplanar case, eccentricity oscillations in binary A are quenched due to relativistic precession, even if $\mathcal{R}_0\sim1$. We note, however, that the purely Newtonian results can be rescaled to other systems (in particular, with larger $a_\mathrm{A}$), in which case relativistic precession in binary A becomes unimportant, whereas the purely Newtonian secular dynamics remain unaffected modulo a rescaling of the KL time-scales. 

In the inclined case, the behaviour of the maximum eccentricity in binary A is more complicated (cf. the right-hand panel of Fig. \ref{fig:sa_PN}). For the lowest $\mathcal{R}_0$, $P_\mathrm{KL,AB}<t_\mathrm{1PN,A}$ as mentioned above. As $\mathcal{R}_0$ is increased, the quantity $e_\mathrm{A,max}$ decreases with increasing $\mathcal{R}_0$, which is due to the increasing relative importance of 1PN precession compared to the torque of binary B. However, the decrease of $e_\mathrm{A,max}$ flattens around $\mathcal{R}_0\approx 2\times10^{-2}$. The latter value of $\mathcal{R}_0$ corresponds to a significant increase of $e_\mathrm{B,max}$. The flattening of $e_\mathrm{A,max}$ as a function of $\mathcal{R}_0$ can be explained by considering that as $e_\mathrm{B,max}$ increases, the KL time-scale for the AB pair decreases (cf. equation~\ref{eq:PK}). Consequently, the latter KL time-scale can become comparable to the 1PN precession time-scale. Here, this is the case for $2\times10^{-2}\lesssim \mathcal{R}_0 \lesssim 10^{-1}$. 

\begin{figure*}
\center
\includegraphics[scale = 0.52, trim = 25mm 0mm 0mm 0mm]{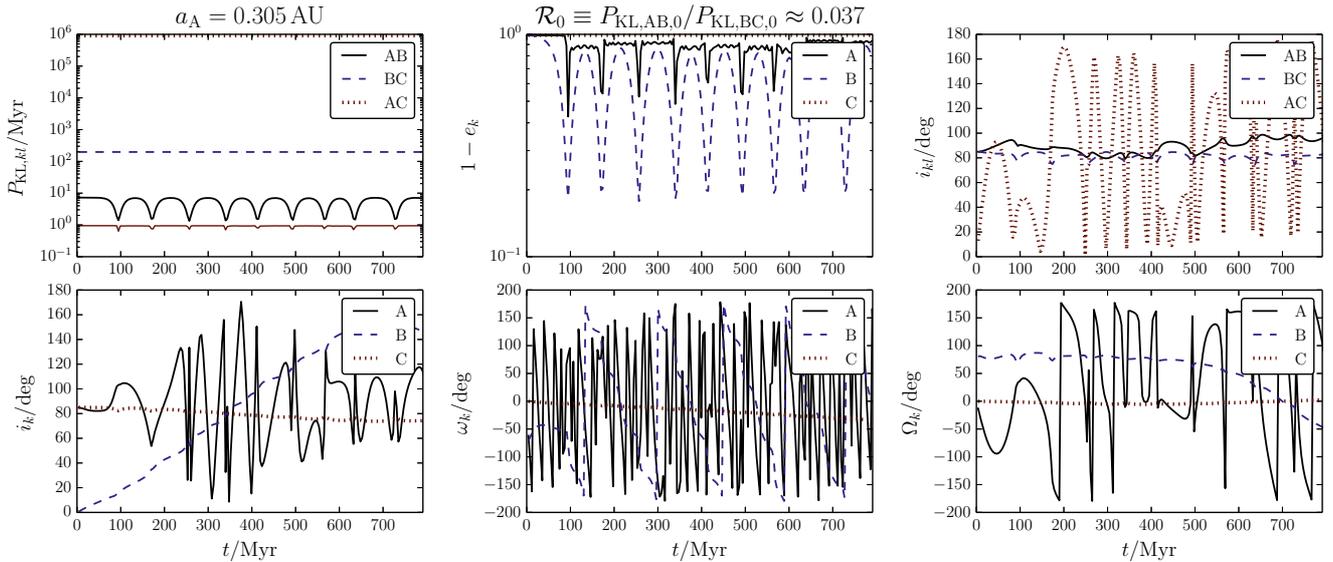}
\caption{\small Evolution for a system taken from Fig. \ref{fig:sa_PN}, demonstrating the effect of `overcoming' 1PN precession by the periodically enhanced eccentricity of system B. 
The assumed initial parameters are semimajor axes $a_\mathrm{B}=0.305\,\mathrm{AU}$, $a_\mathrm{B}=10^2\,\mathrm{AU}$ and $a_\mathrm{C}=5\times10^3\,\mathrm{AU}$, masses $m_1=m_3=m_4=1\,\mathrm{M}_\odot$ and $m_2=0.5\,\mathrm{M}_\odot$, eccentricities $e_\mathrm{A}=e_\mathrm{B}=e_\mathrm{C}=0.01$, inclinations $i_\mathrm{A}=i_\mathrm{C}=85^\circ$ and $i_\mathrm{B}=0^\circ$, arguments of pericentre $\omega_\mathrm{A}=\omega_\mathrm{B}=\omega_\mathrm{C}=0^\circ$ and longitudes of the ascending nodes $\Omega_\mathrm{A}=\Omega_\mathrm{B}=\Omega_\mathrm{C}=0^\circ$. In the top-left panel, the solid red line shows the 1PN precession time-scale in binary A (cf. equation~\ref{eq:t_1PN}). }
\label{fig:inclined:ex:PN}
\end{figure*}

We show an example of this phenomenon in Fig. \ref{fig:inclined:ex:PN}, where $a_\mathrm{A}\approx0.3\,\mathrm{AU}$ and $\mathcal{R}_0\approx 0.04$ (full parameters are given in the caption). At the maxima of $e_\mathrm{B}$, the KL time-scale for the AB binary pair (black solid line in the top-left panel) decreases and becomes comparable to the 1PN precession time-scale in binary A (red solid line in the same panel). This gives rise to increased eccentricities in binary A, to much higher values if $e_\mathrm{B}$ were constant (cf. the top-middle panel). This is a mechanism for -- at least partially -- overcoming the well-known quenching of KL eccentricity cycles induced by 1PN precession. Note, however, that in this case, there is only a narrow region in $\mathcal{R}_0$ for which it is effective: as $\mathcal{R}_0$ increases, $a_\mathrm{A}$ decreases, therefore further decreasing $t_\mathrm{1PN,A}$. 

We note that this phenomenon is general, in the sense that it would also apply if precession in binary A is due to another effect, e.g. tidal effects or mass transfer in stellar systems.

\section{Discussion}
\label{sect:discussion}
\subsection{Application: planetary systems}
\label{sect:discussion:planets}
As mentioned in \S\,\ref{sect:introduction}, the hierarchical configuration considered in this work can be applied to planetary systems consisting of a planet+moon system (binary A) orbiting a central star (in binary B) that is orbited by a more distant and inclined planetary or stellar companion (in binary C). Both binaries A and B are assumed to be initially coplanar and circular. A pertinent question is whether the torque exerted by the fourth body causes the planet+moon system to become inclined with respect to the orbit of the central star, or whether coplanarity is maintained. We note that this is different from the question that has been addressed in the past in which case a different hierarchy was assumed, i.e. all bodies within the stellar binary were assumed to orbit the central star \citep{innanen_ea_97,takeda_rasio_05,takeda_kita_rasio_08}.

Based on the qualitative results presented in \S\,\ref{sect:general:trends}, we expect that coplanarity between binaries A and B is maintained if $P_\mathrm{KL,AB,0}\ll P_\mathrm{KL,BC,0}$, i.e. if the binary companion is distant from the planetary orbit. In addition, we expect KL eccentricity oscillations in the orbit of the planet+moon system with respect to the central star due to the torque of the binary companion to be quenched. This effect could prevent the latter orbit from becoming highly eccentric, i.e. the presence of the moon could `shield' the planet from disruption by the star as a consequence of KL oscillations induced by the binary companion. 

On the other hand, if $P_\mathrm{KL,AB,0}\gg P_\mathrm{KL,BC,0}$, the binary companion is close to the planetary orbit, and the planet+moon system can become inclined with respect to the orbit of the central star. However, in the latter case, the KL time-scale for the AB pair is long compared to that of the BC pair, such that there is no eccentricity excitation in the planet+moon system. In the intermediate regime where $P_\mathrm{KL,AB,0}\sim P_\mathrm{KL,BC,0}$, we expect significant eccentricity oscillations in the planet+moon system. These oscillations could lead to efficient tidal dissipation in cases where this would otherwise not have been important, and, in extreme cases, even to planet+moon collisions. 

We explore in \S\,\ref{sect:discussion:planets:parameter_space} some of the parameter space where significant KL eccentricity oscillations in the planet+moon system are expected, and give a number of examples in \S\,\ref{sect:discussion:planets:examples}. A comprehensive population synthesis study is beyond the scope of this paper.

\begin{figure*}
\center
\includegraphics[scale = 0.5, trim = 20mm 0mm 0mm 0mm]{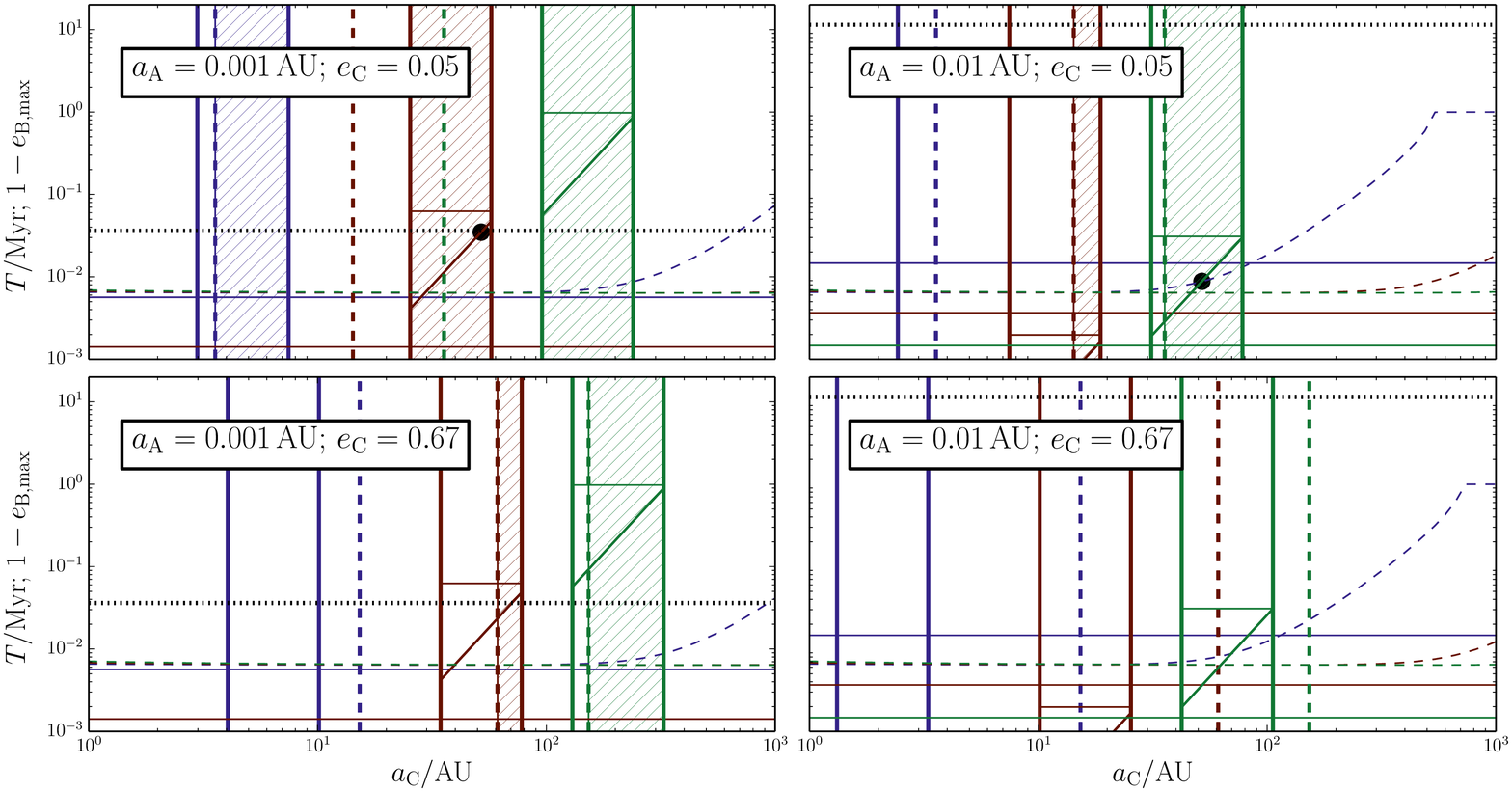}
\caption{\small Exploration of the parameter space where eccentricity oscillations could be induced in a planet+moon system orbiting a central star that is orbited by an inclined binary companion (cf. \S\,\ref{sect:discussion:planets}). Blue, red and green lines correspond to semimajor axes $a_\mathrm{B}$ of the planet+moon system with respect to the central star of 1, 4 and 10 AU, respectively. The ranges of $a_\mathrm{C}$ for which $1<\mathcal{R}_0<20$ are indicated with vertical thick coloured solid lines. Values of $a_\mathrm{C}$ corresponding to dynamical stability of the BC pair (according to the criterion of \citealt{mardling_aarseth_01}) are indicated with vertical thick coloured dashed lines. The regions where we expect that the eccentricity of the planet+moon system is excited, are indicated with hatches. Near the hatches regions, the horizontal solid lines show $P_\mathrm{KL,AB,0}$, whereas the sloped solid lines show $P_\mathrm{KL,BC,0}$. The black horizontal dotted lines indicate the time-scale for relativistic precession in binary A. In addition to these time-scales, we show with non-vertical coloured dashed lines the maximum eccentricity in binary B, computed using the method of \S\,\ref{sect:general:dep_R0:sa}, and assuming $i_\mathrm{BC,0} = 85^\circ$. The horizontal coloured solid lines show the value of $1-e_\mathrm{B}$ for which the planet+moon system is expected to be disrupted by the central star. In the top two panels, the black bullets correspond to the two example systems discussed in \S\,\ref{sect:discussion:planets:examples}. }
\label{fig:discussion:planets}
\end{figure*}

\begin{figure*}
\center
\includegraphics[scale = 0.52, trim = 25mm 8mm 0mm 0mm]{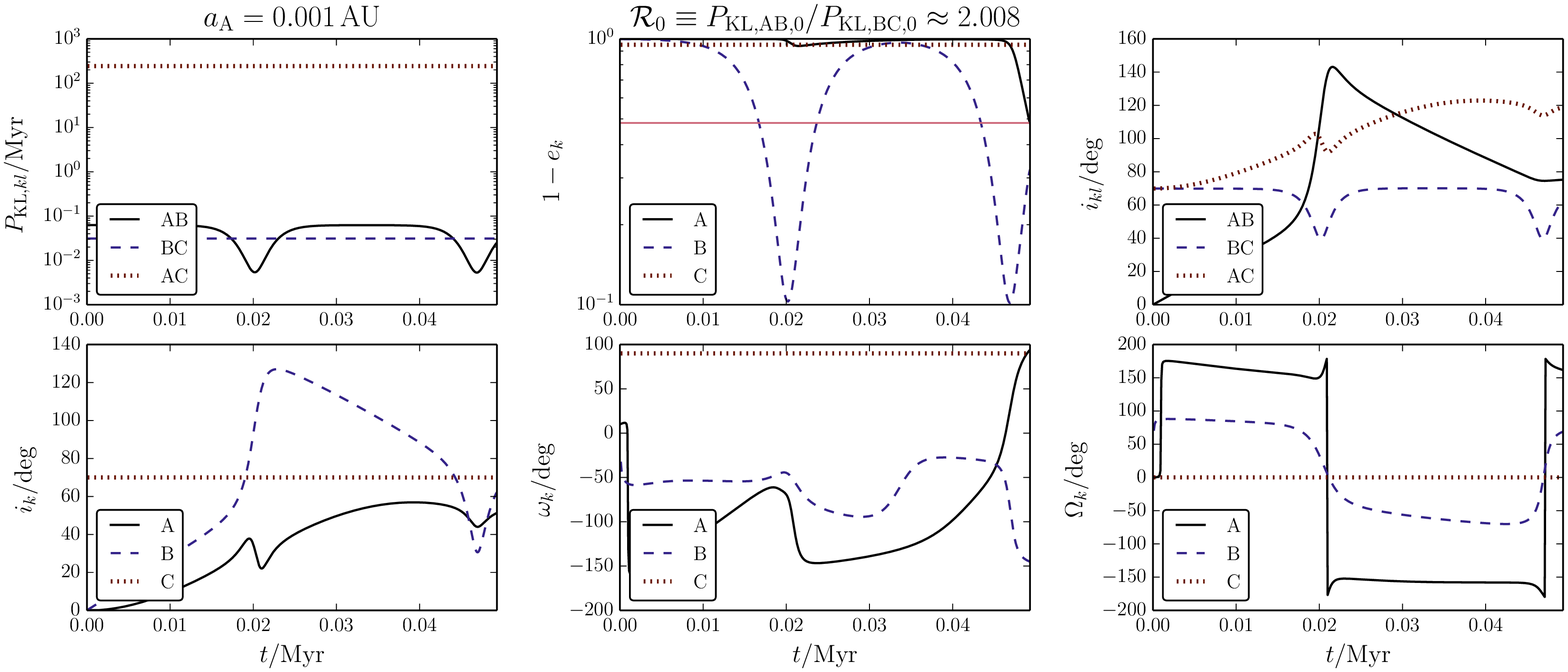}
\includegraphics[scale = 0.52, trim = 25mm 8mm 0mm 0mm]{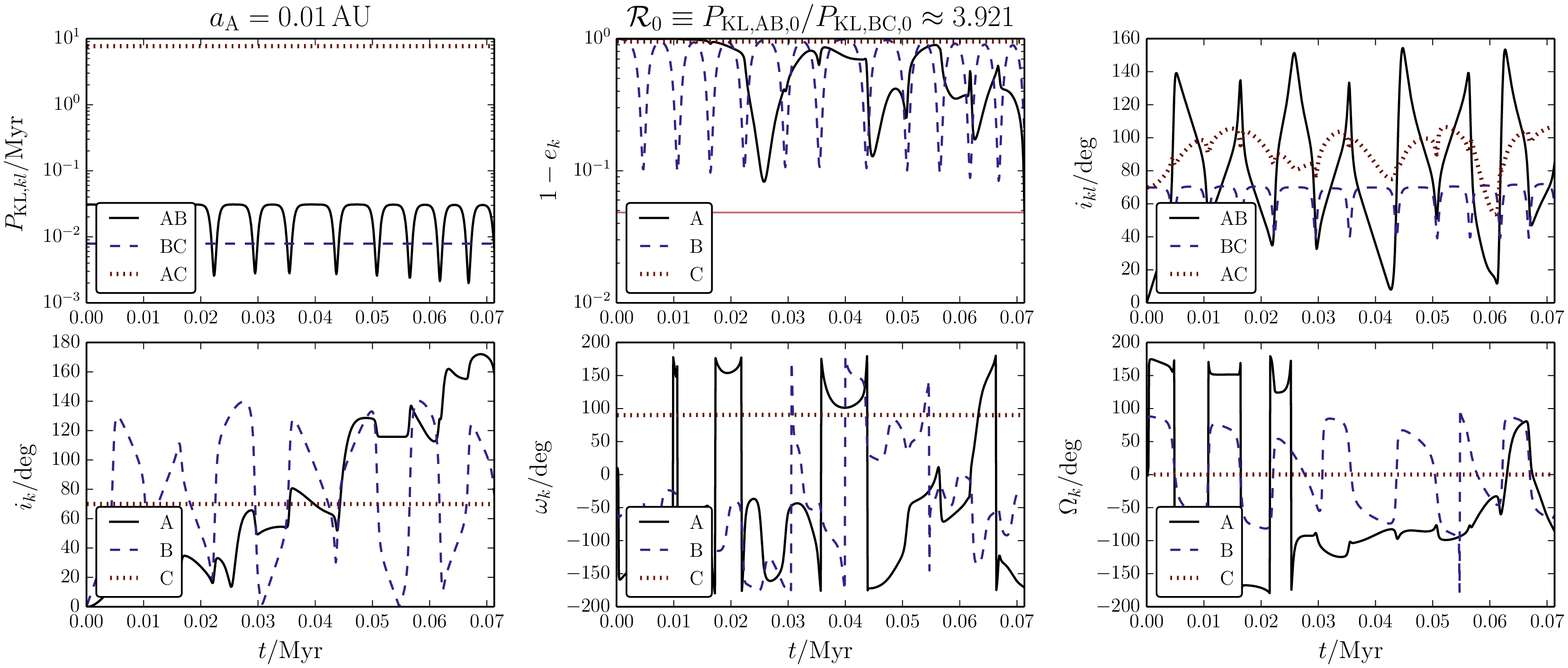}
\caption{\small Evolution of two quadruple systems in the context of planetary systems as discussed in \S\,\ref{sect:discussion:planets:examples}, computed with \textsc{SecularQuadruple}. First two rows: $a_\mathrm{A}=10^{-3}\,\mathrm{AU}$, $a_\mathrm{B}=4\,\mathrm{AU}$ and $a_\mathrm{C}=50\,\mathrm{AU}$; second two rows: $a_\mathrm{A}=10^{-2}\,\mathrm{AU}$, $a_\mathrm{B}=10\,\mathrm{AU}$ and $a_\mathrm{C}=50\,\mathrm{AU}$. Binaries A and B are initially coplanar, whereas $i_\mathrm{BC,0}=70^\circ$. In both examples, the other initial parameters were $m_1=1\,\mathrm{M}_\mathrm{J}$, $m_2=10^{-3}\,\mathrm{M}_\mathrm{J}$, $m_3=1 \, \mathrm{M}_\odot$ and $m_4=0.5 \, \mathrm{M}_\odot$, $e_\mathrm{A} = e_\mathrm{B} = 0.001$ and $e_\mathrm{C} = 0.05$, $i_\mathrm{A}=i_\mathrm{B}=0^\circ$ and $i_\mathrm{C}=70^\circ$, $\omega_\mathrm{A}=10^\circ$, $\omega_\mathrm{B}=40^\circ$ and $\omega_\mathrm{C}=90^\circ$ and $\Omega_\mathrm{A}=\Omega_\mathrm{B}=\Omega_\mathrm{C}=0^\circ$. In the panels showing $e_k$, the horizontal solid red line shows the value of $e_\mathrm{A}$ for which the moon collides with its planet. Here, we assumed a planet radius $R_1 = 1\,R_\mathrm{J}$ and lunar radius $R_2 = 10^{-2} \,R_\mathrm{J}$. The integrations were stopped when $e_\mathrm{A}$ reached this value. }
\label{fig:discussion:planets_examples}
\end{figure*}

\subsubsection{Expectations based on time-scale arguments}
\label{sect:discussion:planets:parameter_space}
We assume a Jupiter-mass planet, $m_1=M_\mathrm{J}$, a moon with mass $m_2 = 10^{-4} \, m_1$ (the order of magnitude of the mass of Jupiter's heaviest moons), a central star with mass $m_3=1\,\mathrm{M}_\odot$, and a binary companion with mass $m_4=0.5\,\mathrm{M}_\odot$. The radii (of interest when considering collisions) are assumed to be $R_1=R_\mathrm{J}$, $R_2=10^{-2} \, R_1$ and $R_3 = 1 \, \mathrm{R}_\odot$.

The semimajor axis of the planet+moon system is assumed to be either $a_\mathrm{A}=10^{-3} \, \mathrm{AU}$ or $a_\mathrm{A}=10^{-2} \, \mathrm{AU}$; the semimajor axis $a_\mathrm{B}$ of the latter system with respect to the central star is either 1, 4 or 10 AU. The eccentricities of binaries A and B are assumed to be $e_\mathrm{A}=e_\mathrm{B}=0.001$; the eccentricity of the orbit of the binary companion is either $e_\mathrm{C}=0.05$ or $e_\mathrm{C}=0.67$. 

In Fig. \ref{fig:discussion:planets}, we show various time-scales of importance as a function of $a_\mathrm{C}$, where in each panel different values are assumed for $a_\mathrm{A}$ and $e_\mathrm{C}$. Quantities pertaining to the three values of $a_\mathrm{B}$ are indicated with blue, red and green lines for values of $a_\mathrm{B}$ of 1, 4 and 10 AU, respectively. The critical values of $a_\mathrm{C}$ corresponding to dynamical stability, computed using the three-body criterion of \citet{mardling_aarseth_01} and where binary A is treated as a point mass, are indicated with vertical dashed lines for each value of $a_\mathrm{B}$. Systems to the left of these lines are expected to be dynamically unstable. 

Extrapolating our results from \S\,\ref{sect:general}, we expect the region in parameter space in which $e_\mathrm{A}$ can be excited (in the absence of relativistic effects and other additional sources of absidal motion), to be approximately $1\lesssim \mathcal{R}_0 \lesssim 20$. The limiting values of $a_\mathrm{C}$, for each value of $a_\mathrm{B}$, are indicated with the vertical solid lines, and between these vertical lines the coloured horizontal (sloped) solid lines indicate the KL time-scales for pair AB (BC). We have indicated with hatched regions the ranges in $a_\mathrm{C}$ satisfying $1< \mathcal{R}_0 < 20$, and the stability constraint. 

In principle, the mechanism for producing high-amplitude oscillations in $e_\mathrm{A}$ in the regime $\mathcal{R}_0\sim1$ can be suppressed if KL oscillations in system B are quenched by relativistic precession in binary B. In all cases in Fig. \ref{fig:discussion:planets}, these time-scales are longer than 10 Myr, and therefore, precession in binary B is not important. Relativistic precession in binary A is of greater importance given the small values of $a_\mathrm{A}$; the associated time-scales are indicated in Fig. \ref{fig:discussion:planets} with black dotted horizontal lines. 

Based on Fig. \ref{fig:discussion:planets}, we expect eccentricity excitation in the planet+moon system for specific ranges in $a_\mathrm{C}$. These ranges strongly depend on $a_\mathrm{A}$, $a_\mathrm{B}$ and $e_\mathrm{C}$. For small semimajor axes of the planet+moon system, i.e. $a_\mathrm{A}=10^{-2}\,\mathrm{AU}$, the criterion of dynamical stability of the orbit of the binary companion does not strongly reduce the parameter space. General relativistic precession is, however, also more important for smaller $a_\mathrm{A}$. Nevertheless, for values of $a_\mathrm{B}$ of 4 and 10 AU, the relativistic precession time-scale in binary A is not much shorter than the KL time-scale for the AB pair. In those cases, there could still be high-eccentricity oscillations in binary A because of the reduction of the KL time-scale for the AB pair as a consequence of the eccentricity oscillations in binary B (cf. \S\,\ref{sect:general:GR}). This is demonstrated below in the first example in \S\,\ref{sect:discussion:planets:examples}.

A larger eccentricity of the binary companion tends to reduce the parameter space of interest. The reason for this decrease is the larger range in $a_\mathrm{C}$ for which the system is not dynamically stable.

As discussed in \S\,\ref{sect:general}, for $\mathcal{R}_0\ll1$, KL eccentricity oscillations in binary B are quenched because of the induced precession from binary A. We have plotted the maximum eccentricity in binary B as a function of $a_\mathrm{C}$ in Fig. \ref{fig:discussion:planets} with dashed lines, computed using the semianalytic method described in \S\,\ref{sect:general:dep_R0:sa}. Here, we assumed $i_\mathrm{BC,0}=85^\circ$ to get a rough upper limit of the maximum eccentricity. The quenching effect is very effective for $a_\mathrm{B}=1\,\mathrm{AU}$ and $a_\mathrm{C}$ larger than a few 100 AU. For large enough $a_\mathrm{C}$, eccentricity oscillations in binary B are completely quenched.

To illustrate the implications of this, we have indicated in Fig. \ref{fig:discussion:planets} with horizontal coloured lines the values of $1-e_\mathrm{B}$ that satisfy $1-e_\mathrm{B}=(a_\mathrm{A}+R_3)/a_\mathrm{B}$, i.e. the eccentricity for which the pericentre distance of the orbit of binary B is equal to $a_\mathrm{A}+R_3$. In the latter case, we expect the planet, the moon, or both, to be disrupted by the central star. For $a_\mathrm{A} = 10^{-2}\,\mathrm{AU}$ and $a_\mathrm{B} = 1\,\mathrm{AU}$, the maximum eccentricity reached in binary B exceeds this value for $a_\mathrm{C}\lesssim 100 \, \mathrm{AU}$. However, for $a_\mathrm{C}\gtrsim 100 \, \mathrm{AU}$, a potentially catastrophic encounter of the planet+moon system with the central star is avoided because of quenching of the KL eccentricity oscillations in binary B. This shows more quantitatively the `shielding' effect mentioned above.

To conclude, we expect that there exist regions in parameter space in which the eccentricity of the planet+moon system is excited, despite initial coplanarity. The region in parameter space is limited, however: the planet should be sufficiently far away from the central star, yet the orbit of the binary companion should also be dynamically stable. In addition, the latter orbit needs to be sufficiently inclined. In contrast, if the orbit of the binary companion is wide, the presence of the moon can prevent the orbit of the planet+moon system around the star from becoming highly eccentric.

\subsubsection{Examples}
\label{sect:discussion:planets:examples}
To further illustrate the planetary system discussed here, we show in Fig. \ref{fig:discussion:planets_examples} two examples of integrations with \textsc{SecularQuadruple}. In the first two rows, $a_\mathrm{A}=10^{-3}\,\mathrm{AU}$, $a_\mathrm{B}=4\,\mathrm{AU}$ and $a_\mathrm{C}=50\,\mathrm{AU}$ (cf. the black bullet in the top-left panel in Fig. \ref{fig:discussion:planets}); in the second two rows, $a_\mathrm{A}=10^{-2}\,\mathrm{AU}$, $a_\mathrm{B}=10\,\mathrm{AU}$ and $a_\mathrm{C}=50\,\mathrm{AU}$ (cf. the black bullet in the top-right panel in Fig. \ref{fig:discussion:planets}). In both cases, we assume $i_\mathrm{BC,0}=70^\circ$ and $e_\mathrm{C,0}=0.05$. For the other parameters, we refer to \S\,\ref{sect:discussion:planets:parameter_space}. In both examples, $R_\mathrm{0} \sim 1$, and high-eccentricity oscillations are expected in the planet+moon system.

The values of $1-e_\mathrm{A}$ corresponding to a collision between the planet and its moon are indicated with horizontal red lines in the corresponding panels in Fig. \ref{fig:discussion:planets_examples}. Such collisions occur in both examples at $\approx 0.05$ and $0.07$ Myr, respectively, and the integrations were subsequently stopped. Note that the eccentricity of binary B does not become high enough for disruption of the planet+moon system by the central star. Particularly in the second example, $e_\mathrm{A}$ shows a complicated behaviour as a function of time, changing rapidly each time $i_\mathrm{AB}$ passes $90^\circ$. 

We remark that tidal dissipation was not included in these examples. This effect is likely important for the small pericentre distances reached during the evolution, therefore possibly not resulting in a collision, but a shrinking of the planet+moon orbit. 

\begin{figure*}
\center
\includegraphics[scale = 0.52, trim = 25mm 8mm 0mm 0mm]{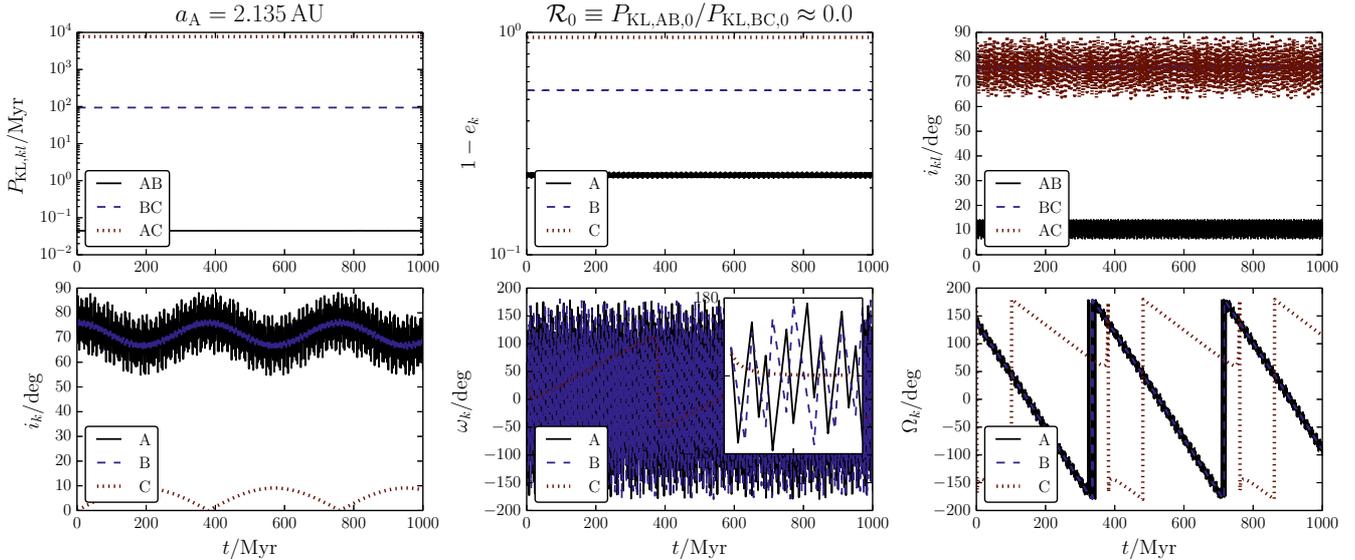}
\caption{\small Example evolution for the quadruple system ADS 1652 as discussed in \S\,\ref{sect:discussion:stellar:ADS_1652}, computed with \textsc{SecularQuadruple}. The parameters are given in Table \ref{table:ADS_1652}; the currently unconstrained parameters pertaining to the outermost orbit, binary C, are $e_\mathrm{C}=0.05$, $i_\mathrm{C}=0^\circ$, $\omega_\mathrm{C}=90.0^\circ$ and $\Omega_\mathrm{C}=130.0^\circ$. In the bottom middle panel, the inset shows a magnification for $t=0$ to 20 Myr; note that both $\omega_\mathrm{A}$ and $\omega_\mathrm{B}$ are undersampled. }
\label{fig:discussion:stellar_ADS_1652}
\end{figure*}

\subsection{Application: observed stellar quadruples}
\label{sect:discussion:stellar}

\subsubsection{ADS 1652}
\label{sect:discussion:stellar:ADS_1652}
The quadruple system ADS 1652 (\citealt{tokovinin_gorynya_morrell_14}, and references therein) is composed of four main-sequence stars in the `3+1' configuration. The system is likely old (age $>\mathrm{Gyr}$) considering the spectral types of its stellar components; the stars in binary A are of spectral type G9V, the star in binary B is of type K5V and the star in binary C is of type G8V. To date, ADS 1652 is one of few quadruple systems for which orbital fits have been obtained for multiple orbits. 

\begin{table*}
\begin{tabular}{ccccccccccccccccccc}
\toprule
$m_1$ & $m_2$ & $m_3$ & $m_4$ & $a_\mathrm{A}$ & $a_\mathrm{B}$ & $a_\mathrm{C}$ & $e_\mathrm{A}$ & $e_\mathrm{B}$ & $e_\mathrm{C}$ & $i_\mathrm{A}$ & $i_\mathrm{B}$ & $i_\mathrm{C}$ & $\omega_\mathrm{A}$ & $\omega_\mathrm{B}$ & $\omega_\mathrm{C}$ & $\Omega_\mathrm{A}$ & $\Omega_\mathrm{B}$ & $\Omega_\mathrm{C}$ \\
\midrule
0.74 & 0.72 & 0.57 & 0.78 & 2.135 & 45.2 & 2500 & 0.769 & 0.45 & -- & 75.9 & 76.0 & -- & 287.9 & 175.0 & -- & 127.2 & 140.0 & -- \\
\bottomrule
\end{tabular}
\caption{ Parameters of the quadruple system ADS 1652 discussed in \S\,\ref{sect:discussion:stellar:ADS_1652}, adopted from \citet{tokovinin_gorynya_morrell_14}, who fitted radial velocity and speckle measurements to the orbits of binaries A and B. The masses are obtained from these orbital fits (cf. the bottom row of table 7 of \citealt{tokovinin_gorynya_morrell_14}), the semimajor axes of the A and B binaries are computed from the fitted orbital periods (cf. table 4 of \citealt{tokovinin_gorynya_morrell_14}) using Kepler's law. The eccentricity and orbital orientation of binary C are unknown (indicated with dashes), and are sampled with a Monte Carlo method in \S\,\ref{sect:discussion:stellar:ADS_1652}. Masses are expressed in solar masses, semimajor axes in AU, and angles in degrees. }
\label{table:ADS_1652}
\end{table*}

Here, we apply the \textsc{SecularQuadruple} algorithm to ADS 1652 to explore its long-term secular dynamical evolution. We adopt the parameters that were obtained by \citet{tokovinin_gorynya_morrell_14}, who fitted radial velocity and speckle measurements to the orbits of binaries A and B, and which are given in Table \ref{table:ADS_1652}. Here, we adopted the component masses obtained from the orbital fits (cf. the bottom row of table 7 of \citealt{tokovinin_gorynya_morrell_14}), and computed the semimajor axes of the A and B binaries from the orbital periods (cf. table 4 of \citealt{tokovinin_gorynya_morrell_14}) using Kepler's law. For the semimajor axis of the C binary, we adopt the observed projected distance of 2500 AU from binary A. Owing to its long orbital period of $\sim10^5 \, \mathrm{yr}$, the eccentricities and orbital angles of binary C are not known. Here, we proceed by sampling these quantities for 500 realizations of the system, where $e_\mathrm{C}$ is sampled from a thermal distribution, $i_\mathrm{C}$ from a distribution uniform in $\cos(i_\mathrm{C})$, and $\omega_\mathrm{C}$ and $\Omega_\mathrm{C}$ from a uniform distribution. In our integrations, we included terms up and including octupole order (excluding the cross term), and the 1PN relativistic precession terms in the three binaries. The integration time is $20\,P_\mathrm{KL,BC}$, which is typically a few Gyr (depending on $e_\mathrm{C}$).

We show in Fig. \ref{fig:discussion:stellar_ADS_1652} the evolution of an example system, where $e_\mathrm{C}=0.05$, $i_\mathrm{C}=0^\circ$, $\omega_\mathrm{C}=90.0^\circ$ and $\Omega_\mathrm{C}=130.0^\circ$. For this value of $e_\mathrm{C}$, $\mathcal{R}_0 \approx 6.7\times10^{-4} \ll 1$, therefore the system is in the regime in which the torque of binary B on binary A dominates compared to the torque of binary C on binary B. Indeed, binaries A and B, which are initially nearly coplanar, remain nearly coplanar during the evolution (cf. the top-right panel in Fig. \ref{fig:discussion:stellar_ADS_1652}). Consequently, the KL eccentricity oscillations in binary A are of a very low amplitude, i.e. $e_\mathrm{A,max}\approx 0.779$, whereas $e_\mathrm{A,0}=0.769$. Furthermore, KL eccentricity oscillations in binary B, which is initially inclined with respect to binary C with $i_\mathrm{BC,0}\approx 70^\circ$, are completely quenched. This can be attributed to the rapid precession induced in B binary by binary A, on the time-scale of $P_\mathrm{KL,AB} \approx 4\times10^{-2} \, \mathrm{Myr} \ll P_\mathrm{KL,BC}\approx 10^2\,\mathrm{Myr}$ (cf. the bottom middle panel of Fig. \ref{fig:discussion:stellar_ADS_1652}). 

\begin{figure}
\center
\includegraphics[scale = 0.32, trim = 15mm 0mm 0mm 0mm]{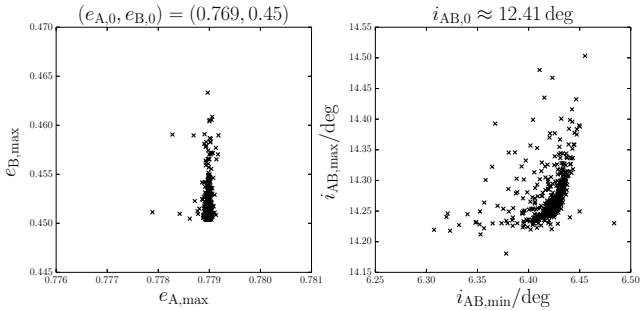}
\caption{\small The maximum eccentricities in the A and B binaries (left-hand panel), and the minimum and maximum inclinations between binaries A and B (right-hand panel), computed from numerical integrations with \textsc{SecularQuadruple} of 500 realizations of ADS 1652, where the parameters of the outermost orbit were varied (except $a_\mathrm{C}$). The dependence on the latter parameters is very weak.}
\label{fig:discussion:stellar_ADS_1652_MC}
\end{figure}

In Fig. \ref{fig:discussion:stellar_ADS_1652_MC}, $e_\mathrm{C,0}=0.05$ was assumed to be low. The quantity $\mathcal{R}_0$ increases with increasing $e_\mathrm{C,0}$ (cf. equation~\ref{eq:R0_def}). Therefore, for larger $e_\mathrm{C,0}$, the system could be in a very different regime in $\mathcal{R}_0$ in which the evolution is very different. This is not the case in our Monte Carlo realizations, however, for which the mean and standard deviations of $\mathcal{R}_0$ are $\approx 1.4 \times 10^{-3}$ and $\approx 1.0\times 10^{-3}$, respectively. In Fig. \ref{fig:discussion:stellar_ADS_1652_MC}, we show for the 500 integrations the maximum eccentricities in the A and B binaries, and the minimum and maximum inclinations between binaries A and B. There is very small spread in all of these quantities, showing that their dependence on $e_\mathrm{C}$, as well as $i_\mathrm{C}$, $\omega_\mathrm{C}$ and $\Omega_\mathrm{C}$, is very weak. 

We conclude that, based on the observed state of ADS 1652, the eccentricities of its orbits will remain very nearly constant for, at least, the remainder of the main-sequence time-scale of its constituents. This conclusion is independent of the currently unknown eccentricity and orientation of the outermost orbit. In particular, even if the latter orbit is highly inclined with respect to the intermediate orbit, any potential KL eccentricity oscillations in the intermediate orbit are efficiently quenched. 

\begin{figure}
\center
\includegraphics[scale = 0.5, trim = 15mm 0mm 0mm 0mm]{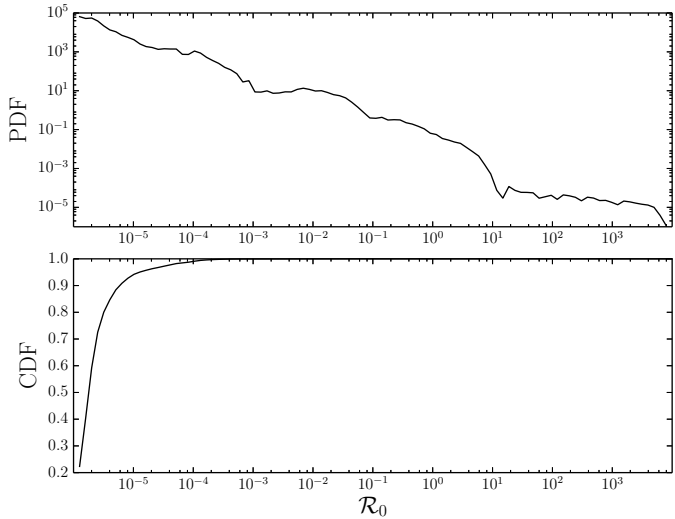}
\caption{\small The distribution of the values of $\mathcal{R}_0$ for the `3+1' quadruple systems in the catalogue of \citet{tokovinin_14a,tokovinin_14b}, obtained by sampling, in 1000 realizations, $e_\mathrm{B}$ and $e_\mathrm{C}$ from a thermal distribution. The probability (cumulative) density function is shown in the top (bottom) panel. For the majority of systems ($\approx 0.9$), $\mathcal{R}_0<10^{-5}$ is small. }
\label{fig:discussion:stellar_tok14}
\end{figure}

\subsubsection{The Tokovinin sample of nearby FG dwarfs}
\label{sect:discussion:stellar:tok}
As mentioned in \S\,\ref{sect:introduction}, 55 of the 4847 observed systems of FG dwarfs in the catalogue of \citet{tokovinin_14a,tokovinin_14b} are quadruple systems. From these, 18 are in the `3+1' configuration, and for 13 of the latter, orbital periods and component masses are known for all three binaries. Here, we briefly explore in which dynamical regimes we expect these systems to be, by computing the associated value of $\mathcal{R}_0$ (cf. \S\,\ref{sect:general:trends}). 

For the 13 systems mentioned above, the orbital elements, apart from the semimajor axes, are unknown. In order to compute $\mathcal{R}_0$, the eccentricities $e_\mathrm{B}$ and $e_\mathrm{C}$ are required (cf. equation~\ref{eq:R0_def}). Therefore, for each of the 13 systems, we sample, in 1000 realizations, $e_\mathrm{B}$ and $e_\mathrm{C}$ from a thermal distribution. Here, we reject sampled eccentricities if either of the AB and BC pair would be unstable according to the dynamical stability criterion of \citet{mardling_aarseth_01}. 

The distribution of the values of $\mathcal{R}_0$ obtained in this approach is shown in Fig. \ref{fig:discussion:stellar_tok14}. The ratio $\mathcal{R}_0$ is typically small; $\approx 0.9$ of the sampled systems have $\mathcal{R}_0<10^{-5}$. This is the regime in which the AB pair is effectively an isolated triple, and where induced precession of binary A on binary B quenches KL eccentricity oscillations in binary B, as a consequence of the torque of binary C. 

We note that one might expect currently observed quadruples not to be in the regime $\mathcal{R}_0\sim1$. If $\mathcal{R}_0\sim 1$, then the large eccentricities in the innermost binary would likely already have strongly affected the system, and possibly have resulted in a merger. Evidently, in this case, the system would not have been observed as a quadruple system, but as a triple system. Conversely, some of the observed quadruple systems may have been quintuple systems in the past, and, triggered by secular dynamical evolution, evolved into quadruple systems through the merging of the stars in (likely) the shortest-period binary.

\section{Conclusions}
\label{sect:conclusions}
We have explored the global gravitational dynamics of hierarchical quadruple systems consisting of a hierarchical triple system orbited by a fourth body. Our main conclusions are as follows.

\medskip \noindent 1. The Hamiltonian for the system has been derived and expanded to up and including fourth order in the ratios of the binary separations $r_\mathrm{A}/r_\mathrm{B}$, $r_\mathrm{B}/r_\mathrm{C}$ and $r_\mathrm{A}/r_\mathrm{C}$ (cf. Fig. \ref{fig:hierarchy}). At each order, we have found three terms that are each mathematically equivalent to the corresponding terms that appear in the hierarchical three-body problem, and that depend on the properties of only two binaries. In addition to these terms, for octupole and higher orders, we have found `cross terms' that depend on properties of all three binaries. Subsequently, we have derived expressions for the orbit-averaged Hamiltonian. A preliminary analysis indicates that the cross terms are typically not important in highly hierarchical systems on short time-scales, i.e. not exceeding time-scales of order $P_\mathrm{KL,BC}$, where $P_\mathrm{KL,BC}$ is the KL time-scale of the BC pair. We have also derived the Hamiltonian for the configuration of two binaries orbiting each other's barycentre (Appendix \ref{app:ham:binary_binary}). 

\medskip \noindent 2. For highly hierarchical systems, i.e. in which the three binaries are widely separated, the global dynamics can be qualitatively described in terms of the (initial) ratio of the KL time-scales of the AB to the BC pairs, $\mathcal{R}_0 \equiv P_\mathrm{KL,AB,0}/P_\mathrm{KL,BC,0}$. 

If $\mathcal{R}_0\ll 1$, the torque of binary B on A dominates compared to the torque of binary C on binary B, and therefore binaries A and B remain coplanar if this was initially the case. If binaries A and B are initially inclined, KL eccentricity oscillations in binary A are not much affected by the presence of the fourth body. Eccentricity oscillations in binary B are efficiently quenched due to short time-scale precession induced on binary B by binary A. 

If $\mathcal{R}_0\gg 1$, the torque of binary C on binary B dominates compared to the torque of binary B on binary A. Initially, the inclination of binary B changes, whereas this is not the case for binary A. This induces a mutual inclination between binaries A and B, even if they are initially not inclined. However, rapid precession of binary B compared to the KL time-scale for the AB pair prevents any significant eccentricity oscillations in binary A, and even quenches KL oscillations if binaries A and B are initially inclined. 

Lastly, if $\mathcal{R}_0\sim 1$, complex KL eccentricity oscillations occur in binary A that are strongly coupled with the KL eccentricity oscillations in binary B. The latter are also affected compared to the situation in which binary A were replaced by a point mass, although this is typically a much smaller effect. Even if binaries A and B are initially coplanar, the induced inclination can result in very high eccentricity oscillations in binary A. These extreme eccentricities could have significant implications for strong interactions such as tidal interactions, gravitational wave dissipation, and collisions and mergers of stars and compact objects.

\medskip \noindent 3. We also included the effects of general relativity, in particular relativistic precession. We have found that the range in the parameter space of the semimajor axis ratios $a_\mathrm{B}/a_\mathrm{A}$ for which KL oscillations are important in binary A can be extended compared to hierarchical triple systems. This is due to a decrease of the KL time-scale of the AB pair when the eccentricity of binary B is at a maximum.

\medskip \noindent 4. We have applied our results to a planetary configuration consisting of a planet+moon system orbiting a central star that is orbited by a more distant and inclined binary companion. We have found that there are regions in parameter space where a planet+moon system that is initially coplanar with respect to the central star, can become inclined and the eccentricity in the planet+moon system can be excited. This could result in significant tidal dissipation and even a collision of the planet with its moon. Furthermore, when the orbit of the binary companion is wide, KL eccentricity oscillations in the orbit of the planet+moon system around the central star can be quenched because of induced precession from the planet+moon system. This effectively shields the planet from high-eccentricity KL oscillations from a binary companion, and, therefore, potential disruption by the central star could be avoided. 

\medskip \noindent 5. Lastly, we applied our results to stellar quadruple systems. In the case of ADS 1652, $\mathcal{R}_0 \sim 10^{-3}$ assuming a thermal distribution of the unknown $e_\mathrm{C}$, and we find almost negligible KL eccentricity oscillations in both the innermost and intermediate orbits, binaries A and B. Even if the outer orbit, binary C, were highly inclined with respect to binary B, any potential KL eccentricity oscillations in binary B would be efficiently quenched. 

For the `3+1' FG stellar quadruples in the catalogue of \citet{tokovinin_14a,tokovinin_14b}, we estimate $\approx 0.9$ of the systems to have $\mathcal{R}_0 < 10^{-5}$. Therefore, we expect that in the majority of these systems, KL eccentricity oscillations in the BC pair are quenched, and, from a secular dynamical point of view, the innermost AB pair can be considered as an isolated triple.

\section*{Acknowledgements}
We thank the referee, Smadar Naoz, for providing very helpful comments that lead to improvement of the paper. This work was initiated during the International Summer-Institute for Modeling in Astrophysics (ISIMA) in 2014, hosted at CITA at the University of Toronto. It was supported by the Netherlands Research Council NWO (grants \#639.073.803 [VICI],  \#614.061.608 [AMUSE] and \#612.071.305 [LGM]) and the Netherlands Research School for Astronomy (NOVA). HBP acknowledges support from the ISF I-CORE programme 1829, The European FP-7 CIG programme `GRAND' (333644), the BSF grant number 2012384 and the Asher foundation.

\bibliographystyle{mnras}
\bibliography{literature}

\onecolumn
\appendix
\section{The Hamiltonian for hierarchical quadruple systems}
\label{app:ham}
The general Newtonian four-body Hamiltonian is given by
\begin{align}
H = T + U = \frac{1}{2} \sum_{i=1}^4 m_i \boldsymbol{V}_i^2 - \frac{1}{2} \sum_{i,j; i\neq j} \frac{Gm_i m_j}{|| \boldsymbol{R}_i-\boldsymbol{R}_j||},
\label{eq:app:Hgen}
\end{align}
where $m_i$, $\boldsymbol{R}_i$ and $\boldsymbol{V}_i$ denote the mass and position and velocity vectors of body $i$. Here, we consider dynamically stable hierarchical configurations. For four bodies, these consist of (1) a hierarchical triple system orbited by a fourth body in an orbit around the triple (discussed in detail in the main text), and (2) two binary systems orbiting each other's centre of mass. Below we discuss both configurations separately.

\subsection{Hierarchical triple system orbited by a fourth body}
\label{app:ham:circumstellar_triple}
In this configuration, we assume that bodies 1 and 2 are bound in binary A, body 3 is bound to the barycentre of bodies 1 and 2 in binary B, and body 4 is bound to the barycentre of bodies 1, 2 and 3 in binary C. It is convenient to define the following separation vectors,
\begin{subequations}
\label{eq:app:rABC}
\begin{align}
\boldsymbol{r}_\mathrm{A} &\equiv \boldsymbol{R}_1 - \boldsymbol{R}_2; \\
\boldsymbol{r}_\mathrm{B} &\equiv \frac{m_1 \boldsymbol{R}_1 + m_2 \boldsymbol{R}_2}{m_1 + m_2} - \boldsymbol{R}_3; \\
\boldsymbol{r}_\mathrm{C} &\equiv \frac{m_1 \boldsymbol{R}_1 + m_2 \boldsymbol{R}_2 + m_3 \boldsymbol{R}_3}{m_1 + m_2 + m_3} - \boldsymbol{R}_4.
\end{align}
\end{subequations}
In addition, we define the centre of mass position of the four-body system, 
\begin{align}
\boldsymbol{r}_\mathrm{CM} \equiv \left [ \sum_{i=1}^4 m_i \right ]^{-1} \sum_{i=1}^4 m_i \boldsymbol{R}_i,
\label{eq:app:rCM}
\end{align}
which satisfies $\dot{\boldsymbol{r}}_\mathrm{CM} = \boldsymbol{0}$. Equations~(\ref{eq:app:rABC}) and (\ref{eq:app:rCM}) are easily inverted to give $\boldsymbol{R}_i$ in terms of $\boldsymbol{r}_\mathrm{A}$, $\boldsymbol{r}_\mathrm{B}$, $\boldsymbol{r}_\mathrm{C}$ and $\boldsymbol{r}_\mathrm{CM}$. Differentiating the resulting relations with respect to time, and assuming that the masses are constant, we find for the kinetic energy
\begin{align}
T = \frac{1}{2} \frac{m_1 m_2}{m_1+m_2} \dot{\boldsymbol{r}}_\mathrm{A}^2 + \frac{1}{2} \frac{(m_1+m_2)m_3}{m_1+m_2+m_3} \dot{\boldsymbol{r}}_\mathrm{B}^2  + \frac{1}{2} \frac{(m_1 + m_2 + m_3)m_4}{m_1+m_2+m_3+m_4} \dot{\boldsymbol{r}}_\mathrm{C}^2.
\end{align}

To find the potential, we similarly invert equations~(\ref{eq:app:rABC}) to give the difference vectors $||\boldsymbol{R}_i-\boldsymbol{R}_j||$ in terms of $\boldsymbol{r}_\mathrm{A}$, $\boldsymbol{r}_\mathrm{B}$ and $\boldsymbol{r}_\mathrm{C}$. Substituting the resulting relations into the potential $U$ yields six terms each of the form 
\begin{align}
\nonumber {||\boldsymbol{r}_\mathrm{C} + \alpha \boldsymbol{r}_\mathrm{B} + \beta \boldsymbol{r}_\mathrm{A}||}^{-1}, \quad {||\boldsymbol{r}_\mathrm{C} + \alpha \boldsymbol{r}_\mathrm{B}||}^{-1} \quad \mathrm{and} \quad {||\boldsymbol{r}_\mathrm{B} + \beta \boldsymbol{r}_\mathrm{A}||}^{-1},
\end{align}
where $\alpha$ and $\beta$, which can be negative, are various mass ratios. We will assume that $r_\mathrm{C} \gg |\alpha| r_\mathrm{B} \gg |\beta| r_\mathrm{A}$ (where $r\equiv ||\boldsymbol{r}||$). In other words, we will assume that the system is sufficiently hierarchical in the sense that it is appropriate to expand the potential in terms of the relative distance ratios $r_\mathrm{A}/r_\mathrm{B}$, $r_\mathrm{B}/r_\mathrm{C}$ and $r_\mathrm{A}/r_\mathrm{C}$, all of which are assumed to be small, and that the mass ratios are not too extreme.
    
We expand the potential in terms of the relative distance ratios using the general expansion
\begin{align}
\nonumber &{||\boldsymbol{r} + \alpha \boldsymbol{r}' + \beta \boldsymbol{r}''||}^{-1} = \frac{1}{r} \left [ 1 -  \alpha \left ( \frac{r'}{r} \right ) \left ( \hat{\boldsymbol{r}}\cdot \hat{\boldsymbol{r}}' \right ) -  \beta \left ( \frac{r''}{r} \right ) \left ( \hat{\boldsymbol{r}}\cdot \hat{\boldsymbol{r}}'' \right ) + \frac{1}{2} \alpha^2 \left ( \frac{r'}{r} \right )^2 \left \{ 3 \left (\hat{\boldsymbol{r}}\cdot \hat{\boldsymbol{r}}' \right )^2 - 1 \right \} + \frac{1}{2} \beta^2 \left ( \frac{r''}{r} \right )^2 \left \{ 3 \left (\hat{\boldsymbol{r}}\cdot \hat{\boldsymbol{r}}'' \right )^2 - 1 \right \} \right . \\
\nonumber &\quad \left. + \alpha \beta \left ( \frac{r'}{r} \right ) \left ( \frac{r''}{r} \right ) \left \{ 3 \left (\hat{\boldsymbol{r}}\cdot \hat{\boldsymbol{r}}' \right) \left (\hat{\boldsymbol{r}}\cdot \hat{\boldsymbol{r}}'' \right ) - \left (\hat{\boldsymbol{r}}'\cdot \hat{\boldsymbol{r}}' \right ) \right \} - \frac{1}{2} \alpha^3 \left ( \frac{r'}{r} \right )^3 \left \{ 5 \left (\hat{\boldsymbol{r}}\cdot \hat{\boldsymbol{r}}' \right )^3 - 3 \left ( \hat{\boldsymbol{r}}\cdot \hat{\boldsymbol{r}}' \right ) \right \} - \frac{1}{2} \beta^3 \left ( \frac{r''}{r} \right )^3 \left \{ 5 \left (\hat{\boldsymbol{r}}\cdot \hat{\boldsymbol{r}}'' \right )^3 - 3 \left ( \hat{\boldsymbol{r}}\cdot \hat{\boldsymbol{r}}'' \right ) \right \} \right. \\
\nonumber &\quad \left. - \frac{1}{2} \alpha^2 \beta \left ( \frac{r'}{r} \right )^2 \left ( \frac{r''}{r} \right ) \left \{ 15 \left (\hat{\boldsymbol{r}}\cdot \hat{\boldsymbol{r}}' \right )^2 \left (\hat{\boldsymbol{r}}\cdot \hat{\boldsymbol{r}}'' \right ) - 3 \left (\hat{\boldsymbol{r}}\cdot \hat{\boldsymbol{r}}'' \right ) - 6 \left (\hat{\boldsymbol{r}}\cdot \hat{\boldsymbol{r}}' \right ) \left (\hat{\boldsymbol{r}}'\cdot \hat{\boldsymbol{r}}'' \right ) \right \} \right. \\
\nonumber &\quad \left. - \frac{1}{2} \alpha \beta^2 \left ( \frac{r'}{r} \right ) \left ( \frac{r''}{r} \right )^2 \left \{ 15 \left (\hat{\boldsymbol{r}}\cdot \hat{\boldsymbol{r}}'' \right )^2 \left (\hat{\boldsymbol{r}}\cdot \hat{\boldsymbol{r}}' \right ) - 3 \left (\hat{\boldsymbol{r}}\cdot \hat{\boldsymbol{r}}' \right ) - 6 \left (\hat{\boldsymbol{r}}\cdot \hat{\boldsymbol{r}}'' \right ) \left (\hat{\boldsymbol{r}}'\cdot \hat{\boldsymbol{r}}'' \right ) \right \} \right. \\
\nonumber &\quad \left. + \frac{1}{8} \alpha^4 \left ( \frac{r'}{r} \right )^4 \left \{3 - 30 \left (\hat{\boldsymbol{r}}\cdot \hat{\boldsymbol{r}}' \right)^2 + 35 \left (\hat{\boldsymbol{r}}\cdot \hat{\boldsymbol{r}}' \right)^4 \right \} + \frac{1}{8} \beta^4 \left ( \frac{r''}{r} \right )^4 \left \{3 - 30 \left (\hat{\boldsymbol{r}}\cdot \hat{\boldsymbol{r}}'' \right)^2 + 35 \left (\hat{\boldsymbol{r}}\cdot \hat{\boldsymbol{r}}'' \right)^4 \right \} \right ] \\
\nonumber &\quad \left. + \frac{1}{2} \alpha^3 \beta  \left ( \frac{r'}{r} \right )^3  \left ( \frac{r''}{r} \right ) \left \{ 35 \left (\hat{\boldsymbol{r}}\cdot \hat{\boldsymbol{r}}' \right )^3 \left (\hat{\boldsymbol{r}}\cdot \hat{\boldsymbol{r}}'' \right ) - 15 \left (\hat{\boldsymbol{r}}\cdot \hat{\boldsymbol{r}}' \right )^2 \left (\hat{\boldsymbol{r}}'\cdot \hat{\boldsymbol{r}}'' \right ) - 15 \left (\hat{\boldsymbol{r}}\cdot \hat{\boldsymbol{r}}' \right ) \left (\hat{\boldsymbol{r}}\cdot \hat{\boldsymbol{r}}'' \right ) + 3 \left (\hat{\boldsymbol{r}}'\cdot \hat{\boldsymbol{r}}'' \right ) \right \}\right. \\
\nonumber &\quad \left. + \frac{1}{2} \alpha \beta^3  \left ( \frac{r'}{r} \right ) \left ( \frac{r''}{r} \right )^3 \left \{ 35 \left (\hat{\boldsymbol{r}}\cdot \hat{\boldsymbol{r}}'' \right )^3 \left (\hat{\boldsymbol{r}}\cdot \hat{\boldsymbol{r}}' \right ) - 15 \left (\hat{\boldsymbol{r}}\cdot \hat{\boldsymbol{r}}'' \right )^2 \left (\hat{\boldsymbol{r}}'\cdot \hat{\boldsymbol{r}}'' \right ) - 15 \left (\hat{\boldsymbol{r}}\cdot \hat{\boldsymbol{r}}' \right ) \left (\hat{\boldsymbol{r}}\cdot \hat{\boldsymbol{r}}'' \right ) + 3 \left (\hat{\boldsymbol{r}}'\cdot \hat{\boldsymbol{r}}'' \right ) \right \} \right. \\
\nonumber &\quad \left. + \frac{1}{4} \alpha^2 \beta^2  \left ( \frac{r'}{r} \right )^2 \left ( \frac{r''}{r} \right )^2 \left \{ 105  \left (\hat{\boldsymbol{r}}\cdot \hat{\boldsymbol{r}}' \right )^2  \left (\hat{\boldsymbol{r}}\cdot \hat{\boldsymbol{r}}'' \right )^2 - 15  \left (\hat{\boldsymbol{r}}\cdot \hat{\boldsymbol{r}}' \right )^2 - 15  \left (\hat{\boldsymbol{r}}\cdot \hat{\boldsymbol{r}}'' \right )^2 - 15  \left (\hat{\boldsymbol{r}}\cdot \hat{\boldsymbol{r}}' \right )  \left (\hat{\boldsymbol{r}}\cdot \hat{\boldsymbol{r}}'' \right )  \left (\hat{\boldsymbol{r}}'\cdot \hat{\boldsymbol{r}}'' \right ) + 3 \left (\hat{\boldsymbol{r}}'\cdot \hat{\boldsymbol{r}}'' \right )^2 + 3 \right \} \right. \\
&\quad \left. + \mathcal{O} \left \{ \left ( \frac{r'}{r}\right)^i \left (\frac{r''}{r}\right)^j \right \} \right ].
\label{eq:app:general_expansion}
\end{align}
Here, $i+j\geq 5$. Substituting this expansion into equation~(\ref{eq:app:Hgen}), we find
\begin{align}
\nonumber H_\mathrm{ts} &= H_\mathrm{bin}(m_1,m_2,\boldsymbol{r}_\mathrm{A},\dot{\boldsymbol{r}}_\mathrm{A}) + H_\mathrm{bin}(m_1+m_2,m_3,\boldsymbol{r}_\mathrm{B},\dot{\boldsymbol{r}}_\mathrm{B}) + H_\mathrm{bin}(m_1+m_2+m_3,m_4,\boldsymbol{r}_\mathrm{C},\dot{\boldsymbol{r}}_\mathrm{C}) \\
\nonumber &\quad + H_\mathrm{quad} (m_1,m_2,m_3,\boldsymbol{r}_\mathrm{A},\boldsymbol{r}_\mathrm{B}) + H_\mathrm{quad} (m_1+m_2,m_3,m_4,\boldsymbol{r}_\mathrm{B},\boldsymbol{r}_\mathrm{C}) + H_\mathrm{quad} (m_1,m_2,m_4,\boldsymbol{r}_\mathrm{A},\boldsymbol{r}_\mathrm{C}) \\
\nonumber &\quad + H_\mathrm{oct} (m_1,m_2,m_3,\boldsymbol{r}_\mathrm{A},\boldsymbol{r}_\mathrm{B}) + H_\mathrm{oct} (m_1+m_2,m_3,m_4,\boldsymbol{r}_\mathrm{B},\boldsymbol{r}_\mathrm{C}) + H_\mathrm{oct} (m_1,m_2,m_4,\boldsymbol{r}_\mathrm{A},\boldsymbol{r}_\mathrm{C}) \\
\nonumber &\quad + H_\mathrm{oct,\,cross}(m_1,m_2,m_3,m_4,\boldsymbol{r}_\mathrm{A},\boldsymbol{r}_\mathrm{B},\boldsymbol{r}_\mathrm{C}) \\
\nonumber &\quad + H_\mathrm{hd}(m_1,m_2,m_3,\boldsymbol{r}_\mathrm{A},\boldsymbol{r}_\mathrm{B}) + H_\mathrm{hd}(m_1+m_2,m_3,m_4,\boldsymbol{r}_\mathrm{B},\boldsymbol{r}_\mathrm{C}) + H_\mathrm{hd}(m_1,m_2,m_4,\boldsymbol{r}_\mathrm{A},\boldsymbol{r}_\mathrm{C}) \\
\nonumber &\quad + H_\mathrm{hd,\,cross,1}(m_1,m_2,m_3,m_4,\boldsymbol{r}_\mathrm{A},\boldsymbol{r}_\mathrm{B},\boldsymbol{r}_\mathrm{C}) + H_\mathrm{hd,\,cross,2}(m_1,m_2,m_3,m_4,\boldsymbol{r}_\mathrm{A},\boldsymbol{r}_\mathrm{B},\boldsymbol{r}_\mathrm{C}) \\
&\quad + \mathcal{O} \left [ \frac{1}{r_\mathrm{C}} \left ( \frac{r_\mathrm{A}}{r_\mathrm{B}}\right)^i \left ( \frac{r_\mathrm{B}}{r_\mathrm{C}}\right)^j \left (\frac{r_\mathrm{A}}{r_\mathrm{C}}\right)^k \right ],
\label{eq:app:H_noav}
\end{align}
where `ts' stands for `triple-star', and $i+j+k\geq 5$. Here, the various functions are given by
\begin{subequations}
\label{eq:app:Hfuncts}
\begin{align}
\label{eq:app:H_bin_non_av}
&H_\mathrm{bin}(m,m',\boldsymbol{r},\dot{\boldsymbol{r}}) = \frac{1}{2} \frac{mm'}{m+m'} \dot{\boldsymbol{r}}^2 - \frac{Gmm'}{r}; \\
\label{eq:app:H_quad_non_av}
&H_\mathrm{quad}(m,m',m'',\boldsymbol{r},\boldsymbol{r}') = -\frac{Gmm'm''}{m+m'} \frac{1}{r'} \left (\frac{r}{r'} \right )^2 \frac{1}{2} \left [ 3 \left (\hat{\boldsymbol{r}}\cdot \hat{\boldsymbol{r}}' \right )^2 - 1 \right ]; \\
\label{eq:app:H_oct_non_av}
&H_\mathrm{oct}(m,m',m'',\boldsymbol{r},\boldsymbol{r}') = -\frac{Gmm'm''(m-m')}{(m+m')^2} \frac{1}{r'} \left (\frac{r}{r'} \right )^3 \frac{1}{2} \left [ 5 \left (\hat{\boldsymbol{r}}\cdot \hat{\boldsymbol{r}}' \right )^3 - 3 \left (\hat{\boldsymbol{r}}\cdot \hat{\boldsymbol{r}}' \right ) \right ]; \\
&H_\mathrm{oct,\,cross}(m_1,m_2,m_3,m_4,\boldsymbol{r}_\mathrm{A},\boldsymbol{r}_\mathrm{B},\boldsymbol{r}_\mathrm{C}) = \frac{Gm_1m_2m_3m_4}{(m_1+m_2)(m_1+m_2+m_3)} \frac{1}{r_\mathrm{C}} \left (\frac{r_\mathrm{A}}{r_\mathrm{C}} \right )^2 \left (\frac{r_\mathrm{B}}{r_\mathrm{C}} \right ) \\
\nonumber &\quad \times \frac{1}{2} \left [ 15 \left ( \hat{\boldsymbol{r}}_\mathrm{B} \cdot \hat{\boldsymbol{r}}_\mathrm{C} \right ) \left ( \hat{\boldsymbol{r}}_\mathrm{A} \cdot \hat{\boldsymbol{r}}_\mathrm{C} \right )^2 - 3 \left ( \hat{\boldsymbol{r}}_\mathrm{B} \cdot \hat{\boldsymbol{r}}_\mathrm{C} \right ) - 6\left (\hat{\boldsymbol{r}}_\mathrm{A} \cdot \hat{\boldsymbol{r}}_\mathrm{C}\right ) \left (\hat{\boldsymbol{r}}_\mathrm{A} \cdot \hat{\boldsymbol{r}}_\mathrm{B}\right ) \right ]; \\
\label{eq:app:H_hd_non_av}
&H_\mathrm{hd}(m,m',m'',\boldsymbol{r},\boldsymbol{r}') = -\frac{Gmm'm''(m^2-mm'+m'^2)}{(m+m')^3} \frac{1}{r'} \left (\frac{r}{r'} \right )^4 \frac{1}{8} \left [ 35 \left (\hat{\boldsymbol{r}}\cdot \hat{\boldsymbol{r}}' \right )^4 - 30  \left (\hat{\boldsymbol{r}}\cdot \hat{\boldsymbol{r}}' \right )^2 + 3 \right ]; \\
\label{eq:app:H_hd_cross1_non_av}
&H_\mathrm{hd,\,cross,1}(m_1,m_2,m_3,m_4,\boldsymbol{r}_\mathrm{A},\boldsymbol{r}_\mathrm{B},\boldsymbol{r}_\mathrm{C}) = -\frac{Gm_1m_2(m_1-m_2)m_3m_4}{(m_1+m_2)^2(m_1+m_2+m_3)} \frac{1}{r_\mathrm{C}} \left (\frac{r_\mathrm{A}}{r_\mathrm{C}} \right )^3 \left (\frac{r_\mathrm{B}}{r_\mathrm{C}} \right ) \\
\nonumber &\quad \times \frac{1}{2} \left [ 3 \left ( \hat{\boldsymbol{r}}_\mathrm{A} \cdot \hat{\boldsymbol{r}}_\mathrm{B} \right ) \left \{ 5 \left ( \hat{\boldsymbol{r}}_\mathrm{A} \cdot \hat{\boldsymbol{r}}_\mathrm{C} \right )^2 - 1 \right \} + 5 \left ( \hat{\boldsymbol{r}}_\mathrm{A} \cdot \hat{\boldsymbol{r}}_\mathrm{C} \right ) \left ( \hat{\boldsymbol{r}}_\mathrm{B} \cdot \hat{\boldsymbol{r}}_\mathrm{C} \right ) \left \{ 3 - 7 \left ( \hat{\boldsymbol{r}}_\mathrm{A} \cdot \hat{\boldsymbol{r}}_\mathrm{C} \right )^2 \right \} \right ]; \\
\label{eq:app:H_hd_cross2_non_av}
&H_\mathrm{hd,\,cross,2}(m_1,m_2,m_3,m_4,\boldsymbol{r}_\mathrm{A},\boldsymbol{r}_\mathrm{B},\boldsymbol{r}_\mathrm{C}) = -\frac{Gm_1m_2m_3^2m_4}{(m_1+m_2)(m_1+m_2+m_3)^2} \frac{1}{r_\mathrm{C}} \left (\frac{r_\mathrm{A}}{r_\mathrm{C}} \right )^2 \left (\frac{r_\mathrm{B}}{r_\mathrm{C}} \right )^2 \\
\nonumber &\quad \times \frac{3}{4} \left [ 1 + 2  \left ( \hat{\boldsymbol{r}}_\mathrm{A} \cdot \hat{\boldsymbol{r}}_\mathrm{B} \right )^2 - 20  \left ( \hat{\boldsymbol{r}}_\mathrm{A} \cdot \hat{\boldsymbol{r}}_\mathrm{B} \right ) \left ( \hat{\boldsymbol{r}}_\mathrm{A} \cdot \hat{\boldsymbol{r}}_\mathrm{C} \right )  \left ( \hat{\boldsymbol{r}}_\mathrm{B} \cdot \hat{\boldsymbol{r}}_\mathrm{C} \right ) - 5  \left ( \hat{\boldsymbol{r}}_\mathrm{B} \cdot \hat{\boldsymbol{r}}_\mathrm{C} \right )^2 + 5  \left ( \hat{\boldsymbol{r}}_\mathrm{A} \cdot \hat{\boldsymbol{r}}_\mathrm{C} \right )^2 \left \{ 7 \left ( \hat{\boldsymbol{r}}_\mathrm{B} \cdot \hat{\boldsymbol{r}}_\mathrm{C} \right )^2-1 \right \} \right ].
\end{align}
\end{subequations}
The function $H_\mathrm{bin}(m,m',\boldsymbol{r},\dot{\boldsymbol{r}})$ is the Hamiltonian for an isolated two-body system with reduced mass $\mu=mm'/(m+m')$. It appears in equation~(\ref{eq:app:H_noav}) for the three binaries A, B and C, and reduces to the binding energy of each these binaries if Kepler orbits are assumed. Therefore, it does not lead to orbital changes. 

The other functions defined in equations~(\ref{eq:app:Hfuncts}) do lead to orbital changes. We associate a term with the `quadrupole' order if the combined power of $(r_\mathrm{A}/r_\mathrm{B})$, $(r_\mathrm{B}/r_\mathrm{C})$ and $(r_\mathrm{A}/r_\mathrm{C})$ is equal to two, to `octupole' order if the combined power is equal to three and `hexadecupole' order if the combined power is equal to four. The functions $H_\mathrm{quad}(m,m',m'',\boldsymbol{r},\boldsymbol{r}')$, $H_\mathrm{oct}(m,m',m'',\boldsymbol{r},\boldsymbol{r}')$ and $H_\mathrm{hd}(m,m',m'',\boldsymbol{r},\boldsymbol{r}')$ are precisely the same functions that appear in the three-body problem. In this case, they each appear in the Hamiltonian three times by replacing the `inner' and `outer' binaries by the combinations AB, BC and AC. 

At the quadrupole level, the three combinations of $H_\mathrm{quad}(m,m',m'',\boldsymbol{r},\boldsymbol{r}')$ are the only terms that appear; the `cross terms' that are present in the expansion in equation~(\ref{eq:app:general_expansion}) cancel. Such a cancellation does not occur at higher orders. At octupole order, we find the `cross term' $H_\mathrm{oct,\,cross}(m_1,m_2,m_3,m_4,\boldsymbol{r}_\mathrm{A},\boldsymbol{r}_\mathrm{B},\boldsymbol{r}_\mathrm{C})$ that is unique to the type of quadruple systems considered here. It depends on all four masses, all three semimajor axes and all three relative orientations between the binary separation vectors, and it is proportional to $(r_\mathrm{A}/r_\mathrm{C})^2(r_\mathrm{B}/r_\mathrm{C})$. At the hexadecupole order, we find two cross terms, $H_\mathrm{hd,\,cross,1} \propto (r_\mathrm{A}/r_\mathrm{C})^3(r_\mathrm{B}/r_\mathrm{C})$ and $H_\mathrm{hd,\,cross,2}\propto (r_\mathrm{A}/r_\mathrm{C})^2(r_\mathrm{B}/r_\mathrm{C})^2$. Equation~(\ref{eq:app:H_noav}) is exact if expanded to infinite order. Here, the expansion is truncated for $i+j+k\geq5$. 

We average the truncated Hamiltonian over the three binary orbits A, B and C, assuming that the Kepler orbit is unperturbed during this time-scale. For orbit $k$, the averaging is defined as
\begin{align}
\langle H \rangle_k \equiv \frac{1}{2\pi} \int_0^{2\pi} H \, \mathrm{d} l_k,
\label{eq:app:avdef}
\end{align}
where $l_k$ is the mean anomaly of orbit $k$. Depending on the sign of the power of $r_k$ it is convenient to use either the true anomaly or eccentric anomaly in equation~(\ref{eq:app:avdef}). If the power of $r_k$ is negative, we use the true anomaly; if it is positive, we use the eccentric anomaly. Formally, averaging the Hamiltonian over the three orbits A, B and C according to equation~(\ref{eq:app:avdef}) is not a canonical transformation. However, applying the Von Zeipel transformation technique to the unaveraged Hamiltonian (\citealt{brouwer_59}), a canonical transformation can be found that eliminates the short-period terms $l_k$ from the Hamiltonian (cf. appendix A2 of \citealt{naoz_ea_13}; the derivation presented there is straightforwardly extended to three, rather than two short-period variables). This transformation leads to a transformed Hamiltonian that is equivalent to the triply-averaged Hamiltonian  $\langle \langle \langle H \rangle_\mathrm{C} \rangle_\mathrm{B} \rangle_\mathrm{A}$ (cf. equation~\ref{eq:app:avdef}; note that the order of integration is arbitrary). Here, the transformed coordinates $\boldsymbol{e}_k^*$ and $\boldsymbol{q}_k^*$ differ from the original ones $\boldsymbol{e}_k$ and $\boldsymbol{q}_k$. However, as noted by \citealt{naoz_ea_13}, the differences between the untransformed and the transformed coordinates contribute to the Hamiltonian only at subleading order. 

We express the angular momenta and orientations of each of the three binaries in terms of the triad of perpendicular orbital state vectors $(\boldsymbol{j}_k,\boldsymbol{e}_k,\boldsymbol{q}_k)$, where $\boldsymbol{q}_k \equiv \boldsymbol{j}_k \times \boldsymbol{e}_k$. Here, $\boldsymbol{j}_k$ is a vector aligned with the angular momentum vector of the orbit with magnitude $j_k=\sqrt{1-e_k^2}$; $\boldsymbol{e}_k$ is the eccentricity or Laplace-Runge-Lenz vector that is aligned with the major axis and with magnitude the orbital eccentricity $e_k$ (see e.g. \citealt{goldstein_75,goldstein_76,goldstein_poolse_safko_02} for historical overviews). In terms of these vectors and the true anomaly, the angle between two instantaneous separation vectors $\boldsymbol{r}_k$ and $\boldsymbol{r}_l$ can be expressed as
\begin{align}
\nonumber \hat{\boldsymbol{r}}_k \cdot \hat{\boldsymbol{r}}_l &= \left [ \cos(f_k) \, \hat{\boldsymbol{e}}_k + \sin(f_k) \, \hat{\boldsymbol{q}}_k \right ] \cdot \left [ \cos(f_l) \, \hat{\boldsymbol{e}}_l + \sin(f_l) \, \hat{\boldsymbol{q}}_l \right ] \\
&= \cos(f_k)\cos(f_l) \left ( \hat{\boldsymbol{e}}_k \cdot \hat{\boldsymbol{e}}_l \right )+ \cos(f_k) \sin(f_l) \left (\hat{\boldsymbol{e}}_k \cdot \hat{\boldsymbol{q}}_l \right ) + \sin(f_k) \cos(f_l) \left ( \hat{\boldsymbol{q}}_k \cdot \hat{\boldsymbol{e}}_l \right ) + \sin(f_k)\sin(f_l) \left ( \hat{\boldsymbol{q}}_k \cdot \hat{\boldsymbol{q}}_l \right ).
\end{align}

Our result of the orbit averaging $\overline{H}_\mathrm{ts} \equiv \langle \langle \langle H_\mathrm{ts} \rangle_\mathrm{C} \rangle_\mathrm{B} \rangle_\mathrm{A}$ is\footnote{The orbit averaging can be carried out in any order of A, B and C, e.g. $\mathrm{C}\rightarrow \mathrm{B} \rightarrow \mathrm{A}$ or $\mathrm{C}\rightarrow \mathrm{A} \rightarrow \mathrm{B}$. Because the integration limits are constants, the result is not affected by the order of integration. }\footnote{We were unable to derive the simplification in $\overline{H}_\mathrm{oct}$ analytically, but we verified it by evaluating both sides numerically. }
\begin{subequations}
\label{eq:app:H_av}
\begin{align}
\label{eq:app:H_av_bin}
&\overline{H}_\mathrm{bin}(m,m',a) = - \frac{Gmm'}{2a}; \\
\label{eq:app:H_av_quad}
&\overline{H}_\mathrm{quad}(m,m',m'',a,a',\boldsymbol{j},\boldsymbol{e},\boldsymbol{j}',\boldsymbol{e}') = \frac{Gmm'm''}{m+m'} \frac{1}{a'} \left (\frac{a}{a'} \right )^2 \frac{1}{8j'^3} \left[ 1-6 e^2 + 15 \left ( \boldsymbol{e} \cdot \unit{j}' \right )^2 - 3 \left(1-e^2\right) \left(\unit{j}\cdot \unit{j}'\right)^2   \right ]; \\
\nonumber &\overline{H}_\mathrm{oct}(m,m',m'',a,a',\boldsymbol{j},\boldsymbol{e},\boldsymbol{j}',\boldsymbol{e}') = -\frac{Gmm'm''(m-m')}{(m+m')^2}  \frac{1}{a'} \left (\frac{a}{a'} \right )^3 \frac{15}{64j'^5} \left [ \left( \boldsymbol{e}\cdot\boldsymbol{e}'\right ) \left \{ 2 e^2 - 9 + 10 \left(1-e^2 \right) \left (\unit{e}\cdot \unit{e}' \right)^2 \right. \right. \\
\nonumber &\quad \quad \quad \left. \left. + \left(10 + 25 e^2\right) \left (\unit{e}\cdot \unit{j}'\right )^2 + 10 \left(1-e^2\right) \left (\unit{e}'\cdot \unit{j}\right)^2 \left [1-\left(\unit{e}\cdot\unit{j}'\right)^2 \right ] + 5 \left(1-e^2\right) \left(\unit{j}\cdot \unit{j}' \right )^2 \left [2 \left(\unit{e}\cdot\unit{e}'\right)^2-1\right ] \right \} \right. \\
\nonumber &\quad \quad \left. - 10\left(1-e^2\right) \left [2 \left(\unit{e} \cdot \unit{e}'\right)^2-1 \right ] \left(\boldsymbol{e}\cdot \unit{j}' \right ) \left ( \boldsymbol{e}'\cdot \unit{j} \right ) \left ( \unit{j} \cdot \unit{j}'\right) \right ] \\
\nonumber &\quad =-\frac{Gmm'm''(m-m')}{(m+m')^2}  \frac{1}{a'} \left (\frac{a}{a'} \right )^3 \frac{15}{64j'^5} \left [ \left ( \boldsymbol{e} \cdot \boldsymbol{e}' \right ) \left \{ 1-8e^2 + 35 \left(\boldsymbol{e}\cdot \unit{j}'\right)^2 \right. \right. \\
&\quad \quad \left. \left. - 5 \left ( 1-e^2 \right ) \left( \unit{j}\cdot \unit{j}' \right )^2 \right \} - 10 \left(1-e^2\right ) \left ( \boldsymbol{e} \cdot \unit{j}' \right ) \left ( \boldsymbol{e}' \cdot \unit{j} \right ) \left ( \unit{j} \cdot \unit{j}' \right ) \right ];
\label{eq:app:H_av_oct}
\end{align}
\begin{align}
\nonumber &\overline{H}_\mathrm{oct,\,cross}(m_1,m_2,m_3,m_4,a_\mathrm{A},a_\mathrm{B},a_\mathrm{C},\boldsymbol{j}_\mathrm{A},\boldsymbol{e}_\mathrm{A},\boldsymbol{j}_\mathrm{B},\boldsymbol{e}_\mathrm{B},\boldsymbol{j}_\mathrm{C},\boldsymbol{e}_\mathrm{C}) = -\frac{Gm_1m_2m_3m_4}{(m_1+m_2)(m_1+m_2+m_3)} \frac{1}{a_\mathrm{C}} \left ( \frac{a_\mathrm{A}}{a_\mathrm{C}} \right )^2 \left ( \frac{a_\mathrm{B}}{a_\mathrm{C}} \right ) \frac{9}{32j_\mathrm{C}^5} \\ 
\nonumber &\quad \times \left [ 2\left(1-e_\mathrm{A}^2\right) \left ( \boldsymbol{e}_\mathrm{B} \cdot \unit{j}_\mathrm{A} \right ) \left ( \boldsymbol{e}_\mathrm{C} \cdot \unit{j}_\mathrm{A} \right ) \left \{4 - 5 \left ( \unit{e}_\mathrm{A} \cdot \unit{j}_\mathrm{C} \right )^2 \right \} -10 \left(\unit{e}_\mathrm{A}\cdot \boldsymbol{e}_\mathrm{C}\right) \left(\unit{e}_\mathrm{A}\cdot \unit{j}_\mathrm{C}\right) \left \{ \left(1+4e_\mathrm{A}^2\right) \left(\boldsymbol{e}_\mathrm{B}\cdot \unit{j}_\mathrm{C} \right) \right. \right. \\
\nonumber &\quad \quad \left. \left. - \left(1-e_\mathrm{A}^2\right) \left(\boldsymbol{e}_\mathrm{B} \cdot \unit{j}_\mathrm{A}\right ) \left(\unit{j}_\mathrm{A}\cdot \unit{j}_\mathrm{C}\right)	\right \} + \left( \boldsymbol{e}_\mathrm{B}\cdot \boldsymbol{e}_\mathrm{C} \right) \left \{ -\left(1-6e_\mathrm{A}^2\right) - 10 \left(1-e_\mathrm{A}^2\right) \left( \unit{e}_\mathrm{C} \cdot \unit{j}_\mathrm{A} \right )^2 - 5 \left (\unit{e}_\mathrm{A} \cdot \unit{j}_\mathrm{C} \right)^2 \right. \right. \\
\nonumber &\quad \quad \left. \left. \times \left [ 5e_\mathrm{A}^2 - 2 \left(1-e_\mathrm{A}^2\right) \left(\unit{e}_\mathrm{C}\cdot \unit{j}_\mathrm{A} \right)^2 \right ] - 20 \left(1-e_\mathrm{A}^2\right) \left(\unit{e}_\mathrm{A}\cdot \unit{e}_\mathrm{C} \right ) \left ( \unit{e}_\mathrm{A}\cdot \unit{j}_\mathrm{C} \right ) \left ( \unit{e}_\mathrm{C}\cdot \unit{j}_\mathrm{A} \right ) \left ( \unit{j}_\mathrm{A}\cdot \unit{j}_\mathrm{C} \right ) + 5 \left(1-e_\mathrm{A}^2\right) \left ( \unit{j}_\mathrm{A}\cdot \unit{j}_\mathrm{C} \right )^2  \right. \right. \\
\nonumber &\quad \quad \left. \left. - 10 \left(1-e_\mathrm{A}^2\right) \left ( \unit{e}_\mathrm{A}\cdot \unit{e}_\mathrm{C} \right )^2 \left [1-\left ( \unit{j}_\mathrm{A}\cdot \unit{j}_\mathrm{C} \right )^2 \right ] \right \} +10 \left( \unit{e}_\mathrm{A}\cdot \boldsymbol{e}_\mathrm{B} \right ) \left \{ \left(1-e_\mathrm{A}^2\right) \left ( \unit{e}_\mathrm{A}\cdot \unit{j}_\mathrm{C} \right ) \left ( \boldsymbol{e}_\mathrm{C}\cdot \unit{j}_\mathrm{A} \right ) \left ( \unit{j}_\mathrm{A}\cdot \unit{j}_\mathrm{C} \right ) \right. \right. \\
&\quad \quad \left. \left. + \left(\unit{e}_\mathrm{A}\cdot \boldsymbol{e}_\mathrm{C} \right ) \left [1-\left(1-e_\mathrm{A}^2\right) \left(\unit{j}_\mathrm{A}\cdot \unit{j}_\mathrm{C}\right )^2 \right ] \right \} \right ]; 
\end{align}
\begin{align}
\label{eq:app:H_av_hd}
\nonumber &\overline{H}_\mathrm{hd}(m,m',m'',a,a',\boldsymbol{j},\boldsymbol{e},\boldsymbol{j}',\boldsymbol{e}') = \frac{Gmm'm''(m^2-mm'+m'^2)}{(m+m')^3}  \frac{1}{a'} \left (\frac{a}{a'} \right )^4 \frac{3}{1024j'^7} \\
\nonumber &\quad \times \left [ 262 + 423 e'^2 - 
      40 e^2 \left\{-38 - 3 e'^2 + 6 e^2 \left ( 9 + 2 e'^2\right ) \right \} - 
      280 \left \{ \left (\unit{e}\cdot \unit{j}' \right )^2  + (\unit{e}'\cdot \unit{j})^2 \right \} + 
      5 \left \{ -441 e^4 \left (2 + e'^2 \right ) \left (e\cdot \unit{j}' \right )^4 \right. \right. \\
      \nonumber &\quad \quad \left. \left. + 
         8 \left \{ 7 e^2 \left (-5 + 6 e^2 \right ) + 
            3 \left (-4 + e^2 + 3 e^4 \right ) e'^2\right \} \left (\unit{e}'\cdot \unit{j} \right )^2 + 
         56 \left (-1 - 5 e^2 + 6 e^4 \right ) \left (2 + 
            3 e'^2\right ) \left ( \unit{e}\cdot \unit{e}'\right) \left(\unit{e}\cdot \unit{j}'\right) \left( \unit{e}'\cdot \unit{j}\right) \left ( \unit{j}\cdot \unit{j}' \right ) \right. \right. \\
            \nonumber &\quad \quad \left. \left. - 
         112 \left (-1 - 5 e^2 + 6 e^4 \right ) \left (2 + 
            3 e'^2\right ) \left (\unit{e}\cdot \unit{e}'\right)^3 \left( \unit{e}\cdot \unit{j}'\right) \left(\unit{e}'\cdot \unit{j}\right) \left(\unit{j}\cdot \unit{j}'\right) - 
         2 \left(1 - e^2\right) \left \{ 10 + 39 e'^2 + 
            6 e^2 \left (38 + 11 e'^2 \right) \right. \right. \right. \\
            \nonumber &\quad \quad \left. \left. \left. - 
            42 \left (1 - e^2 \right) e'^2 \left(\unit{e}'\cdot \unit{j}\right)^2 \right \} \left (\unit{j}\cdot \unit{j}'\right )^2 - 
         21 \left (1 - e^2\right)^2 \left(2 + e'^2\right) \left(\unit{j}\cdot \unit{j}'\right)^4 - 
         56 \left(-1 - 5 e^2 + 6 e^4\right) \left(2 + 
            3 e'^2\right) \left(\unit{e}\cdot \unit{e}'\right)^4 \right. \right. \\
            \nonumber &\quad\quad\quad \left. \left. \times \left \{1 - \left(\unit{j}\cdot \unit{j}'\right)^2\right\} + 
         14 \left(\unit{e}\cdot \unit{j}'\right)^2 \left \{ -26 e^2 + 84 e^4 + 
            3 \left(-2 - e^2 + 6 e^4\right) e'^2 - \left(1 - 
               e^2\right) \left [ \left \{-4 - 6 e'^2 \right. \right. \right. \right. \right. \\
               \nonumber &\quad \quad \left. \left. \left. \left. \left. + 
                  6 e^2 \left(-4 + e'^2\right )\right \} \left (\unit{e}'\cdot \unit{j}\right)^2 - 
               21 e^2 \left(2 + e'^2\right) \left (\unit{j}\cdot \unit{j}' \right )^2 \right ] \right \} + 
         28 \left(\unit{e}\cdot \unit{e}'\right)^2 \left [-2 \left(-1 - 5 e^2 + 6 e^4\right) \left(2 + 
               3 e'^2\right) \left(\unit{e}'\cdot \unit{j}\right)^2 \right. \right. \right. \\
               \nonumber &\quad \quad \left. \left. \left.  + \left(\unit{e}\cdot \unit{j}'\right)^2 \left \{ 4 + 20 e^2 - 
               24 e^4 + 
               3 \left(2 + 10 e^2 + 9 e^4\right) e'^2 + 
               2 \left(-1 - 5 e^2 + 6 e^4\right) \left(2 + 
                  3 e'^2\right) \left(\unit{e}'\cdot \unit{j} \right)^2 \right \} \right. \right. \right. \\
                  &\quad \quad \left. \left. \left. + 
            3 \left \{-2 - 3 e'^2 + 
               2 e^2 \left[-5 + 6 e^2 - 
                  5 \left(1 - e^2\right) e'^2 \right] + \left(1 - e^2\right) \left [ 2 + 
                  3 e'^2 + 
                  e^2 \left(12 + 
                    11 e'^2\right) \right ] \left (\unit{j}\cdot \unit{j}' \right )^2\right\} \right ] \right \}  \right ].
\end{align}
\end{subequations}
In order to simplify the expressions in equation~(\ref{eq:app:H_av}) we repeatedly used a vector identity for the dot product of two vector products, $(\boldsymbol{a}\times\boldsymbol{b})\cdot(\boldsymbol{c}\times\boldsymbol{d}) = (\boldsymbol{a}\cdot \boldsymbol{c})(\boldsymbol{b}\cdot\boldsymbol{d}) - (\boldsymbol{b}\cdot \boldsymbol{c})(\boldsymbol{a}\cdot \boldsymbol{d})$, and the scalar product of two scalar triple products,
\begin{align}
\label{eq:vec_id}
\left [ \left ( \boldsymbol{a}\times \boldsymbol{b} \right ) \cdot \boldsymbol{c} \right ]\left [ \left ( \boldsymbol{d}\times \boldsymbol{e} \right ) \cdot \boldsymbol{f} \right ] = \mathrm{det} \left | \begin{array}{ccc}
\boldsymbol{a} \cdot \boldsymbol{d} & \boldsymbol{a} \cdot \boldsymbol{e} & \boldsymbol{a} \cdot \boldsymbol{f} \\
\boldsymbol{b} \cdot \boldsymbol{d} & \boldsymbol{b} \cdot \boldsymbol{e} & \boldsymbol{b} \cdot \boldsymbol{f} \\
\boldsymbol{c} \cdot \boldsymbol{d} & \boldsymbol{c} \cdot \boldsymbol{e} & \boldsymbol{c} \cdot \boldsymbol{f} 
\end{array} \right |.
\end{align}
Furthermore, we have omitted the explicit expressions for $\overline{H}_\mathrm{hd,cross,1}$ and $\overline{H}_\mathrm{hd,cross,2}$ because they are excessively long. The expressions for $\overline{H}_\mathrm{quad}$ and $\overline{H}_\mathrm{oct}$ are identical to those of \citet{boue_fabrycky_14}, who also adopted a description in terms of vectorial vectors.
     
\subsection{Two binaries orbiting each other}
\label{app:ham:binary_binary}
In this configuration, we assume that bodies 1 and 2 are bound in binary A, bodies 3 and 4 are bound in binary B, and the barycentres of binaries A and B are bound in binary C. Although not explored in the main text, here, we present the formalism that can be used to study this hierarchy. The derivation closely parallels that of \S\,\ref{app:ham:circumstellar_triple}.  We define the following instantaneous separations,
\begin{subequations}
\label{eq:app:rABC_bb}
\begin{align}
\boldsymbol{r}_\mathrm{A} &\equiv \boldsymbol{R}_1 - \boldsymbol{R}_2; \\
\boldsymbol{r}_\mathrm{B} &\equiv \boldsymbol{R}_3 - \boldsymbol{R}_4; \\
\boldsymbol{r}_\mathrm{C} &\equiv \frac{m_1 \boldsymbol{R}_1 + m_2 \boldsymbol{R}_2}{m_1+m_2} - \frac{m_3 \boldsymbol{R}_3 + m_4 \boldsymbol{R}_4}{m_3 + m_4}.
\end{align}
\end{subequations}
Again assuming that the centre of mass (cf. equation~\ref{eq:app:rCM}) is constant, the kinetic energy is given by
\begin{align}
T = \frac{1}{2} \frac{m_1 m_2}{m_1+m_2} \dot{\boldsymbol{r}}_\mathrm{A}^2 + \frac{1}{2} \frac{m_3 m_4}{m_3+m_4} \dot{\boldsymbol{r}}_\mathrm{B}^2  + \frac{1}{2} \frac{(m_1 + m_2)(m_3+m_4)}{m_1+m_2+m_3+m_4} \dot{\boldsymbol{r}}_\mathrm{C}^2.
\end{align}
In contrast to \S\,\ref{app:ham:circumstellar_triple}, we do not assume that $r_\mathrm{A}\ll r_\mathrm{B}\ll r_\mathrm{C}$. Instead, we assume that $r_\mathrm{A}\ll r_\mathrm{C}$ and $r_\mathrm{B}\ll r_\mathrm{C}$, without making any (explicit) assumptions on the relation between $r_\mathrm{A}$ and $r_\mathrm{B}$. Using the general expansion equation~(\ref{eq:app:general_expansion}) we find for the Hamiltonian (cf. equation~\ref{eq:app:Hgen}) in terms of the variables in equation~(\ref{eq:app:rABC_bb})
\begin{align}
\nonumber H_\mathrm{bb} &= H_\mathrm{bin}(m_1,m_2,\boldsymbol{r}_\mathrm{A},\dot{\boldsymbol{r}}_\mathrm{A}) + H_\mathrm{bin}(m_3,m_4,\boldsymbol{r}_\mathrm{B},\dot{\boldsymbol{r}}_\mathrm{B}) + H_\mathrm{bin}(m_1+m_2,m_3+m_4,\boldsymbol{r}_\mathrm{C},\dot{\boldsymbol{r}}_\mathrm{C}) \\
\nonumber &\quad + H_\mathrm{quad} (m_1,m_2,m_3+m_4,\boldsymbol{r}_\mathrm{A},\boldsymbol{r}_\mathrm{C}) + H_\mathrm{quad} (m_3,m_4,m_1+m_2,\boldsymbol{r}_\mathrm{B},\boldsymbol{r}_\mathrm{C}) \\
\nonumber &\quad + H_\mathrm{oct} (m_1,m_2,m_3+m_4,\boldsymbol{r}_\mathrm{A},\boldsymbol{r}_\mathrm{C}) + H_\mathrm{oct} (m_4,m_3,m_1+m_2,\boldsymbol{r}_\mathrm{B},\boldsymbol{r}_\mathrm{C}) \\
\nonumber &\quad + H_\mathrm{hd}(m_1,m_2,m_3+m_4,\boldsymbol{r}_\mathrm{A},\boldsymbol{r}_\mathrm{C}) + H_\mathrm{hd}(m_4,m_3,m_1+m_2,\boldsymbol{r}_\mathrm{B},\boldsymbol{r}_\mathrm{C})  + H_\mathrm{hd,\,cross,bb}(m_1,m_2,m_3,m_4,\boldsymbol{r}_\mathrm{A},\boldsymbol{r}_\mathrm{B},\boldsymbol{r}_\mathrm{C}) \\
&\quad + \mathcal{O} \left [ \frac{1}{r_\mathrm{C}} \left ( \frac{r_\mathrm{A}}{r_\mathrm{B}}\right)^i \left ( \frac{r_\mathrm{B}}{r_\mathrm{C}}\right)^j \left (\frac{r_\mathrm{A}}{r_\mathrm{C}}\right)^k \right ],
\label{eq:app:H_noav_bb}
\end{align}
where `bb' stands for `binary-binary', and $i+j+k\geq 5$. Here, the various functions are the same as in equation~(\ref{eq:app:Hfuncts}). The cross term $H_\mathrm{hd,\,cross,bb}$ is unique to the binary-binary configuration, and is given by
\begin{align}
\nonumber &H_\mathrm{hd,\,cross,bb}(m_1,m_2,m_3,m_4,\boldsymbol{r}_\mathrm{A},\boldsymbol{r}_\mathrm{B},\boldsymbol{r}_\mathrm{C}) = - \frac{Gm_1m_2m_3m_4}{(m_1+m_2)(m_3+m_4)} \frac{1}{r_\mathrm{C}} \left ( \frac{r_\mathrm{A}}{r_\mathrm{C}} \right )^2 \left ( \frac{r_\mathrm{B}}{r_\mathrm{C}} \right )^2 \\
&\quad \times \frac{3}{4} \left [ 1 - 5 \left( \unit{r}_\mathrm{A} \cdot \unit{r}_\mathrm{C} \right )^2 - 5 \left( \unit{r}_\mathrm{B} \cdot \unit{r}_\mathrm{C} \right )^2 + 35 \left( \unit{r}_\mathrm{A} \cdot \unit{r}_\mathrm{C} \right )^2 \left( \unit{r}_\mathrm{B} \cdot \unit{r}_\mathrm{C} \right )^2 + 2 \left( \unit{r}_\mathrm{A} \cdot \unit{r}_\mathrm{B} \right )^2 -20 \left( \unit{r}_\mathrm{A} \cdot \unit{r}_\mathrm{B} \right ) \left( \unit{r}_\mathrm{A} \cdot \unit{r}_\mathrm{C} \right ) \left( \unit{r}_\mathrm{B} \cdot \unit{r}_\mathrm{C} \right )\right ].
\label{eq:app:H_hd_cross_bb}
\end{align}
As might be expected, equation~(\ref{eq:app:H_hd_cross_bb}) is invariant under interchange of the A and B binaries, i.e. $(m_1,m_2,\boldsymbol{r}_\mathrm{A}) \leftrightarrow (m_3,m_4,\boldsymbol{r}_\mathrm{B})$. 

It is interesting to compare the (non-averaged) Hamiltonian between the `triple-single' and `binary-binary' configurations. For the latter, the Hamiltonian is simpler, in the sense that at each order, there are only two, rather than three, terms that depend on the properties of two binaries ($H_\mathrm{quad}$, $H_\mathrm{oct}$ and $H_\mathrm{hd}$ -- for each of these, the combinations AC and BC occur). In addition, `cross terms' appear at octupole order in the `triple-single' configuration, whereas for the `binary-binary' configuration, the lowest order at which these terms appear is the next higher, hexadecupole, order.

The orbit-averaged Hamiltonian is obtained directly by appropriate substitutions in equation~(\ref{eq:app:H_noav_bb}), using the results of equations~(\ref{eq:app:H_av_bin}), (\ref{eq:app:H_av_quad}), (\ref{eq:app:H_av_oct}) and (\ref{eq:app:H_av_hd}). The orbit-average of the cross term equation~(\ref{eq:app:H_hd_cross_bb}) is excessively long, and not included here.

\section{Test of the \textsc{SecularQuadruple} algorithm for three-body systems}
\label{app:three_body_test}
\begin{figure*}
\center
\includegraphics[scale = 0.55, trim = 10mm 0mm 0mm 0mm]{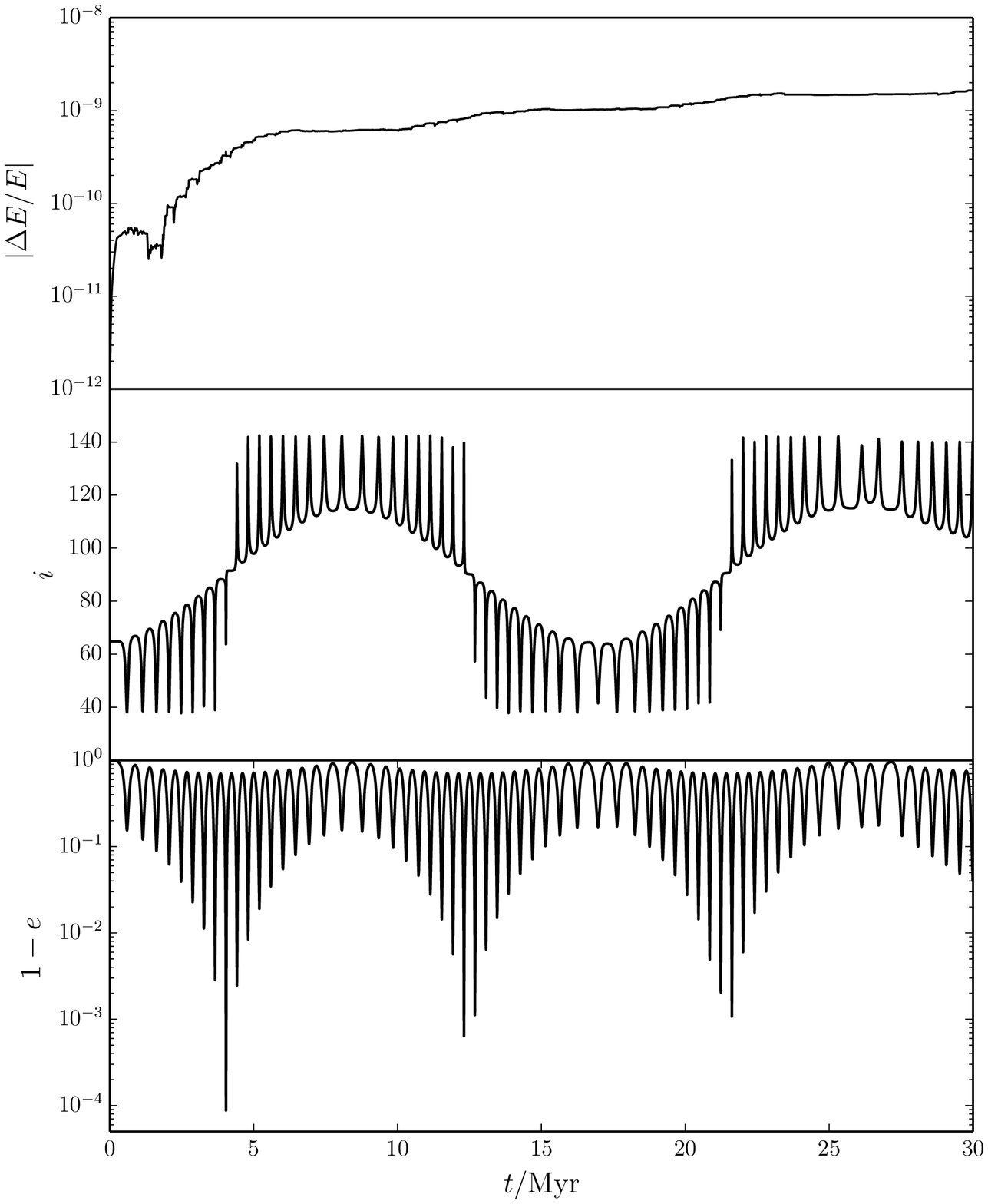}
\caption{\small Test of the \textsc{SecularQuadruple} algorithm for a hierarchical three-body system; the parameters are set to mimic the system of fig. 3 of \citet{naoz_ea_13}. The inner binary consists of a star of mass $1 \, \mathrm{M}_\odot$ and a planet of mass $1\, M_\mathrm{J}$ with semimajor axis $a_\mathrm{in}=6 \, \mathrm{AU}$ and initial eccentricity $e_\mathrm{in} = 0.001$. The outer object is a brown dwarf with mass $40\,M_\mathrm{J}$ and the outer binary has semimajor axis $a_\mathrm{out} = 100 \, \mathrm{AU}$ and initial eccentricity $e_\mathrm{out}  = 0.6$. The binary orbits are initially inclined by $65^\circ$. }
\label{fig:appendix:three_body_test}
\end{figure*}

We tested part of the \textsc{SecularQuadruple} algorithm by comparing to previously obtained integrations for three-body systems. We show an example in Fig. \ref{fig:appendix:three_body_test}, where we assumed the same hierarchical three-body system as in fig. 3 of \citet{naoz_ea_13}. Here, we have both tested applying the triple parameters to the AB systems, choosing a very large value of $a_\mathrm{C}$ (essentially making AB an isolated triple), and to the BC systems, making $a_\mathrm{A}$ very small (essentially reducing binary A to a point mass). The results from \textsc{SecularQuadruple} agree very well with those of \citet{naoz_ea_13}, who performed both integrations based on the orbit-averaged equations, and direct-$N$ body integrations. The maximum relative error of the ODE variables between time-steps was set to $10^{-15}$. Consequently, the Hamiltonian is conserved to high accuracy; $|\Delta E/E|<2\times10^{-9}$ for the integration shown in Fig. \ref{fig:appendix:three_body_test}.

\label{lastpage}
\end{document}